\documentclass[aps,prd,twocolumn,superscriptaddress,nofootinbib]{revtex4-1}

\usepackage{graphicx,color}
\usepackage{latexsym,amsmath,amssymb,graphicx,booktabs}
\usepackage{hyperref}
\definecolor{MyBlue}{rgb}{0.15,0.15,0.70}
\hypersetup{
colorlinks=true,
citecolor=MyBlue,
linkcolor=MyBlue,
urlcolor=MyBlue
}

\newcommand{\nn}{\nonumber}

\newcommand{\iBox}{\Box^{-1}}

\renewcommand\({\left(}
\renewcommand\){\right)}
\renewcommand\[{\left[}
\renewcommand\]{\right]}

\newcommand{\ra}{\rightarrow}

\def\lsim{\raise 0.4ex\hbox{$<$}\kern -0.8em\lower 0.62
ex\hbox{$\sim$}}

\def\gsim{\raise 0.4ex\hbox{$>$}\kern -0.7em\lower 0.62
ex\hbox{$\sim$}}

\def\lbar{{\hbox{$\lambda$}\kern -0.7em\raise 0.6ex
\hbox{$-$}}}

\newcommand\eq[1]{eq.~(\ref{#1})}
\newcommand\eqs[2]{eqs.~(\ref{#1}) and (\ref{#2})}
\newcommand\Eq[1]{Equation~(\ref{#1})}

\newcommand\eqss[3]{eqs.~(\ref{#1}), (\ref{#2}) and (\ref{#3})}

\newcommand\pa{\partial}
\newcommand\p{\partial}

\newcommand\ee{\end{equation}}
\newcommand\be{\begin{equation}}
\def\bea{\begin{array}}
\def\eea{\end{array}}\def\ea{\end{array}}
\newcommand\ees{\end{eqnarray}}
\newcommand\bees{\begin{eqnarray}}
\def\nn{\nonumber}

\def\s{\sigma}

\def\d{\delta}

\def\dslash{\hspace{-1mm}\not{\hbox{\kern-2pt $\partial$}}}
\def\Dslash{\not{\hbox{\kern-2pt $D$}}}
\def\pslash{\not{\hbox{\kern-2.1pt $p$}}}
\def\kslash{\not{\hbox{\kern-2.3pt $k$}}}
\def\qslash{\not{\hbox{\kern-2.3pt $q$}}}

\newcommand{\vk}{{\bf k}}
\newcommand{\vx}{{\bf x}}

\def\p1{{\bf p}_1}
\def\p2{{\bf p}_2}
\def\k1{{\bf k}_1}
\def\k2{{\bf k}_2}

\newcommand{\emn}{\eta_{\mu\nu}}

\newcommand{\gmn}{g_{\mu\nu}}

\newcommand{\hmn}{h_{\mu\nu}}

\newcommand{\hTTij}{h_{ij}^{\rm TT}}

\newcommand{\hatn}{\hat{\bf n}}

\newcommand{\Gmn}{G_{\mu\nu}}

\newcommand{\dddM}{\kern 0.2em \raise 1.9ex\hbox{$...$}\kern -1.0em \hbox{$M$}}
\newcommand{\dddQ}{\kern 0.2em \raise 1.9ex\hbox{$...$}\kern -1.0em \hbox{$Q$}}
\newcommand{\dddI}{\kern 0.2em \raise 1.9ex\hbox{$...$}\kern -1.0em\hbox{$I$}}
\newcommand{\dddJ}{\kern 0.2em \raise 1.9ex\hbox{$...$}\kern-1.0em
\hbox{$J$}}
\newcommand{\dddcalJ}{\kern 0.2em \raise 1.9ex\hbox{$...$}\kern-1.0em
\hbox{${\cal J}$}}

\newcommand{\dddO}{\kern 0.2em \raise 1.9ex\hbox{$...$}\kern -1.0em
\hbox{${\cal O}$}}
\def\dddz{\raise 1.5ex\hbox{$...$}\kern -0.8em \hbox{$z$}}
\def\dddd{\raise 1.8ex\hbox{$...$}\kern -0.8em \hbox{$d$}}
\def\dddbd{\raise 1.8ex\hbox{$...$}\kern -0.8em \hbox{${\bf d}$}}
\def\ddbd{\raise 1.8ex\hbox{$..$}\kern -0.8em \hbox{${\bf d}$}}
\def\dddx{\raise 1.6ex\hbox{$...$}\kern -0.8em \hbox{$x$}}

\newcommand{\Sch}{Schwarzschild }

\newcommand{\mplr}{m_{\rm Pl}}

\newcommand{\ode}{\Omega_{\rm DE}}
\newcommand{\oma}{\Omega_{M}}

\newcommand{\ola}{\Omega_{\Lambda}}

\newcommand{\rde}{\rho_{\rm DE}}
\newcommand{\rgw}{\rho_{\rm gw}}
\newcommand{\wde}{w_{\rm DE}}

\graphicspath{ {./Graphs/} }

\begin{document}


\title{Modified gravitational-wave  propagation and standard sirens }


\author{Enis Belgacem}
\affiliation{D\'epartement de Physique Th\'eorique and Center for Astroparticle Physics, Universit\'e de Gen\`eve, 24 quai Ansermet, CH--1211 Gen\`eve 4, Switzerland}

\author{Yves Dirian}
\affiliation{D\'epartement de Physique Th\'eorique and Center for Astroparticle Physics, Universit\'e de Gen\`eve, 24 quai Ansermet, CH--1211 Gen\`eve 4, Switzerland}

\author{Stefano Foffa}
\affiliation{D\'epartement de Physique Th\'eorique and Center for Astroparticle Physics, Universit\'e de Gen\`eve, 24 quai Ansermet, CH--1211 Gen\`eve 4, Switzerland}

\author{Michele Maggiore}
\affiliation{D\'epartement de Physique Th\'eorique and Center for Astroparticle Physics, Universit\'e de Gen\`eve, 24 quai Ansermet, CH--1211 Gen\`eve 4, Switzerland}



\begin{abstract}

Studies of dark energy  at advanced gravitational-wave (GW) interferometers normally focus on the dark energy equation of state $\wde(z)$. However,  modified gravity theories that predict a non-trivial dark energy equation of state generically also predict deviations from general relativity in the propagation of GWs across cosmological distances, even in theories where the speed of gravity is equal to $c$.  We   find that, in generic modified gravity models, the effect of modified 
GW propagation dominates over that of $\wde(z)$, making modified GW propagation a crucial observable for dark energy studies with standard sirens. We present a convenient parametrization of the effect  in terms of two parameters $(\Xi_0,n)$, analogue to the $(w_0,w_a)$ parametrization of the   dark energy equation of state, and we give  a  limit  from the LIGO/Virgo measurement of $H_0$ with the neutron star binary GW170817.
We then perform a  Markov Chain Monte Carlo analysis to
estimate the sensitivity of the Einstein Telescope (ET) to the cosmological parameters, including $(\Xi_0,n)$, both using only  standard sirens,  and combining them with other cosmological datasets. In particular, the Hubble parameter can be measured with an accuracy better than $1\%$ already using only standard sirens while, when combining ET with current CMB+BAO+SNe data, $\Xi_0$ can be measured to  $0.8\%$ .
We discuss the predictions for modified GW propagation of a specific nonlocal modification of gravity,  recently developed by our group, and we show that they are within the reach of ET. 
Modified GW propagation also affects the GW transfer function,  and therefore  the tensor contribution to the ISW effect.
\end{abstract}


\maketitle

\section{Introduction}

In the last few years the spectacular observations of the gravitational waves (GWs) from binary black-hole coalescences by the LIGO/Virgo collaboration \cite{Abbott:2016blz,Abbott:2016nmj,Abbott:2017vtc,Abbott:2017gyy,Abbott:2017oio}, as well as the observations of the GWs from the binary neutron star merger GW170817~\cite{TheLIGOScientific:2017qsa}, of the associated $\gamma$-ray burst 
\cite{Goldstein:2017mmi,Savchenko:2017ffs,Monitor:2017mdv}, and the follow-up studies of the electromagnetic counterpart 
\cite{GBM:2017lvd} have opened the way for gravitational-wave astrophysics and cosmology.
 
It has long been recognized \cite{Schutz:1986gp} that the detection of GWs from coalescing compact binaries allows us to obtain an absolute measurement of their luminosity distance. Therefore coalescing compact binaries are the GW analogue  of standard candles, or ``standard sirens", as they are usually called.
The standard expression of the luminosity distance as a function of redshift, $d_L(z)$, is 
\be\label{dLem}
d_L(z)=\frac{1+z}{H_0}\int_0^z\, 
\frac{d\tilde{z}}{E(\tilde{z})}\, ,
\ee
where
\be\label{E(z)}
E(z)=\sqrt{\oma (1+z)^3+\rde(z)/\rho_0 }\, ,
\ee
and, as usual, $\rho_0=3H_0^2/(8\pi G)$, $\rde(z)$ is the DE density and $\oma$ is the present matter density fraction (and we have neglected the contribution of radiation, which is negligible at the redshifts relevant for standard sirens). In the limit $z\ll  1$ we recover the Hubble law $d_{L}(z)\simeq H^{-1}_0z$, so from a measurement at such redshifts we can only get information on $H_0$.
This is the case of GW170817, which is at $z\simeq  0.01$. Indeed, from the observation of 
GW170817 has been extracted a value $H_0=70.0^{+12.0}_{-8.0}\,\, {\rm km}\, {\rm s}^{-1}\, {\rm Mpc}^{-1}$~(\cite{Abbott:2017xzu}; see also the updated analysis in~\cite{Abbott:2018wiz}), that rises to $H_0=75.5^{+11.6}_{-9.6}\,\, {\rm km}\, {\rm s}^{-1}\, {\rm Mpc}^{-1}$ if one includes in the analysis a modeling of the broadband X-ray to radio emission to constrain the inclination of the source, as well as a different  estimate of the peculiar velocity of the host galaxy~\cite{Guidorzi:2017ogy}. The cosmological significance of this measurement can be traced to the discrepancy between the local $H_0$ measurement~\cite{Riess:2016jrr,2018ApJ...855..136R} and the value obtained from the {\em Planck} cosmic microwave background (CMB) data~\cite{Planck_2015_CP}, that are in tension at  the $3.7\sigma$ level. While the local measurement is a direct measurement of $H_0$, independent of the cosmological model, the CMB data can be translated into a measurement  of $H_0$ only by assuming a cosmological model, and  performing Bayesian parameter estimation for the parameters of the given model. The $3.7\sigma$ discrepancy occurs if one  assumes a standard  $\Lambda$CDM model. Thus, this tension could be a signal of deviations from $\Lambda$CDM. 
The current accuracy on $H_0$ from the measurement with the single standard siren GW170817 is not accurate enough to discriminate between the local measurements and the {\em Planck}  value. However, each standard siren provides an independent measurement of $H_0$, so with $N$ standard sirens with comparable signal-to-noise ratio the error scales approximately as $1/\sqrt{N}$. The analysis of  \cite{Chen:2017rfc,Feeney:2018mkj} indicates that with about 50-100 standard sirens one could discriminate between the local measurement and the {\em Planck}/$\Lambda$CDM value.

The next generation of GW interferometers, such as the space interferometer LISA~\cite{Audley:2017drz}, which is expected to fly by 2034, as well as third-generation ground-based interferometers currently under study, such as the Einstein Telescope (ET) \cite{Sathyaprakash:2012jk} in Europe and  Cosmic Explorer in the US,  will have the ability to detect standard sirens at much higher redshifts. The information that one could get is then potentially much richer,  since the result is now in principle sensitive to  the dark energy (DE) density $\rde(z)$ or, equivalently, to the   DE equation of state (EoS) $\wde(z)$. Several studies have been performed to investigate the accuracy that one could obtain in this way on the DE EoS
~\cite{Dalal:2006qt,MacLeod:2007jd,Nissanke:2009kt,Cutler:2009qv,Sathyaprakash:2009xt,Zhao:2010sz,DelPozzo:2011yh,Nishizawa:2011eq,Taylor:2012db,Camera:2013xfa,Tamanini:2016zlh,Caprini:2016qxs,Cai:2016sby}.

In $\Lambda$CDM $\rde(z)/\rho_0=\ola$ is a constant, while in a generic modified gravity model it will be a non-trivial function of $z$. The evolution of the DE density is determined by its EoS function $\wde(z)$ through the conservation equation
\be
\dot{\rho}_{\rm DE}+3H(1+\wde)\rde=0\, ,
\ee
which implies 
\be\label{4rdewdeproofs}
\rde(z)/\rho_0  =\Omega_{\rm DE}\exp\left\{ 3\int_{0}^z\, \frac{d\tilde{z}}{1+\tilde{z}}\, [1+\wde(\tilde{z})]\right\}\, ,
\ee
where $\Omega_{\rm DE}=\rde(0)/\rho_0$.
Studies of standard sirens usually assume a simple phenomenological parametrization of $\wde(z)$, given just by a constant $\wde(z)=w_0$, resulting in the $w$CDM model, or  use the $(w_0,w_a)$ parametrization \cite{Chevallier:2000qy,Linder:2002et}
\be\label{w0wa}
w_{\rm DE}(z)= w_0+\frac{z}{1+z} w_a\, ,
\ee 
and then provide forecasts on  $w_0$, or on  $(w_0,w_a)$~\cite{Dalal:2006qt,MacLeod:2007jd,Nissanke:2009kt,Cutler:2009qv,Sathyaprakash:2009xt,Zhao:2010sz,DelPozzo:2011yh,Nishizawa:2011eq,Taylor:2012db,Camera:2013xfa,Tamanini:2016zlh,Caprini:2016qxs}, or else try to reconstruct the whole function $\wde(z)$~\cite{Cai:2016sby}.

In this paper, elaborating on results presented in \cite{Belgacem:2017ihm} (see also \cite{Belgacem:2017cqo}) we perform a more complete analysis of the predictions of generic modified gravity models for standard sirens. In general, if the dark energy sector of a theory differs from a simple cosmological constant, this affects both the background evolution and the cosmological perturbations. The change in the background evolution is expressed by a non-trivial DE EoS $\wde(z)$. Cosmological perturbations will also be affected,  both in the scalar and in the tensor sector (vector perturbations usually only have decaying modes and are  irrelevant, in GR as well as in typical  modified gravity models).  In particular, modifications in the tensor sector can be very important for standard sirens and, as we will see, their effect on the luminosity distance can be more easily observable than that due to a non-trivial DE  EoS. 

The paper is organized as follows. In Section~\ref{sect:modprop} we discuss the structure of cosmological perturbations in modified gravity theories, studying in particular the tensor sector. 
We will see, following previous works, that  the effect of modified GW propagation is described by a function $\delta(z)$ that modifies the friction term  in the propagation equation of GWs over a cosmological background.
We will study in some detail this effect  using as an example an explicit modified gravity model, the RR model, proposed and developed in the last few years by our group (based on the addition of a nonlocal term to the quantum effective action), which is very predictive and fits very well the current cosmological observations, but the arguments that we will present are more general. 
This explicit example will also allow us to propose a simple parametrization of the function $\delta(z)$, or, better yet, directly of the  ratio between the  gravitational and electromagnetic luminosity distances,  in terms of  a pair of parameters $(\Xi_0,n)$, that complements the pair $(w_0,w_a)$ that parametrizes the modification of the background evolution. We will see that, among these four parameters, $\Xi_0$ can be the most important for observational purposes with standard sirens, so in a theory with modified GW propagation a minimal truncation of this parameter space should be to the pair $(\Xi_0,w_0)$. 
In Section~\ref{sect:Geff} we discuss the relation between modified GW propagation and the effective Newton constant that appears in the equation for scalar perturbations in modified gravity.
In Section~\ref{sect:GW170817} we show that standard sirens at low redshift are sensitive to the value of $\delta(z=0)$, and we find that the LIGO/Virgo measurement of the luminosity distance to  GW170817 already gives a limit on this quantity.
In Section~\ref{sect:w0waXi0},  using  the Markov Chain Monte Carlo (MCMC) method,
we study the accuracy with which we can  measure $w_0$, $w_a$, $\Xi_0$ and $n$, in different combinations,  using the estimated sensitivity of ET  and combining it with  {\em Planck\,} CMB data, supernovae (SNe) and baryon acoustic oscillations (BAO) to reduce the degeneracies between these parameters and $H_0$ and $\oma$. 
In  Section~\ref{sect:RRatET} we turn to a concrete model,  rather than just a phenomenological parametrization, studying   the perspectives for discriminating our nonlocal modification of gravity from $\Lambda$CDM, as a function of the number of standard sirens observed. In Section~\ref{sect:transfer} we will study further observable effects related to modified GW propagation, due to the modification of the transfer function that connects a primordial GW spectrum to that observed at later epochs.
Section~\ref{sect:concl} contains our conclusions. We use units $c=1$ and the signature $\emn=(-,+,+,+)$.

\section{Cosmological perturbations and GW propagation in modified gravity} \label{sect:modprop}

Models such as $w$CDM, or the $(w_0,w_a)$ parametrization (\ref{w0wa}), are not fundamental theories of dark energy but just phenomenological parametrizations, assumed to catch the effects coming from some fundamental theory that modifies general relativity (GR) at cosmological scales. Once a fundamental theory is  specified, however, its consequences are much richer. First of all, the theory will modify the cosmological evolution at the background level. This will usually be equivalent to introducing an extra form of energy density $\rde(z)$, or equivalently a DE EoS $\wde(z)$, and this might (or might not) be caught by a simple expression  such as the $(w_0,w_a)$ parametrization.  On top of this, a specific modified gravity model will also generate cosmological perturbations that differ from those of $\Lambda$CDM.

In GR,  scalar perturbations are expressed in terms of the two gauge-invariant Bardeen potentials $\Psi(t,\vx)$ and $\Phi(t,\vx)$, or, equivalently, in terms of their Fourier modes $\Psi_{\vk}(t)$ and $\Phi_{\vk}(t)$. In a modified gravity theory the  equations obeyed by these potential are modified, and, much as one introduces $\wde(z)$ at the background level, one can define some functions that describe the deviation of the evolution of scalar perturbations from that in GR. However, at the level of perturbations, these functions could in principle depend both on time $t$ (or, equivalently, on redshift $z$) and on   $k=|\vk|$, where $\vk$ is  the wavenumber $\vk$ of the mode. For instance, a commonly used  parametrization is \cite{Daniel:2010ky,Amendola:2007rr}
\bees
\Psi_{\vk}(z)&=&[1+\mu(z;k)]\Psi_{\vk}^{\rm GR}(z)\, ,\label{defmu}\\
\Psi_{\vk}(z)-\Phi_{\vk}(z)&=&[1+\Sigma(z;k)] [\Psi_{\vk}(z)-\Phi_{\vk}(z)]^{\rm GR}\label{defSigma}\, ,
\ees
where $\mu$ and $\Sigma$ are, a priori,  functions of  $z$ and $k$, and  the superscript ``GR" denotes the same quantities computed in GR, assuming a $\Lambda$CDM model with the same value of $\oma$ as the modified gravity model. This parametrization is convenient because it  separates  the modifications to the motion of non-relativistic particles, which is described by $\mu$, from the modification to light propagation, which is encoded in 
$\Sigma$. Therefore $\mu$ affects  cosmological structure formation and $\Sigma$ affects  lensing. 
Alternatively, one can use an effective Newton constant $G_{\rm eff}(z;k)$, defined so that, for modes well inside the horizon,  the Poisson equation for $\Phi$ becomes formally the same as in GR, with $G$ replaced by $G_{\rm eff}(z;k)$, together with a second indicator such as $[\Psi_{\vk}(z)+\Phi_{\vk}(z)]/\Phi_{\vk}(z)$ that, in GR, in the absence of anisotropic stresses, vanishes, but can be non-vanishing in modified gravity theories~ \cite{Amendola:2007rr,Kunz:2012aw}. For modes well inside the horizon, for typical modified gravity models the functions $\mu(z;k)$, $\Sigma(z;k)$ or $G_{\rm eff}(z;k)$  are actually independent of $k$. Indeed, in the absence of another explicit length scale in the cosmological model,  for dimensional reasons all dependence on $k$ in the recent cosmological epoch will be through the ratio $\lambda_k/H^{-1}_0$ (where $\lambda_k=2\pi/k$), which is extremely small. Thus, any dependence of $\mu(z;k)$ and $\Sigma(z;k)$ on $k$ can be expanded in powers of $\lambda_k/H^{-1}_0$ [or, in fact, more typically in powers of $(\lambda_k/H^{-1}_0)^2$], and for modes well inside the horizon we can stop  to the zeroth-order term.

When studying standard sirens we are rather interested in the modification of the perturbation equations in the tensor sector, i.e. in the modification of the propagation equation of GWs over the cosmological background. Let us recall that, in GR,  tensor perturbations over a Friedmann-Robertson-Walker (FRW) background satisfy 
\be\label{4eqtensorsect}
\tilde{h}''_A+2{\cal H}\tilde{h}'_A+k^2\tilde{h}_A=16\pi G a^2\tilde{\s}_A\, ,
\ee
where $\tilde{h}_A(\eta, \vk)$ are  the Fourier modes of the GW amplitude, $A=+,\times$ labels the two polarizations, the prime denotes the derivative with respect to cosmic time $\eta$,  $a(\eta)$ is the scale factor,
${\cal H}=a'/a$, and the source term $\tilde{\s}_A(\eta, \vk)$ is related to the helicity-2 part of the anisotropic stress tensor (see e.g.~\cite{Maggiore:2018zz}). In a generic modified gravity model each term in this equation could a priori be modified by a function of redshift and wavenumber. A modification to the source term  would induce a change in the production mechanism and therefore in the phase of an inspiralling binary. To understand the effect of  changes in the terms $2{\cal H}\tilde{h}'_A$ or $k^2\tilde{h}_A$ let us consider first   the free propagation in GR (we follow  the discussion in \cite{Belgacem:2017ihm,Belgacem:2017cqo}).  
We then set $\tilde{\s}_A=0$ and we  introduce a field $\tilde{\chi}_A(\eta, \vk)$ from
\be\label{4defhchiproofs}
\tilde{h}_A(\eta, \vk)=\frac{1}{a(\eta)}  \tilde{\chi}_A(\eta, \vk)\, .
\ee
Then \eq{4eqtensorsect} becomes
\be\label{4propchiproofs1}
\tilde{\chi}''_A+\(k^2-\frac{a''}{a}\) \tilde{\chi}_A=0\, .
\ee
For modes well inside the horizon, such as the GWs targeted by ground-based and space-born detectors, the term $a''/a$ is totally negligible with respect to $k^2$,\footnote{More precisely, for GWs from astrophysical sources with frequencies in the range of ground-based  interferometers, $(a''/a) k^{-2}$ corresponds to 
an effective change of the propagation speed $\Delta c/c=O(10^{-41})$ [with a weak time dependence: since, in MD, $a''/a\propto \eta^{-2}\propto 1/a=1+z$, $\Delta c/c$ changes  by a factor $1+z=O(1)$ in the  propagation from the source at redshift $z$ to us]. Even if this gives rise to an integrated effect over the propagation time, still this is totally negligible compared to the bound $|c_{\rm gw}-c|/c< O(10^{-15})$ from GW170817, which of course also comes from an effect integrated over the propagation time. For wavelengths comparable to the horizon size, for which the term  $(a''/a) k^{-2}$ is not so small,  one can use a WKB approximation, as in \cite{Nishizawa:2017nef}.} and we get a standard wave equation that tells us that GWs propagate at the speed of light.   

In contrast, the factor $1/a$ in \eq{4defhchiproofs}, that was inserted to get rid of the ``friction" term proportional to $\tilde{h}'_A$ in \eq{4eqtensorsect}, 
tells us how the GW amplitude decreases in the propagation across cosmological distances, from the source to the observer. In particular,  for inspiraling binaries this factor combines with other factors coming from the transformation of masses and frequency from the source frame to the detector frame, to produce  the usual  dependence of the GW amplitude from the luminosity distance,
\be
\tilde{h}_A(\eta, \vk)\propto \frac{1}{d_L(z)}\, ,
\ee
see e.g.  Section 4.1.4 of \cite{Maggiore:1900zz}. 
From this discussion we see that
tampering with the coefficient of the $k^2\tilde{h}_A$ term in \eq{4eqtensorsect} is  very dangerous, since this would modify the speed of propagation of GWs. This is by now excluded,
at the level  $|c_{\rm gw}-c|/c< O(10^{-15})$, by the observation of  GW170817/GRB~170817A \cite{Monitor:2017mdv}, and indeed this has ruled out a large class of scalar-tensor and vector-tensor modifications of GR~\cite{Creminelli:2017sry,Sakstein:2017xjx,Ezquiaga:2017ekz,Baker:2017hug}.
We next study the effect of modifying the coefficient of the friction term
$2{\cal H}\tilde{h}'_A$.
We then consider the propagation equation 
\be\label{prophmodgrav}
\tilde{h}''_A  +2 {\cal H}[1-\delta(\eta)] \tilde{h}'_A+k^2\tilde{h}_A=0\, ,
\ee
where $\delta(\eta)$ is a function that parametrizes the deviation from GR, and that we have taken to be  independent of the wavenumber.  In this case, to eliminate the friction term, we must  introduce $\tilde{\chi}_A(\eta, \vk)$ from 
\be\label{4defhchiproofsRR}
\tilde{h}_A(\eta, \vk)=\frac{1}{\tilde{a}(\eta)}  \tilde{\chi}_A(\eta, \vk)\, ,
\ee
where 
\be\label{deftildea}
\frac{\tilde{a}'}{\tilde{a}}={\cal H}[1-\delta(\eta)]\, .
\ee
Then we get 
\be
\tilde{\chi}''_A+\(k^2-\frac{\tilde{a}''}{\tilde{a}}\) \tilde{\chi}_A=0\, .
\ee 
Once again, inside the horizon the term $\tilde{a}''/\tilde{a}$ is totally negligible, so   GWs propagate at the speed of light. However, now the amplitude of $\tilde{h}_A$ is proportional to $1/\tilde{a}$ rather than $1/a$. As a result,
rather than being just proportional to $1/d_L(z)$, the GW amplitude observed today, after the propagation from the source to the observer, will have decreased by a factor 
\be\label{tildeaem}
\frac{\tilde{a}_{\rm emis}}{\tilde{a}_{\rm obs}}\equiv \frac{\tilde{a}(z)}{\tilde{a}(0)}
\ee
 instead of a factor $a_{\rm emis}/a_{\rm obs}=a(z)/a(0)$, where the labels refer to the emission time (at redshift $z$) and the observation time, at redshift zero, respectively. Therefore
\be\label{dLtilde}
\tilde{h}_A\propto  \frac{\tilde{a}(z)}{\tilde{a}(0) }\, \frac{a(0)}{a(z) }\, 
\frac{1}{d_L(z)}=
 \frac{\tilde{a}(z)}{a(z)} \frac{1}{d_L(z)}
\, ,
\ee
where $d_L(z)\equiv d_L^{\,\rm em}(z)$ is the usual notion of luminosity distance appropriate for electromagnetic signals  and, since only the ratios $\tilde{a}(z)/\tilde{a}(0)$ and $a(z)/a(0)$ enter, without loss of generality we can choose the normalizations $\tilde{a}(0)=a(0)=1$.
Then, we see that in such a modified gravity model we must in general distinguish between the usual 
luminosity distance appropriate for electromagnetic signal, $d_L^{\,\rm em}(z)$, which is given by
\eqs{dLem}{E(z)},  and a GW luminosity distance $d_L^{\,\rm gw}(z)$, 
with 
\be\label{dgwaadem}
d_L^{\,\rm gw}(z)=\frac{a(z)}{\tilde{a}(z)}\, d_L^{\,\rm em}(z)\, .
\ee
Standard sirens measure $d_L^{\,\rm gw}(z)$ rather than $d_L^{\,\rm em}(z)$.
\Eq{deftildea} can be rewritten as
 \be
 (\log a/\tilde{a})'=\delta(\eta) {\cal H}(\eta)\, ,
 \ee 
 whose integration gives~\cite{Belgacem:2017ihm,Belgacem:2017cqo}
\be\label{dLgwdLem}
d_L^{\,\rm gw}(z)=d_L^{\,\rm em}(z)\exp\left\{-\int_0^z \,\frac{dz'}{1+z'}\,\delta(z')\right\}\, .
\ee
To sum up, in modified gravity all terms in \eq{4eqtensorsect} can in principle be different from GR.
A modification of the source term  affects the phase of the binary waveforms; the recent BH--BH observations, in particular of
GW150914 and GW151226, have set some limit on such modifications, although for the moment not very stringent~\cite{TheLIGOScientific:2016pea}.
A modification of the $k^2\tilde{h}_A$ term changes the speed of gravity, and is now basically excluded. 
A modification of the $2{\cal H}\tilde{h}'_A$ term changes the amplitude of the GW signal received from a source at cosmological distance. This is particularly interesting because it implies that the luminosity distance measured with standard sirens is in principle different from that measured with standard candles or other electromagnetic probes such as CMB or BAO, and this could
provide a ``smoking gun" signature of modified gravity.

\begin{figure}[t]
\includegraphics[width=0.4\textwidth]{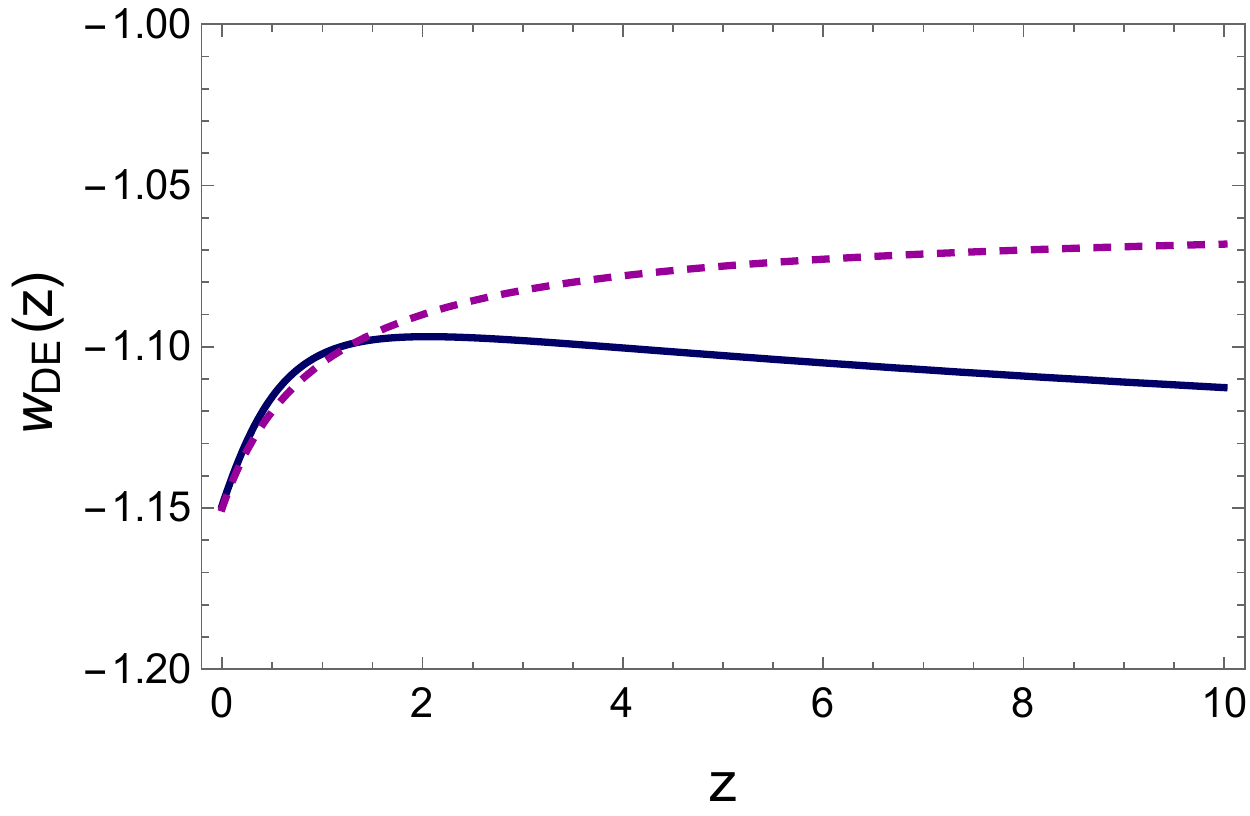}
\caption{The DE EoS in  the RR nonlocal model with $u_0=0$ (blue solid line). By comparison  we also show  the $(w_0,w_a)$ parametrization with the values $w_0=-1.15$ and $w_a=0.09$ (magenta dashed line), obtained by fitting  the function $\wde(a)$ to the parametrization
(\ref{w0wa}) near $a=1$.}
\label{fig:wdeRR}
\end{figure}

One might wonder how likely it is that a modified gravity theory predicts a non-vanishing  $\delta(\eta)$, while still leaving untouched the term $k^2\tilde{h}_A$. 
Actually, as observed in  \cite{Belgacem:2017cqo,Belgacem:2017ihm}, this is just what happens   in  an explicit model in which gravity is modified by the addition of a nonlocal term proportional to $m^2R\Box^{-2}R$, the so-called RR model~\cite{Maggiore:2013mea,Maggiore:2014sia}. The field theoretical motivations for this model, as well as its cosmological predictions,  have been reviewed in detail in refs.~\cite{Maggiore:2016gpx,Belgacem:2017cqo}, whom we refer the interested reader. We will come back to this model in Section~\ref{sect:RRatET}. For the present discussion, the relevant point is that the model features a nonlocal term  (which is assumed to emerge, at the level of the quantum effective action, because of  infrared quantum effects), that behaves effectively as a dark energy, with a phantom equation of state shown in Fig.~\ref{fig:wdeRR}. The model is quite predictive and (in its simpler version, with a parameter $u_0$ set to zero, see Section~\ref{sect:RRatET}) has the same number of parameters as $\Lambda$CDM, with the cosmological constant replaced by the mass parameter $m$ that appears in the term $m^2R\Box^{-2}R$;  assuming flatness, both the parameter $m$ in the RR model and the cosmological constant in $\Lambda$CDM can be taken as derived parameters, so the free parameters in the two models are eventually the same. The cosmological perturbations of the model have been worked out~\cite{Dirian:2014ara} and turn out to be stable, and quite close to those of $\Lambda$CDM. The perturbations have then been inserted in a modified Einstein-Boltzmann code and compared to CMB, BAO, SNe and structure formation data~\cite{Dirian:2014bma,Dirian:2016puz,Dirian:2017pwp,Belgacem:2017cqo}. The result is that   the model fits the cosmological observations at a level statistically indistinguishable from $\Lambda$CDM. Furthermore, parameter estimation gives a larger value of the Hubble parameter, that significantly reduces  the tension  with local $H_0$ measurements, and provides a prediction for the sum of the neutrino masses consistent with oscillation experiments.

\begin{figure}[t]
\includegraphics[width=0.4\textwidth]{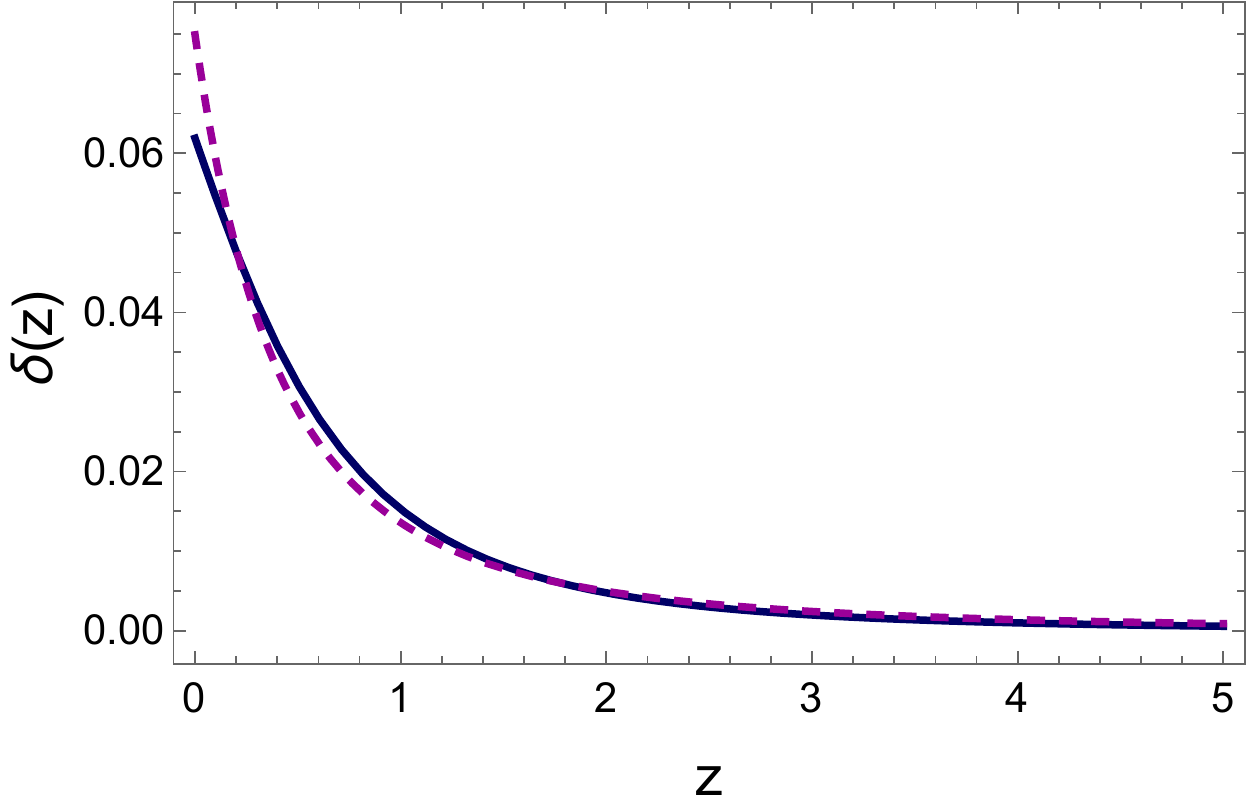}
\caption{The function $\delta(z)$ in  the RR nonlocal model (blue solid line) and the fit given by \eq{paramdeltaz}.}
\label{fig:delta_vs_z}
\end{figure}

In the  RR model the equation governing the tensor perturbations turns out to be~\cite{Dirian:2016puz,Belgacem:2017ihm,Belgacem:2017cqo}
\be\label{prophmodgravsource}
\tilde{h}''_A  +2 {\cal H}[1-\delta(\eta)] \tilde{h}'_A+k^2\tilde{h}_A=16\pi G a^2\tilde{\s}_A\, .
\ee
We see that the speed of gravity is unchanged.  The function $\delta(\eta)$ is predicted explicitly by the model (see \cite{Belgacem:2017ihm,Belgacem:2017cqo}), and is shown  as a function of redshift as the blue solid line  in Fig.~\ref{fig:delta_vs_z}. At large redshifts it goes to zero because, in this model, the modifications from GR only appear close to the recent cosmological epoch. 

\begin{figure}[t]
\includegraphics[width=0.4\textwidth]{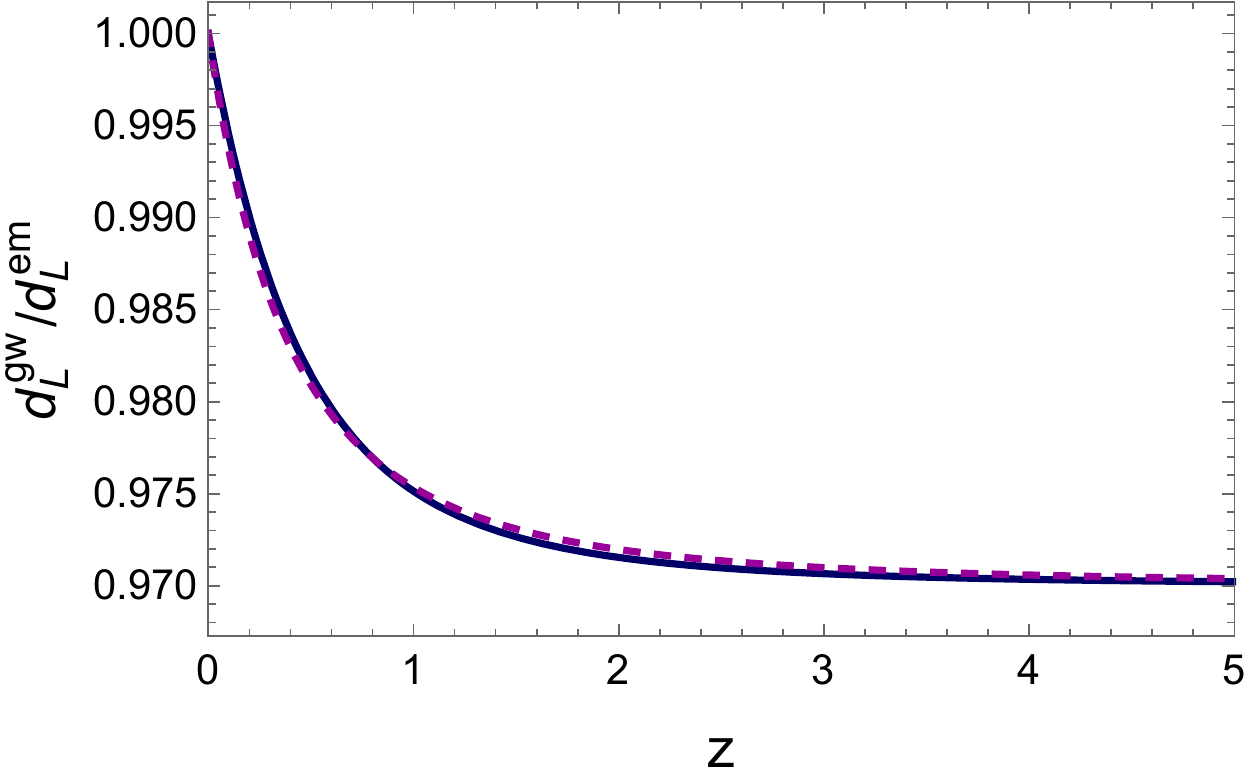}
\caption{The ratio $d_L^{\,\rm gw}(z)/d_L^{\,\rm em}(z)$ in  the RR nonlocal model, as a function of redshift (blue solid line) and the fit given by \eq{eq:fitz}.}
\label{fig:dLgw_over_dLem}
\end{figure}

The corresponding ratio of the gravitational and electromagnetic luminosity distances, obtained from
\eq{dLgwdLem}, is shown as the blue solid line in Fig.~\ref{fig:dLgw_over_dLem}. The fact that $\delta(z)$ goes to zero at large $z$ implies that $d_L^{\,\rm gw}(z)/d_L^{\,\rm em}(z)$ saturates to a constant value, while $d_L^{\,\rm gw}(z=0)/d_L^{\,\rm em}(z=0)=1$ because there can be no effect from modified propagation when the redshift of the source goes to zero.
The red dashed line in  Fig.~\ref{fig:dLgw_over_dLem}  shows the  simple fitting function
\be\label{eq:fitz}
\frac{d_L^{\,\rm gw}(z)}{d_L^{\,\rm em}(z)}
=\Xi_0 +\frac{1-\Xi_0}{(1+z)^n}
\, .
\ee
with $n=5/2$ and $\Xi_0=0.970$. Equivalently, in terms of the value of the scale factor corresponding to the source redshift, we have ${d_L^{\,\rm gw}(a)}/{d_L^{\,\rm em}(a)}=\Xi_0 +a^n (1-\Xi_0)$.

Observe that the parametrization (\ref{eq:fitz}) is such that,  at large redshift, 
$d_L^{\,\rm gw}(z)/d_L^{\,\rm em}(z)$ goes to the constant value $\Xi_0$, while,  
at $z=0$, $d_L^{\,\rm gw}(z)/d_L^{\,\rm em}(z)=1$. This simple parametrization reproduces the exact function extremely well [indeed, in this model it performs much better than the $(w_0,w_a)$ parametrization for the equation of state, compare with Fig.~\ref{fig:wdeRR}]. 

One can find the corresponding  parametrization for the function $\delta(z)$ observing,  from \eq{dLgwdLem}, that
\be
\delta(z)=-(1+z)\frac{d}{dz}\log\( \frac{d_L^{\,\rm gw}(z)}{d_L^{\,\rm em}(z)} \)
\, .
\ee
Then, in terms of $\delta(z)$, the parametrization (\ref{eq:fitz})  reads
\be\label{paramdeltaz}
\delta(z)=\frac{n  (1-\Xi_0)}{1-\Xi_0+ \Xi_0 (1+z)^n}
\, .
\ee
This fitting function is shown as the red dashed line in Fig.~\ref{fig:delta_vs_z}, for the same values of $\Xi_0$ and $n$ as in Fig.~\ref{fig:dLgw_over_dLem}.
Note  that, for this parametrization, 
\be\label{deltazeta0}
\delta(z=0)=n(1-\Xi_0)\, ,
\ee
but near $z=0$ the fitting formula for $\delta(z)$ is no longer very accurate.
We observe that it is more convenient to parametrize directly the ratio
${d_L^{\,\rm gw}(z)}/{d_L^{\,\rm em}(z)}$, rather than the function $\delta(z)$. Indeed, 
the ratio of the luminosity distances is the quantity that can be directly compared to observations, and, at least in the RR model, it is fitted with remarkable accuracy  by the very simple expression given in \eq{eq:fitz}.

Note that this parametrization  also includes the case in which $\delta(z)$ is actually independent of  time (i.e. of the source redshift), which is obtained by setting   $\Xi_0=0$. Then, from \eq{paramdeltaz}, the dependence on $(1+z)$ disappears and $\delta(z)=n$ while, from \eq{eq:fitz}, 
\be\label{paramXi0zero}
\frac{d_L^{\,\rm gw}(z)}{d_L^{\,\rm em}(z)}=\frac{1}{(1+z)^n}\, ,\qquad (\Xi_0=0)\, .
\ee 
The case of constant $\delta$ has been considered in the literature, at the phenomenological level~\cite{Nishizawa:2017nef}. Note however that in our model, as in any other modified gravity model in which an effective dark energy  density only kicks in near the recent cosmological epoch, $\delta(z)$ goes to zero at large redshift, and $d_L^{\,\rm gw}(z)/d_L^{\,\rm em}(z)$ saturates to a non-vanishing value $\Xi_0$.

%
%

A modified propagation equation of the form (\ref{prophmodgrav}) has also been previously found in other modified gravity models. To the best of our knowledge this was first observed 
in~\cite{Deffayet:2007kf} within the 
DGP model~\cite{Dvali:2000hr} (which, in its more interesting self-accelerated branch,  is  now ruled out by 
instabilities at the level of cosmological perturbations~\cite{Luty:2003vm,Nicolis:2004qq,Gorbunov:2005zk,Charmousis:2006pn}). In this case the effect is due to the fact that, at large scales,
gravity  leaks into extra dimensions, and this affects the $1/a$ behavior of the amplitude of a gravitational signal. In this case, however, because of the loss of gravitons to the extra dimensions, a GW source would appear dimmer and therefore we would have $d_L^{\,\rm gw}(z)> d_L^{\,\rm em}(z)$.
In \cite{Deffayet:2007kf} this effect has been assumed to have the form 
$d_L^{\,\rm gw}(z)/d_L^{\,\rm em}(z)=[1+(d_L^{\,\rm em}(z)/R_c)^{n/2}]^{1/n}$, where $R_c$ is a cross-over scale and $n$ determines the steepness of the transition between the two regimes. Note that with this ansatz  $d_L^{\,\rm gw}(z)/d_L^{\,\rm em}(z)$ does not saturate to a constant value at large $z$.
A  recent  discussion of  the modification to the GW  luminosity distance induced by the  leaking of gravity into extra dimension is given in \cite{Pardo:2018ipy}, where this effect is used to put constraints on the number of extra dimensions or on the associated crossover scale.\footnote{The fact that, in brane models,  a gravitational signal can travel along geodesics in the  extra dimensions, while electromagnetic signals are  confined to the (3+1)-dimensional brane, can also lead to delays between the arrival time of a GW  and the associated  electromagnetic signal~\cite{Chung:1999xg,Caldwell:2001ja}. Bounds on this effect from GW170817 are discussed in \cite{Visinelli:2017bny}.}
Modification of the propagation equation has also been  found  in Einstein-Aether models and in scalar-tensor   theories of the Horndeski class 
 in  \cite{Saltas:2014dha,Lombriser:2015sxa,Arai:2017hxj,Amendola:2017ovw,Linder:2018jil}. This indicates that a modified propagation equation for the tensor modes of the form (\ref{prophmodgrav})   is quite generic in alternatives to $\Lambda$CDM; see
also \cite{Gleyzes:2014rba} for a discussion within  the  effective field theory approach to dark energy, and 
\cite{Nishizawa:2017nef} for a general formalism for testing gravity with GW propagation.

Observe also that, in \eq{prophmodgrav}, we could have inserted a function $\delta(z;k)$. The RR model actually predicts a function $\delta(z)$ independent of wavenumber, for modes well inside the horizon. More generally,  the GWs relevant for ground-based or space-borne GW interferometers have a wavelength so small, compared to the horizon size, that no significant dependence on momentum should be expected in typical modified gravity models, for the same reason discussed  above for the functions $\mu(z)$ and $\Sigma(z)$ that parametrize the modifications in the scalar sector.

Observe also that, when $\delta(z)>0$ at all redshifts, as in the RR model, or more generally,  when 
\be
\int_0^z \frac{dz'}{1+z'} \delta(z') >0\, ,
\ee 
we have 
\be
d_L^{\rm gw}(z)<d_L^{\rm em}(z)\, .
\ee 
Since the GW amplitude is proportional to $1/d_L^{\rm gw}(z)$, in this case a GW source is actually magnified, with respect to the GR prediction, and can therefore be seen to larger distances.

\section{Modified GW propagation and effective Newton's constant}\label{sect:Geff}

It is  interesting to observe that, in some modified gravity theories, the function $\delta(z)$ that  parametrizes the deviations from $\Lambda$CDM in the tensor sector can be related to the effective Newton constant $G_{\rm eff}(z)$ that parametrizes the modification of the growth of structure. Indeed, it has been observed in \cite{Linder:2018jil} that, in a subclass of Horndeski theories called  ``no slip gravity'', \eq{dLgwdLem} can be rewritten as
\be\label{dLgwdLemGeff}
d_L^{\rm gw}(z)=d_L^{\rm em}(z)\, \sqrt{\frac{G_{\rm eff}(z) }{G_{\rm eff}(0)}}\, .
\ee
Quite remarkably, this relation holds also in the RR model [for modes well inside the horizon, where the dependence on wavenumber of $G_{\rm eff}(z)$ disappears], as found in \cite{Belgacem:2017ihm}. 
In turn, \eq{eq:fitz} provides a simple fitting formula for $G_{\rm eff}(z)/G_{\rm eff}(0)$. 

This result can be understood as follows. First of all, let us recall how the usual scaling for the GW energy density $\rgw\propto a^{-4}$ comes out from the cosmological evolution in GR (we follow Section~19.5.1 of \cite{Maggiore:2018zz}). From \eqs{4defhchiproofs}{4propchiproofs1} it follows that, for modes well inside the horizon, for which $k\eta\ll 1$,
\be\label{tildehAvsa}
\tilde{h}_A(\eta, \vk)= \frac{1}{a(\eta)} h_A\sin(k\eta+\alpha_A)\, ,
\ee
where $h_A$ is an amplitude and $\alpha_A$ is a phase. Then, again for modes well inside the horizon,  
\be
h'_A(\eta,\vk)\simeq  \frac{h_Ak\cos(k\eta+\alpha_A)}{a(\eta)}\, ,
\ee 
since the time derivative of the factor $1/a(\eta)$ gives a term proportional to $a'/a^2$, which for modes well inside the horizon is negligible compared to $k/a$.
Using $\dot{h}_A=(1/a)h'_A$ we get
\be\label{dothGR}
\dot{h}_A\simeq   \frac{h_Ak\cos(k\eta+\alpha_A)}{a^2}\, .
\ee
The energy density of GWs propagating in a cosmological background is given by
\be\label{rhogwGR}
\rho_{\rm gw}(t)=\frac{1}{16\pi G}\langle \dot{h}_+(t)^2+\dot{h}_{\times}(t)^2\rangle\, ,
\ee
where the angular bracket denotes an average over several periods (see e.g. Section~1.4 of \cite{Maggiore:1900zz}). Plugging \eq{dothGR} into \eq{rhogwGR}  the term $\cos^2(k\eta+\alpha_A)$, averaged over several periods, simply gives a factor $1/2$, and it then 
follows that $\rho_{\rm gw}\propto a^{-4}$. This is the result that is expected for any form of radiation, and allows an interpretation of the energy density in terms of gravitons, with a number density $n$ that, with respect to physical coordinates, scales as $1/a^3$ (i.e.  the number density with respect to comoving coordinates is conserved), times an energy that, for massless particles, scales as $1/a$, leading to the overall $1/a^4$ scaling. Note that the interpretation in terms of gravitons only emerges when the mode is well inside the horizon.

Consider now the scaling of $\rgw$ with $a$ in the RR model.
In the equations of motion of the RR model, the Einstein tensor $\Gmn$ is multiplied by a factor $[1-(m^2/6)S]$, where $S$ is an auxiliary field used to write the theory in a local form~\cite{Maggiore:2014sia}. On cosmological scales, the time-dependence of $S$ therefore gives rise to an effective time-dependent Newton constant $G_{\rm eff}(t)$. Note that, since the factor $[1-(m^2/6)S]$ multiplies the full covariant Einstein tensor, one has the same effective Newton constant both  for  scalar  perturbations and for tensor perturbations.\footnote{It should be stressed  that, in general,  the time-dependence of $S$  only holds on cosmological scales. On solar system scales, the RR model admits a static solution that smoothly reduces to the standard \Sch solution of GR~\cite{Kehagias:2014sda}. An interesting and non-trivial issue is whether the cosmological and \Sch solutions match continuously, without inducing any further time dependence on $S$ on solar-system scales~\cite{Barreira:2014kra}.\label{foot:solarsystem}} Then, the energy density of GWs propagating in a cosmological background will be proportional to
\be
\rho_{\rm gw}(t)=\frac{1}{16\pi G_{\rm eff}(t)}\langle \dot{h}_+(t)^2+\dot{h}_{\times}(t)^2\rangle\, ,
\ee
in which, in the standard GR expression,
$G$ has been replaced by $G_{\rm eff}(t)$. At the same time, from \eq{4defhchiproofsRR}
we see that modified GW propagation implies that\footnote{Observe that the overall normalization of the GW amplitude in a modified gravity model will be the same as in GR only if a screening mechanism ensures that, in the local source zone, the theory is governed by the same Newton's constant as in GR, rather than by an effective Newton's constant. As discussed in footnote~\ref{foot:solarsystem}, for the RR model this is not obvious, since it depends on whether the large-distance cosmological solution can be matched to the static \Sch solution without leaving a residual time dependence in the auxiliary field $S$. In the following we will focus on  how  GWs propagate across cosmological distances, rather than on the possible modifications in the local source zone.}
\be\label{tildehAvstildea}
\tilde{h}_A(\eta, \vk)= \frac{1}{\tilde{a}(\eta)} \, h_A\sin(k\eta+\alpha)\, ,
\ee
Therefore, proceeding as above,
\be\label{dothRR}
\dot{h}_A\simeq   \frac{a(t)}{\tilde{a}(t)}\,   \frac{h_Ak\cos(k\eta+\alpha)}{a^2(t)}\, .
\ee
[Note that the passage from derivative with respect to conformal time $\eta$ to derivatives with respect to $t$ still brings a factor $1/a$, as in \eq{dothGR}].
Thus,  with respect to the GR scaling $\rho_{\rm gw}\propto a^{-4}(t)$, in the RR model 
$\rho_{\rm gw}$ apparently acquires  an extra  $t$-dependent factor proportional to
\be
\frac{G}{G_{\rm eff}(t)}\,  \[\frac{a(t)}{\tilde{a}(t)}\]^2\, \, .
\ee
Using \eq{dgwaadem} we can rewrite this as
\be\label{GoverGeffdLdL}
\frac{G}{G_{\rm eff}(t)}\, \[ \frac{d_L^{\,\rm gw}(t)}{d_L^{\,\rm em}(t)} \]^2\, .
\ee
However, \eq{dLgwdLemGeff} ensures that this factor is actually time independent.
Thus, we see that \eq{dLgwdLemGeff} ensures the standard scaling $\rho_{\rm gw}(t)\propto 1/a^{4}(t)$. In turn, this allows an interpretation of the GWs in the modified gravity theory in terms of a collection of gravitons whose comoving number density is conserved, just as in GR. 

From this discussion we understand that the relation (\ref{dLgwdLemGeff}) is very general, since it expresses the conservation of the graviton number. However, one can certainly imagine modified gravity theories that violate it. For instance, theories with extra dimensions where, from the point of view of a four-dimensional observer, gravitons can  be lost to a higher-dimensional  bulk, would generically violate it. The same holds e.g. for scalar-tensor theories with an unstable  graviton that can decay into scalar particles. In those cases, $\rgw$ will decrease faster than $1/a^4$ with time, i.e. one will have the inequality
\be
\frac{d}{dt}\[ \frac{G}{G_{\rm eff}(t)}\, \( \frac{d_L^{\,\rm gw}(t)}{d_L^{\,\rm em}(t)}\)^2 \]<0\, .
\ee
It is actually possible to show that the relation  (\ref{dLgwdLemGeff}) holds in any 
modified gravity theory with an action of the form
\be\label{SAR}
S=\frac{1}{8 \pi G}\int \sqrt{-g} A R + \dots\,,
\ee
which is minimally coupled to matter. Here the dots indicate other possible gravitational interaction terms, including possible extra fields  in the gravitational sector (such as the auxiliary fields in the nonlocal model, or dynamical scalar  fields in scalar-tensor theories)
but do not contain terms that are purely quadratic in the gravitational field $\hmn$ nor interactions with ordinary matter,
and $A$ can be a nontrivial functional of fields in the gravitational  sector: for instance in the RR model we have $A=[1-(m^2/6)S]$.
Clearly, this model has an effective Newton constant given by
$G_{\rm eff}=G/A$,
and so
\be\label{GGAA}
\sqrt{\frac{G_{\rm eff}(z)}{G_{\rm eff}(0)}}=\sqrt{\frac{A(0)}{A(z)}}\,.
\ee 
In order to show that the ratio $d^{\,\rm gw}_L/d^{\,\rm em}_L$ is given by the same expression
we introduce a new  metric
\be\label{gAg}
\tilde{g}_{\mu\nu}=A g_{\mu\nu}\, .
\ee
In terms of this metric the  action reads
\be\label{StildegR}
S=\frac{1}{8 \pi G}\int \sqrt{-\tilde{g}} \tilde{R} + \dots\,, 
\ee
i.e. $\tilde{g}_{\mu\nu}$ is the ``Einstein frame" metric. Using conformal time, in FRW at the background level we have $\gmn=a^2\emn$, and therefore $\tilde{g}_{\mu\nu}=Aa^2\emn=\tilde{a}^2\emn$ with 
\be\label{aAa}
\tilde{a}^2 \equiv A a^2\, .
\ee
In conformal time, the spatial
perturbations of the original metric are defined as usual by
\be
g_{ij}=a^2 (\d_{ij}+h_{ij})\, ,
\ee
while the spatial
perturbations of the  metric in the Einstein frame are defined by
\be
\tilde{g}_{ij}=\tilde{a}^2 (\d_{ij}+\tilde{h}_{ij})\, .
\ee
\Eq{gAg}, together with \eq{aAa}, then implies that 
\be\label{htildeh}
h_{ij}=\tilde{h}_{ij}\, .
\ee
Since  \eq{StildegR} is formally the same as the usual Einstein-Hilbert action of GR,  
and the terms denoted by dots corresponds to possible interaction terms that do not affect the free propagation of GWs,  
the propagation equation   for the tensor perturbation $\tilde{h}_{ij}$ over the background metric $\tilde{a}^2\emn$ is the same as the standard GR equation for $h_{ij}$ propagating over the background $a^2\emn$, and therefore 
$\tilde{h}_{ij}$ scales ``normally" with respect to the new scale factor $\tilde{a}$, 
$\tilde{h}_{ij}\propto {1}/{\tilde{a}}$.
Then, \eq{htildeh} shows that also the original metric perturbation $h_{ij}$ scales as $1/\tilde{a}$, i.e. with a modified scale factor, so, as in \eq{dLtilde}, for a coalescing binary
\be
h_{ij}\propto\frac{\tilde{a}(z)}{\tilde{a}(0)}\frac{a(0)}{a(z)}\frac{1}{d^{\, \rm em}_L(z)}=\sqrt{\frac{A(z)}{A(0)}}\frac{1}{d_L^{\, \rm em}(z)}\equiv \frac{1}{d_L^{\, \rm gw}(z)}\,.
\ee
Therefore
\be
\frac{d_L^{\, \rm gw}(z)}{d_L^{\, \rm em}(z)}=\sqrt{\frac{A(0)}{A(z)}}
=\sqrt{\frac{G_{\rm eff}(z)}{G_{\rm eff}(0)}}\, ,
\ee
where in the second equality we used \eq{GGAA}. This proves that \eq{dLgwdLemGeff} is a general property of modified gravity theories of the form (\ref{SAR}).
The previous discussion then shows that for theories of this form the comoving graviton number is conserved during the GW propagation over cosmological distances (i.e. the physical graviton number scales as $1/a^3$).

\section{Limits on modified GW propagation from GW170817}\label{sect:GW170817}

We first compare the above results with the limits on modified GW propagation that can already be obtained by the single standard siren provided by the coincident detection of the GWs from the neutron star binary GW170817 \cite{TheLIGOScientific:2017qsa} with the $\gamma$-ray burst GRB170817A~\cite{Goldstein:2017mmi,Savchenko:2017ffs,Monitor:2017mdv}.  From the identification of the electromagnetic 
counterpart~\cite{GBM:2017lvd}, the redshift of the source is  $z=0.00968(79)$, so in this case we are at very small redshift. 
After correcting for the peculiar velocity of the host galaxy, one finds a  cosmological redshift $z=0.00980(79)$~\cite{Hjorth:2017yza}.
To first order in $z$, \eq{dLgwdLem} becomes
\be\label{eq:fitlowz}
\frac{d_L^{\,\rm gw}(z)}{d_L^{\,\rm em}(z)}=1-z \delta(0)+{\cal O}(z^2)\, ,
\ee
so in this limit we are actually sensitive to $\delta(0)\equiv \delta(z=0)$. In  the $(\Xi_0,n)$ parametrization  $\delta(0)=n(1-\Xi_0)$, but in fact the analysis for sources at such a low redshift can be carried out independently of any parametrization of the function $\delta(z)$. Note, however, that the deviation of $d_L^{\,\rm gw}(z)/d_L^{\,\rm em}(z)$ from 1 is also proportional to $z$. It is therefore clear that, with a single standard siren at a redshift $z\simeq 10^{-2}$, we cannot get a stringent limit on modified GW propagation. In any case, it is methodologically interesting to carry out this exercise more quantitatively. 
As we will now discuss, there are two different ways out carrying  this test, one based on a comparison of the Hubble constant extracted from standard sirens with the Hubble constant extracted from standard candles, and one based on the comparison of electromagnetic and GW luminosity distances for the same source.

\subsection{Comparison of the Hubble parameter}

We begin by comparing the value of the Hubble parameter obtained from GW170817/GRB170817A with that obtained from a set of standard candles.
To this purpose, we assume that the correct value of $H_0$ is the one obtained from  local electromagnetic measurements~\cite{2018ApJ...855..136R} 
\be\label{H0Riess}
H_0=73.48\pm 1.66 \, ,
\ee 
(here and in the rest of the paper $H_0$ is given in  units of ${\rm km}\, {\rm s}^{-1}\, {\rm Mpc}^{-1}$), that updates the value $H_0=73.24\pm 1.74$ found in 
\cite{Riess:2016jrr}.\footnote{We use the local measurement of $H_0$, rather than the {\em Planck} value, since we want to compare standard sirens to  standard candles at comparable redshifts. In 
this logic, the discrepancy between the local measurement of $H_0$ and  the {\em Planck} value would be due to the fact that local (electromagnetic) measurements are independent of the cosmological model, while 
to translate the CMB observations  into a value of $H_0$ we need a cosmological model. A $3.7\sigma$ discrepancy takes place if one assumes $\Lambda$CDM, while in modified gravity model, in particular with a phantom De EoS, this tension is reduced or eliminated. In particular, in the (minimal) RR nonlocal model 
$H_0=69.49\pm 0.80$~\cite{Belgacem:2017cqo}, so
the tension with the updated value (\ref{H0Riess}) is reduced to the $2.2\sigma$ level.}
The value of $H_0$ obtained from GW170817/GRB170817A in \cite{Abbott:2017xzu}, assuming no modification in the GW propagation, is $H_0^{\,\rm gw}=70.0^{+12.0}_{-8.0}$, where we have added the superscript ``gw'' to stress that the measurement is obtained with standard sirens. 
This value has been recently updated in \cite{Abbott:2018wiz},  in a reanalysis of GW170817 that uses the known source location, improved modeling and recalibrated Virgo data, to 
\be\label{H0LIGOhigh}
H_0^{\,\rm gw}=70^{+13}_{-7}\, ,
\ee 
when using a high spin prior, and 
\be\label{H0LIGOlow}
H_0^{\,\rm gw}=70^{+19}_{-8}\, ,
\ee 
when using a low spin prior. 
As we mentioned in the Introduction, this   value rises to 
\be\label{H0Guidorzi}
H_0^{\,\rm gw}=75.5^{+11.6}_{-9.6}\, ,
\ee 
if one includes in the analysis a modeling of the broadband X-ray to radio emission to constrain the inclination of the source, as well as a different  estimate of the peculiar velocity of the host galaxy~\cite{Guidorzi:2017ogy}. 
\Eq{H0Riess} is obtained from electromagnetic probes, and then using  
$H_0=z/d_L^{\,\rm em}(z)+{\cal O}(z^2)$. In contrast,  
\eqss{H0LIGOhigh}{H0LIGOlow}{H0Guidorzi}  
are obtained from the measurement of the GW luminosity distance of this source, and then  
evaluating the quantity $H_0^{\,\rm gw}\equiv z/d_L^{\,\rm gw}(z)+{\cal O}(z^2)$. This is the same as 
$H_0$  only if there is no modified GW propagation, so that $d_L^{\,\rm gw}(z)=d_L^{\,\rm em}(z)$. 
If we rather take into account the possibility of modified GW propagation, the correct value of $H_0$ obtained from a standard siren at low redshift is rather
\bees
H_0&\equiv &\frac{z}{d_L^{\,\rm em}(z)} +{\cal O}(z^2)\nn\\
&=& \[1-z \delta(0)\]\frac{z}{d_L^{\,\rm gw}(z)}
+{\cal O}(z^2)\nn\\
&=& \[1-z \delta(0)\]H_0^{\,\rm gw}
+{\cal O}(z^2)
\, ,\label{H0correctedHgw}
\ees
and therefore, for a source for which we can neglect the ${\cal O}(z^2)$ terms, such as GW170817,
\be\label{n1menoXi0}
\delta(0)=\frac{H_0^{\,\rm gw}-H_0}{H_0^{\,\rm gw}z}\, .
\ee
Using in \eq{n1menoXi0} the value of   $H_0$ given  in \eq{H0Riess}, the value of $H_0^{\,\rm gw}$ in \eq{H0LIGOhigh} obtained using the high-spin prior,   and the cosmological redshift of the source $z=0.00980(79)$, we get\footnote{According to \eq{n1menoXi0}, 
the separate errors 
$\Delta H_0$,  $\Delta H_0^{\,\rm gw}$ and 
$\Delta z$ induce an error on $\delta(0)$ given, respectively, by $\Delta H_0/(H_0^{\,\rm gw}z)$,
$(H_0/H_0^{\,\rm gw})\Delta H_0^{\,\rm gw}/(H_0^{\,\rm gw}z)$ and $|H_0^{\,\rm gw}-H_0| (\Delta z/z)/ (H_0^{\,\rm gw}z)$.
We have computed  the error  on $\delta(0)$ by adding these errors in quadrature, both for the upper limit and for the lower limit. In principle there could be a correlation between the $\Delta z/z$ and $\Delta H_0^{\,\rm gw}$, since the uncertainty in the cosmological redshift has already been used in \cite{Abbott:2017xzu} when determining $\Delta H_0^{\,\rm gw}$. However, numerically $|H_0^{\,\rm gw}-H_0| (\Delta z/z)/ (H_0^{\,\rm gw}z)$ is actually negligible with respect to $\Delta H_0/(H_0^{\,\rm gw}z)$ and
$\Delta H_0^{\,\rm gw}/(H_0^{\,\rm gw}z)$, so the issue here is irrelevant.}
\be\label{limitLIGOhigh}
\delta(0)=-5.1^{+20}_{-11}\, .
\ee
Using instead  the result (\ref{H0LIGOlow}) obtained with a low spin prior, we get
\be\label{limitLIGOlow}
\delta(0)=-5.1^{+29}_{-12}\, ,
\ee
and, using the value in \eq{H0Guidorzi},
\be\label{limitLIGO2}
\delta(0)=2.7^{+15.4}_{-12.8}\, .
\ee
By comparison,   the RR model   predicts $\delta(0)\simeq 0.062$ [note that, for 
$n=5/2$ and $\Xi_0=0.970$, we have  $n(1-\Xi_0)\simeq 0.075$, but we see from 
Fig.~\ref{fig:dLgw_over_dLem} that close to $z=0$ the parametrization (\ref{paramdeltaz}) of $\delta(z)$ is no longer accurate].\footnote{Actually, the value  $\delta(0)\simeq 0.062$ is the prediction of the ``minimal" version of the RR model, where a parameter $u_0$ is set to zero,  see Section~\ref{sect:RRatET}.}
As expected, the limits (\ref{limitLIGOhigh})--(\ref{limitLIGO2}), that are obtained from a single standard siren at very low redshift, are not  stringent. However, with $N$ standard sirens with comparable accuracy the error scales as $1/\sqrt{N}$, so with $100$ standard sirens at advanced LIGO/Virgo, at a  redshift comparable to that of GW170817,
 the error on $\delta(0)$ would become ${\cal O}(1)$. Furthermore, sources at higher redshift (that should be available at advanced LIGO/Virgo at design sensitivity) could significantly improve these limits since, as we saw in \eq{eq:fitlowz}, in the low-$z$ regime the deviation of $d_L^{\,\rm gw}(z)/d_L^{\,\rm em}(z)$ from 1 is proportional to $z$. 
 
\subsection{Source-by-source comparison of the luminosity distance} 
 
Even if, for this single standard siren at low redshift, the limits on $\delta(0)$ cannot be stringent, it is methodologically interesting to examine another way of placing such a limit. In the previous analysis
we have compared the value of $H_0^{\,\rm gw}$ obtained from this standard siren  to the value of $H_0$ obtained from other, different, local probes, such as type Ia SNe. Actually, what one directly measures with standard sirens  is the GW luminosity distance, and this is translated into a value of $H_0^{\,\rm gw}$ by using the information on the cosmological redshift of the source. In  turn, this requires subtracting the peculiar velocity of the source from the measured redshift, which, for a source at this small redshift, introduces a non-negligible error. Another option is to compare directly 
the measured GW luminosity distance of this standard siren with its own  electromagnetic luminosity distance, which is obtained by determining the distance to the host galaxy NGC4993.\footnote{We thank the referee for suggesting this comparison.} The GW luminosity distance to GW170817, determined by the LIGO/Virgo observation, is~\cite{Abbott:2017xzu}
\be\label{dgw}
d^{\,\rm gw}_L=43.8^{+2.9}_{-6.9}\,\, {\rm Mpc}\, .
\ee
To measure accurately the electromagnetic luminosity distance to a galaxy, such as NGC4993, which is too distant for resolving its individual stars, there are only a few methods that can provide high precision distance measurements~\cite{Freedman:2010xv}. As discussed in \cite{Cantiello:2018ffy}, for NGC4993 three methods (Cepheids, tip of the red giant branch, and Tully-Fisher relations) are impractical or impossible, while  water masers or  type Ia SNe have not been observed in NGC4993. This leaves only surface brightness fluctuations (SBF) as the only viable method. In the SBF method one measures the pixel-to-pixel variance of the integrated stellar luminosity. For early-type galaxies, such as NGC4993, the contribution to this variance is dominated by stars on the red-giant branch. For pixels of fixed angular size, the ratio of the variance to the mean surface brightness  scales as $(1/d_L^{\,\rm em})^2$. The distance is then obtained by calibrating the absolute magnitude on the mean properties of the stellar population.
With the resolution available at space-based telescopes such as the Hubble Space Telescope, this is the most accurate method for distance measurement for early-type galaxies in the range $10-100$~Mpc. Using this method, the electromagnetic luminosity distance to NGC4993 has been determined in \cite{Cantiello:2018ffy} to be
\be\label{dem}
d^{\,\rm em}_L=40.7 \pm 1.4 \, {\rm (stat)} \pm 1.9\,  {\rm (sys)} \, \, {\rm Mpc}\, .
\ee
From \eq{eq:fitlowz}, neglecting again the ${\cal O}(z^2)$ term,
\be\label{delta0dmenod}
\delta(0)=\frac{d^{\,\rm em}_L-d^{\,\rm gw}_L}{z d^{\,\rm em}_L}\, .
\ee 
Using \eqs{dgw}{dem} we then find
\be\label{limitSBF}
\delta(0)=-7.8^{+9.7}_{-18.4}\, ,
\ee
where we have combined  in quadrature the  statistic and systematic errors in \eq{dem}, and  the upper or, respectively,  lower errors in \eq{dgw} (as well as  the error on the redshift, which however contributes negligibly).
Independently of the method used, or of the assumptions used in the determination of $H_0^{\,\rm gw}$, we find that  $\delta(0)$ is always well consistent with zero within  $1\sigma$, as we expected at this level of accuracy. 

Observe  that the technique based on the direct comparison of the gravitational and electromagnetic luminosity distances,  at least in the case of GW170817,  gives a slightly smaller error, at least compared to \eqs{limitLIGOhigh}{limitLIGOlow}. More importantly, the method based on the comparison of $H_0^{\,\rm gw}$ obtained from a standard siren (or from  a set of standard sirens) with the value of $H_0$ obtained with a set of standard candles, and the method based on the comparison of $d^{\,\rm em}_L$ with $d^{\,\rm gw}_L$ on a source-by-source basis have  different systematic errors; 
for instance,  the latter method
circumvents the problem of determining the peculiar velocity of the host galaxy. Therefore, whenever possible, a comparison of the results of the  two methods would increase the confidence in the analysis. 
However, the surface brightness fluctuations method is accurate only for galaxies at distances less than about 100~Mpc. For sources at larger distances  we expect that it will in general become  more effective the comparison between the inferred value of $H_0^{\, \rm gw}$ and the value of $H_0$ from local electromagnetic measurements (furthermore, increasing the redshift, the uncertainty due to the peculiar velocity becomes less relevant). 

Eventually, for sources with redshift for which the simple Hubble law approximation to the luminosity distance is no longer appropriate, and the full dependence of $d_L$ on $\oma$ and on the dark energy density enters, we need a full cosmological analysis, in combination with other datasets that reduce the degeneracies among these parameters, as we will discuss in the next section.

\section{Measuring  $\{w_0, w_a, \Xi_0,n\}$ with standard sirens at large redshift} \label{sect:w0waXi0}

We next discuss the prospects for measuring modified GW propagation and the dark energy EoS for sources at larger redshift, as appropriate for LISA and ET. While a large part of our discussion is general, the specific comparison with observations will be made using  sources (binary neutron stars) and the sensitivity appropriate for ET.

\subsection{Understanding the role of degeneracies}\label{sect:under}

\begin{figure}[t]
\includegraphics[width=0.4\textwidth]{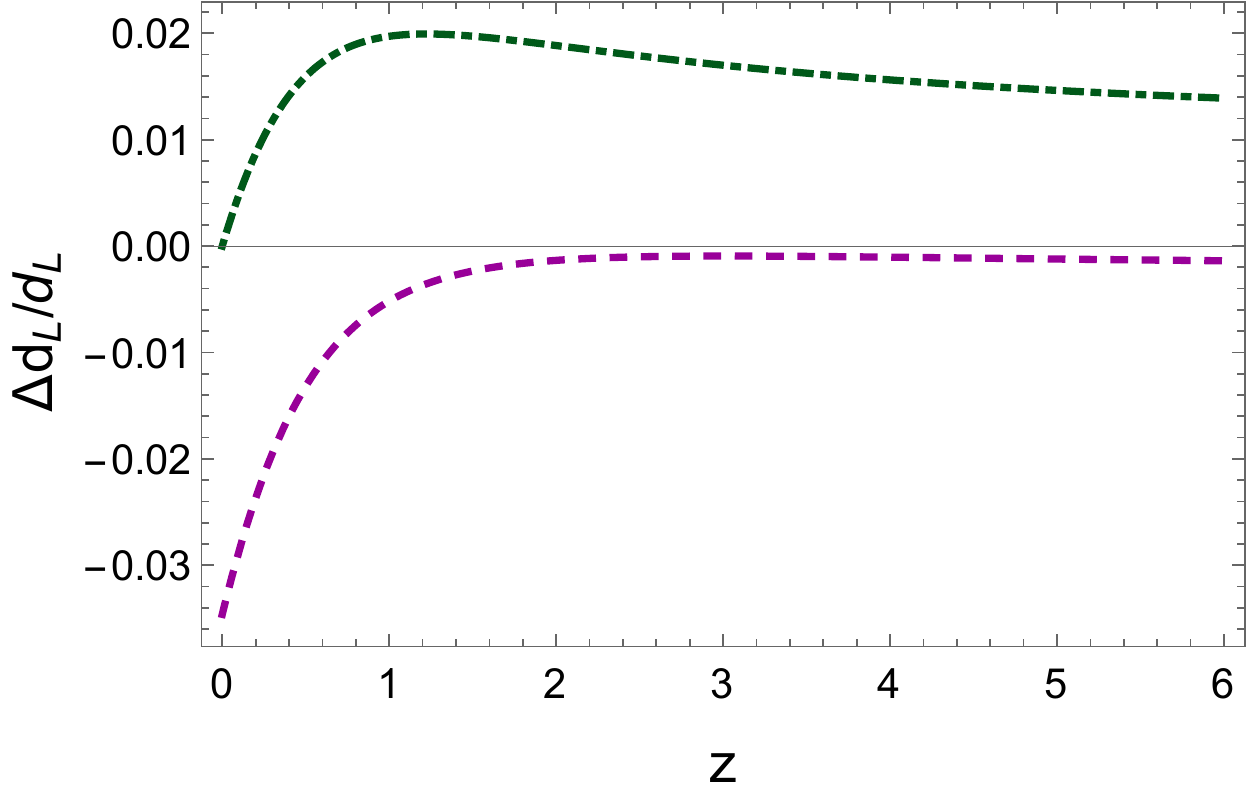}
\caption{The relative difference $\Delta d_L/d_L$ between  $w$CDM with $w=-1.1$ and $\Lambda$CDM. Green, dot-dashed curve: using the same values of $\oma,H_0$ for the two models.
Magenta, dashed curve: using in each model its own best-fit values of $\oma,H_0$.}
\label{fig:wCDM11}
\end{figure}

\begin{figure}[t]
\includegraphics[width=0.4\textwidth]{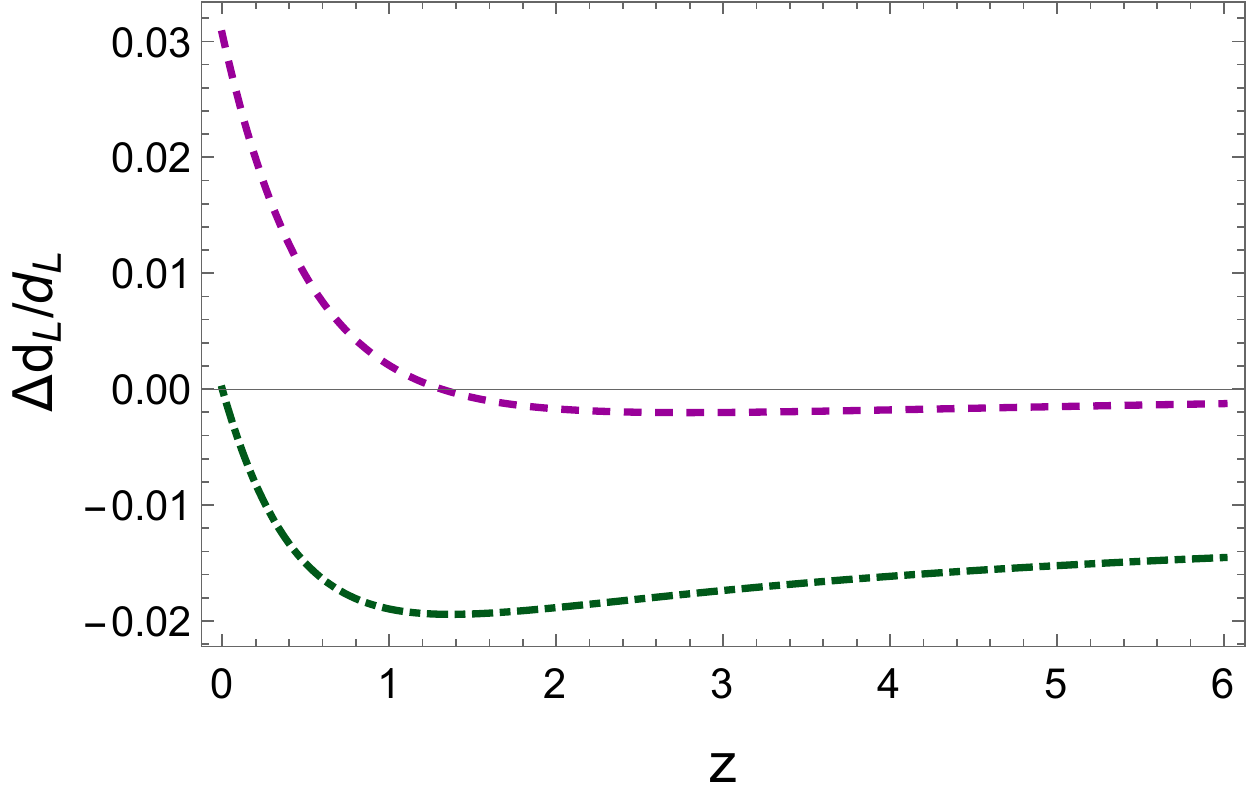}
\caption{The relative difference $\Delta d_L/d_L$ between  $w$CDM with $w=-0.9$ and $\Lambda$CDM. Green, dot-dashed curve: using the same values of $\oma,H_0$ for the two models.
Magenta, dashed curve: using in each model its own best-fit values of $\oma,H_0$.}
\label{fig:wCDM09}
\end{figure}

Before moving directly to the results of our MCMC runs, it is useful to understand in physical terms why 
the parameter $\Xi_0$ [or, more generally,  the function $\delta(z)$ that describes modified GW propagation] can be  more relevant than $w_0$ [or, more generally, of the DE EoS $\wde(z)$], for studies of DE with standard sirens. To this purpose, let us first start from a simple
$w$CDM model, with a fixed value of $w_0$, and let us ask how  a set of measurements of the luminosity distances with standard sirens could help in discriminating it from $\Lambda$CDM.
For $w$CDM, $\wde(z)=w_0$ is constant and \eq{4rdewdeproofs} gives
\be
\rde(z)/\rho_0  =\ode (1+z)^{3(1+w_0)}\, ,
\ee
where $\ode$ is fixed in terms of $\oma$ by the flatness condition, $\oma+\ode=1$. Thus,
\bees\label{dLemw0}
d_L(z;H_0,\oma,w_0)&=&\frac{1+z}{H_0}\, \\
&&\hspace*{-25mm}\times\int_0^z \frac{d\tilde{z}}{\sqrt{ \oma (1+z)^3 +(1-\oma) (1+z)^{3(1+w_0)}}   } \, ,\nn
\ees
where we have written explicitly the dependence of $d_L$ 
on the cosmological parameters.  Let us first consider 
\be\label{rel1}
\frac{\Delta d_L}{d_L}\equiv \frac{d_L(z;H_0,\oma,w_0)-d^{\Lambda\rm CDM}_L(z;H_0,\oma)}{d^{\Lambda\rm CDM}_L(z;H_0,\oma)}\, .
\ee
This is the relative difference between the luminosity distance in $w$CDM with a given value of $w_0$, and the luminosity distance in $\Lambda$CDM (where $w_0=-1$), at fixed $\oma,H_0$. For $w_0=-1.1$ this quantity is shown as the green, dot-dashed curve in Fig.~\ref{fig:wCDM11}, while for $w_0=-0.9$ it is given by  the green, dot-dashed curve in Fig.~\ref{fig:wCDM09}.

However, this is not the quantity  relevant to observations.
The only way of obtaining reasonably accurate values of both $H_0$ and $\oma$, currently, is to use the {\em Planck} CMB data, in combination with other datasets such as baryon acoustic oscillations (BAO) and supernovae, {\em assume a cosmological model}, and determine $H_0$ and $\oma$ through Bayesian parameter estimation in that model.\footnote{At least for $H_0$, one might argue that one can use the value from local measurements, which is independent of the cosmology.  In any case, no sufficiently accurate measurement of $\oma$ can be currently obtained without using the CMB data, and therefore without assuming a cosmological model.}  
In other words, if we want to compute the prediction of $w$CDM (with a given $w_0$) for the luminosity distance,
we must not only use the chosen value of $w_0$ in \eq{dLemw0}, but we must also use  the predictions of $w$CDM for $\oma$ and $H_0$, that are obtained by comparing $w$CDM (with the chosen value of $w_0$) to CMB+BAO+SNe and performing the corresponding parameter estimation. This must be compared with the prediction of $\Lambda$CDM, which is of course also obtained using in \eq{dLemw0} the values of $\oma$ and $H_0$ obtained by fitting the same dataset to $\Lambda$CDM. Thus the relevant quantity, when comparing the predictions of 
$w$CDM with a given $w_0$ to the predictions of $\Lambda$CDM, is not the relative difference given in \eq{rel1}, but rather 
\be\label{rel2}
\overline{\(\frac{\Delta d_L}{d_L}\)}\equiv \frac{d_L(z;H^{w_0}_0,\oma^{w_0},w_0)-d^{\Lambda\rm CDM}_L(z;H_0,\oma)}{d^{\Lambda\rm CDM}_L(z;H_0,\oma)}\, ,
\ee
where we have denoted by $H^{w_0}_0,\oma^{w_0}$ the values obtained from parameter estimation in $w$CDM with the given $w_0$, and by $H_0,\oma$ the values obtained in $\Lambda$CDM (more precisely, one should use the relative priors in the two models; in this discussion we will use the best-fit values for making the presentation simpler). 
One might think that this is a point of details, and that it will not change the order of magnitudes involved. However, this is not true, and the effect of parameter estimation is quite significant. This can be understood by observing that, when we perform Bayesian parameter estimation, we are  basically comparing the predictions of a model to a set of fixed distance indicators, such as  the scales given by the peaks of the CMB, or  the BAO scale, or the observed luminosity distances of type Ia SNe. Thus, the best-fit values of the cosmological parameters in a modified gravity  model change, compared to their values in $\Lambda$CDM, just in the direction necessary to compensate   the change in luminosity distance (or in comoving distance, or in angular diameter distance) 
induced by the non-trivial DE EoS, so that in the end the luminosity distance at large redshift retains basically the same value. 

In order to check and quantify this statement, we have run a series of Markov Chain Monte Carlo (MCMC), fitting both $\Lambda$CDM and $w$CDM (with $w_0=-1.1$ and with $w_0=-0.9$) to the same dataset of cosmological observations. In particular, we use the CMB+BAO+SNe dataset described in detail in section~3.3.1 of \cite{Belgacem:2017cqo}, that includes {\em Planck} CMB data for temperature and polarization, a compilation of isotropic and anisotropic BAO measurements, and the JLA supernovae dataset.
For such datasets, Bayesian parameter estimation for $\Lambda$CDM gives the best-fit parameters 
\be
H_0=67.64\, ,\qquad \oma=0.3087\, .
\ee
In contrast,  for $w$CDM with $w_0=-1.1$ we get 
\be\label{H0omaw11}
H_0=70.096\, ,\qquad \oma =0.2908\, , 
\ee
while, for $w$CDM with $w_0=-0.9$, we find 
\be\label{H0omaw09}
H_0=65.658\, ,\qquad \oma =0.32406\, .
\ee
The magenta dashed curves in Figs.~\ref{fig:wCDM11} and \ref{fig:wCDM09} show the relative difference in luminosity distance (\ref{rel2}), obtained using for each model its own best-fit values of $H_0$ and $\oma$. We see two important effects. 

\begin{enumerate}

\item At redshifts $z\, \gsim \, (1-2)$ the  relative difference of luminosity distances becomes much smaller (in absolute value) than that obtained by keeping $\oma$ and $H_0$ fixed (and given by the green  dot-dashed curves), and this suppression is of about one order of magnitude.  For instance, for $w_0=-1.1$, keeping fixed $\oma$ and $H_0$, the relative difference of luminosity distances at $z=2$ is $1.77\%$, while, once parameter estimation in the respective models is taken into account, this becomes  $-0.16\%$, with a drop in absolute value of about a factor of 10. 

\item As $z\ra 0$, the green curves in Figs.~\ref{fig:wCDM11} and \ref{fig:wCDM09} go to zero. This is of course a consequence of the fact that, for $z\ll 1$, \eqs{dLem}{E(z)} reduce to $d_{L}(z)\simeq H^{-1}_0z$, and to compute the green curves we have used the same value of $H_0$ in the two models. In contrast, the magenta curves do not go to zero, since for each model we are using its own value of $H_0$. Observe that the fact that the relative difference (\ref{rel2}) does not go to zero at $z=0$ 
is precisely the reason that allows the LIGO/Virgo measurement of $H_0$ to have potentially interesting cosmological consequences. Bayesian parameter estimation to the CMB data in different cosmological models predict different values of $H_0$, and therefore a local measurement of $H_0$, whether with standard candles or with standard sirens, can discriminate among different cosmological models. 

\end{enumerate}

\begin{figure}[t]
\includegraphics[width=0.4\textwidth]{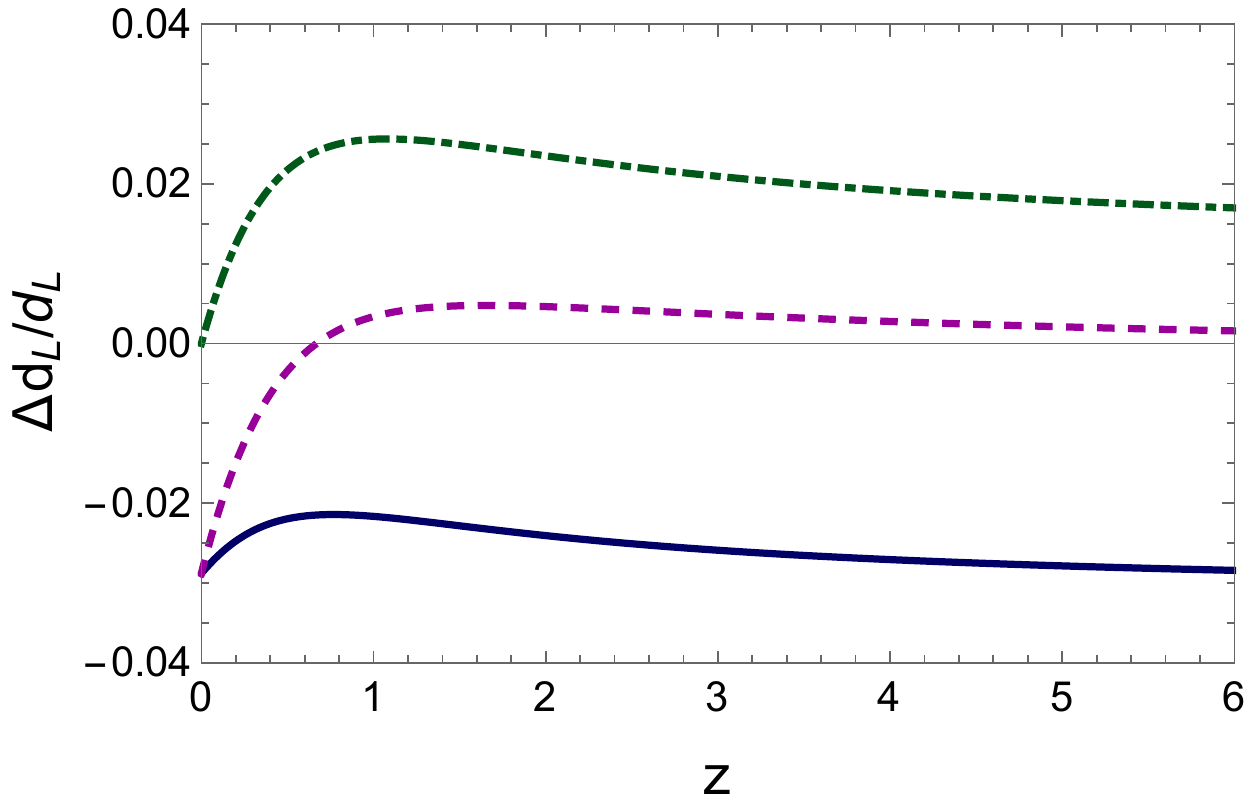}
\caption{The relative difference $\Delta d_L/d_L$ between  the nonlocal RR model and $\Lambda$CDM. Green, dot-dashed curve: using the same values of $\oma,H_0$ for the two models.
Magenta, dashed curve: using in each model its own best-fit values of $\oma,H_0$. Blue solid curve: the relative difference using, for the RR model, the GW luminosity distance (and the respective best-fit values of $\oma,H_0$ for the two models).}
\label{fig:DeltadRR}
\end{figure}

If GW propagation is the same as in GR, the ``signal" that we must detect with standard sirens, to distinguish the modified gravity model from $\Lambda$CDM,  is then given by the magenta lines in Fig.~\ref{fig:wCDM11} and \ref{fig:wCDM09} and is much smaller, in absolute value, than the signal that would be obtained if $H_0$ and $\oma$ were externally fixed quantities, determined independently  of the cosmological model. Let us
assume now that, in the modified gravity theory, on top of a modified DE EoS, there is also a modified GW propagation. Then, for standard sirens, the relevant quantity is the GW luminosity distance 
$d_L^{\,\rm gw}(z)$ given by \eq{dLgwdLem}. In particular, for models (such as the RR model) where the parametrization (\ref{eq:fitz}) is valid, at the redshifts $z\, \gsim \, 0.5-1$ relevant for LISA and ET,  in the modified gravity model
$d_L^{\,\rm gw}(z)$ basically differs from $d_L^{\,\rm em}(z)$ by the factor $\Xi_0$. In contrast, in $\Lambda$CDM the luminosity distance for standard sirens is the same as the standard electromagnetic luminosity distance. Thus, the relevant quantity for discriminating a modified gravity model from $\Lambda$CDM is now
\be\label{rel1mod}
\overline{\(\frac{\Delta d_L}{d_L}\)}^{\rm gw}\equiv \frac{d_L^{\, m,\rm gw}(z;H^m_0,\oma^m)-d^{\Lambda\rm CDM}_L(z;H_0,\oma)}{d^{\Lambda\rm CDM}_L(z;H_0,\oma)}\, .
\ee
where the superscript ``$m$"  denotes the quantities relative to the modified gravity model. Writing
$d_L^{\, m,\rm gw}\simeq \Xi_0 d_L^{\, m,\rm em}$, we get
\bees
\overline{\(\frac{\Delta d_L}{d_L}\)}^{\rm gw}&&\simeq (\Xi_0-1)\frac{d_L^{\, m,\rm em}(z;H^m_0,\oma^m)}{d^{\Lambda\rm CDM}_L(z;H_0,\oma)}\nn\\
&&\phantom{xx}+ \frac{d_L^{\, m,\rm em}(z;H^m_0,\oma^m)-d^{\Lambda\rm CDM}_L(z;H_0,\oma)}{d^{\Lambda\rm CDM}_L(z;H_0,\oma)}\nn\\
&&\simeq (\Xi_0-1)+\overline{\(\frac{\Delta d_L}{d_L}\)}
\, .
\ees
The last term is the relative difference in the electromagnetic luminosity distances introduced  in \eq{rel2}.
We have seen in the example of $w$CDM with $w_0=-0.9$ or $-1.1$ that, even if $w_0$ differ from $-1$ by $10\%$, eventually, because of the compensating effect of $H_0$ and $\oma$, this term in absolute value is only of order $(0.1-0.2)\%$. Thus, if $\Xi_0$ differs from one by more than this, it will dominate the signal. 

This is indeed what happens in the RR model, where $\wde(z)$ differs from $-1$ by about $10\%$, similarly to  the  $w$CDM model with $w_0=-1.1$, and
$|\Xi_0-1|\simeq 3\%$, as we see from Fig.~\ref{fig:dLgw_over_dLem}. The situation for the relative difference of luminosity distances in  the RR model
is illustrated in Fig.~\ref{fig:DeltadRR}, where the green and magenta curves are obtained as in Fig.~\ref{fig:wCDM11} and \ref{fig:wCDM09}, while the blue solid curve is the relative difference of luminosity distances (\ref{rel1mod}) where, for the RR model, we use the GW luminosity distance (while, of course, for $\Lambda$CDM the GW luminosity distance is the same as the electromagnetic one). We see that the ``signal" that allows us to distinguish the RR model from $\Lambda$CDM, represented by the blue curve, is much larger, in absolute value, than that obtained neglecting modified GW propagation, represented by the dashed magenta curve. 

These results indicate that, in a generic modified gravity theory where GW propagation differs from GR, the measurement of luminosity distances of standard sirens should in general  be more sensitive to modified propagation  than to the DE EoS. It also implies that the prospects for detecting deviations from $\Lambda$CDM are better than previously expected.\footnote{Of course, in a given specific modified gravity model, the function $\delta (z)$ could simply be zero, or anyhow such that $|\delta(z)|\ll |1+\wde(z)|$, in which case the main effect would come from $\wde(z)$; what our argument shows is that, in a generic modified gravity  model where the deviation of $\d(z)$ from zero and the deviation of $\wde(z)$ from $-1$ are of the same order, the effect of $\d(z)$ dominates; see also the discussion in \cite{Belgacem:2017ihm}.}

\subsection{Standard sirens and modified gravity with  the Einstein Telescope}\label{sect:pred}

We now wish to determine more quantitatively the prospects for studying dark energy and  modified gravity with future GW experiments, using the MCMC method for standard sirens combined with CMB+BAO+SNe data. For definiteness we will focus on ET.
At its projected sensitivity, ET could have access to binary neutrons star (BNS) mergers up to redshifts $z\sim 8$, corresponding to $10^5-10^6$ events per year~\cite{Sathyaprakash:2009xt}. However, only a fraction of the GW events will have an observed  associated $\gamma$-ray burst. Estimates of the probability of observing the $\gamma$-ray burst are quite uncertain, depending on the opening angle of the jet (typically estimated between $5^{\circ}$ and $20^{\circ}$) and of the efficiency of the network of  existing and future $\gamma$-ray telescopes~\cite{Regimbau:2014nxa}. A typical working hypothesis is that ET might observe ${\cal O}(10^3)$ to ${\cal O}(10^4)$ BNS with electromagnetic counterpart  over a three-year period~\cite{Sathyaprakash:2009xt,Zhao:2010sz,Regimbau:2014nxa}.\footnote{Information of the redshift could also be obtained statistically, by exploiting the narrowness of the neutron star mass function (see \cite{Taylor:2012db} and references therein) or by using tidal effects in neutron stars~\cite{Messenger:2011gi}. In this paper we restrict to sources with detected counterpart.\label{note:statmeth}}

We then proceed as follows. We generate a catalog of BNS detections for ET, with $N_s=10^3$ source, all taken to have an electromagnetic counterpart. We choose a fiducial model, that we take to be $\Lambda$CDM with $H_0=67.64$ and $\oma=0.3087$,
and we generate our simulated catalog of events by  assuming that, for a source at redshift $z_i$, the actual luminosity distance will be $d^{\Lambda\rm CDM}_L(z_i;H_0,\oma)$. The measured value of the luminosity distance is then randomly extracted from a Gaussian distribution centered on $d^{\Lambda\rm CDM}_L(z_i;H_0,\oma)$, and  with a width 
$\sigma_i\equiv \Delta d_L(z_i)$ obtained from an estimate of the error on the luminosity distance at ET. 
For this error, we add in quadrature the instrumental error and the error due to lensing,
\be\label{errortot}
\frac{\Delta d_L(z)}{d_L(z)}=\[ \(\frac{\Delta d_L(z)}{d_L(z)}\)^2_{\rm ET}+
\(\frac{\Delta d_L(z)}{d_L(z)}\)^2_{\rm lensing} \]^{1/2}\, .
\ee
For the instrumental error we
assume the expression given in \cite{Zhao:2010sz},
\be\label{errorET}
\(\frac{\Delta d_L(z)}{d_L(z)}\)_{\rm ET}\simeq 0.1449 z-0.0118 z^2+0.0012 z^3\, ,
\ee
while for  the error due to lensing we use the estimate~\cite{Sathyaprakash:2009xt,Zhao:2010sz}
\be\label{errorlensing}
\(\frac{\Delta d_L(z)}{d_L(z)}\)_{\rm lensing}\simeq 0.05 z\, .
\ee 
Observe that, at ET,  the error due to lensing  is subdominant compared to the estimate of the instrumental error in  \eq{errorET}.
 
To generate our catalog of events we use the standard expression of the number density of the observed events between redshift $z$ and $z+dz$, which is given by  $f(z)dz$, where
\be\label{deffz}
f(z)=\frac{4\pi {\cal N}  r(z) d_L^2(z)}{H(z)(1+z)^3}\, ,
\ee
(see e.g. \cite{Zhao:2010sz}) and  $r(z)$ is the coalescence rate at redshift $z$.\footnote{\Eq{deffz} is derived by observing that the comoving volume between comoving distances $d_c(z)$ and $d_c(z)+d (d_c)$ is 
$4\pi d_c^2 (z) d (d_c)$. One then uses $d (d_c)=[ d(d_c)/ dz] dz= dz/H(z)$, and $d_c(z)=d_L(z)/(1+z)$. Thus the observed event distribution is proportional to $4\pi  (dn/dt_{\rm obs}) d_L^2(z)/[H(z)(1+z)^2]$ where $dn/dt_{\rm obs}$ is the number of events per unit time in the observer frame. Time in the observer frame, $t_{\rm obs}$, is related to  time  in the source frame, $t_s$,  by $dt_{\rm obs}=(1+z)dt_s$, which provides the extra factor of $(1+z)$ at the denominator. Thus $r(z)$ is the number of event per unit time, with respect to the unit of time relevant at redshift $z$.} The normalization constant ${\cal N}$ is determined by  requiring that the total number of sources $N_s$ be given by
\be
N_s=\int_{z_{\rm min}}^{z_{\rm max}}dz\, f(z)\, .
\ee
We take $z_{\rm max}=2$, as in \cite{Zhao:2010sz}, to have a typical signal-to-noise ratio above 8, and we also use a lower cutoff $z_{\rm min}$  to exclude sources for which a modelisation of the local Hubble flow is necessary, before including them in the analysis. This cutoff can be estimated by observing that typical uncertainties on the recessional velocities of galaxies are around $150-250$~km/s~\cite{Chen:2017rfc}. In the small-$z$ regime, where  $d_L(z) H_0\simeq z$ and $z\simeq v/c$, the error on the Hubble parameter is given by
\be
\(\frac{\Delta H_0}{H_0}\)^2\simeq \(\frac{\Delta v}{v}\)^2+\(\frac{\Delta d_L}{d_L}\)^2\, .
\ee
Setting $\Delta v=200$~km/s, $v=zc$, and using \eq{errorET} (that, in the low-$z$ regime, can be approximated by the first term), we get
\be
\(\frac{\Delta H_0}{H_0}\)^2\simeq \(\frac{6.67\times 10^{-4}}{z}\)^2+(0.15 z)^2\, .
\ee
Observe that $\Delta v/v$ goes as $1/z$, dominating at low redshifts, while   
$\Delta d_L/d_L$ is basically proportional to $z$  (because it is  proportional to the inverse of the signal-to-noise ratio). Therefore there is an optimal redshift where  $\Delta H_0/H_0$ is minimum, whose value depends of course on the sensitivity of the detector. For the sensitivity given in \eq{errorET}, we get $z_{\rm min}\simeq 0.07$, corresponding 
to a distance $d_L\simeq 290 \, (0.7/h_0)$~Mpc, where $h_0$ is defined as usual by $H_0=100\, h_0{\rm km}\, {\rm s}^{-1}\, {\rm Mpc}^{-1}$.

\begin{figure}[t]
\includegraphics[width=0.4\textwidth]{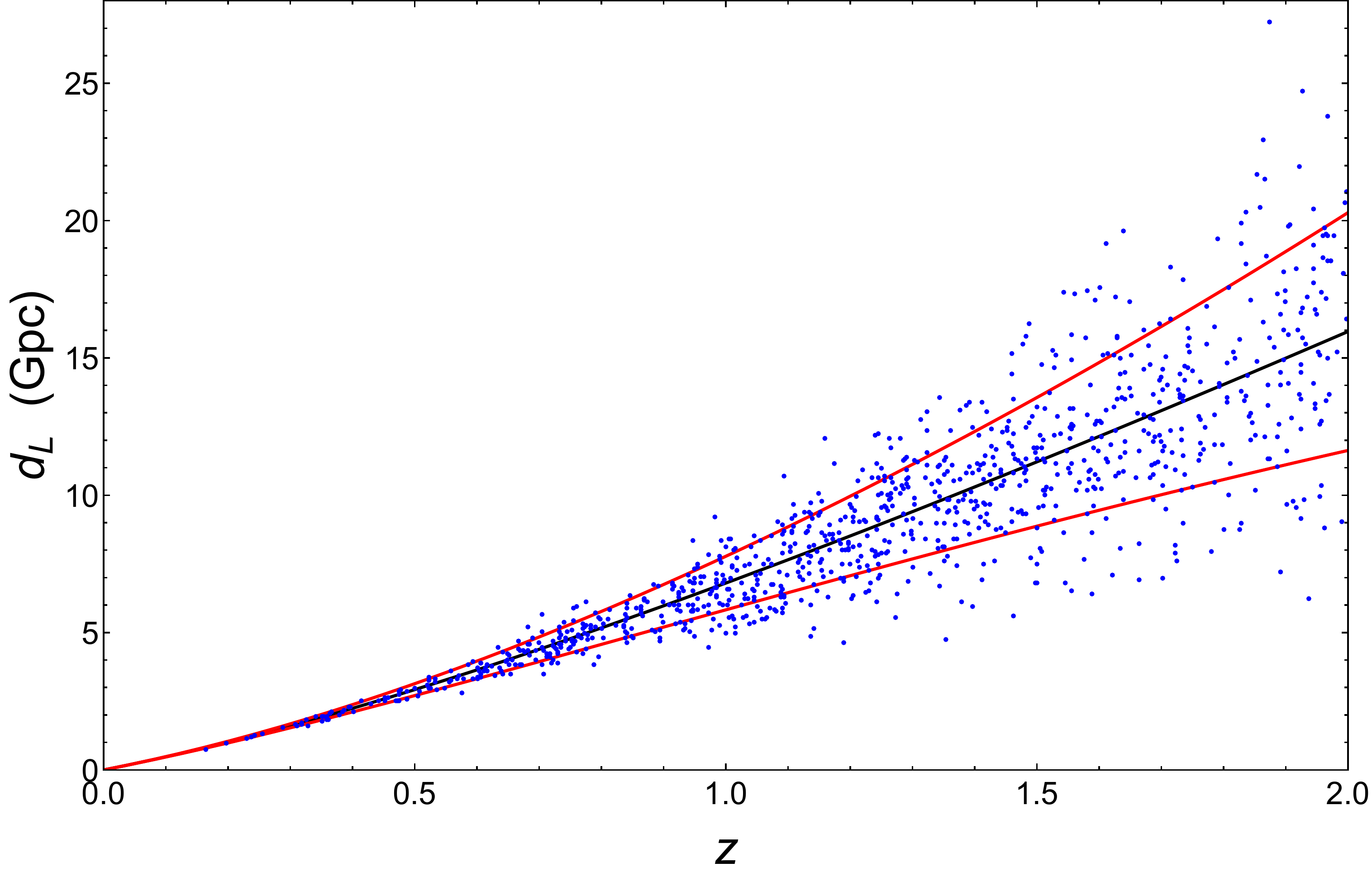}
\caption{A sample of 1000 sources distributed in redshift according to  \eq{deffz}, and scattered in $d_L(z)$ according to the ET error estimate (\ref{errorET}). The black and red curves show the theoretical prediction for $d_L(z)$ and  the $1\sigma$ ET error, respectively. The cosmological model assumed is $\Lambda$CDM with $\Omega_M=0.3087$ and  
$H_0=67.64$.}
\label{fig:sources}
\end{figure}

\begin{figure}[t]
\includegraphics[width=0.4\textwidth]{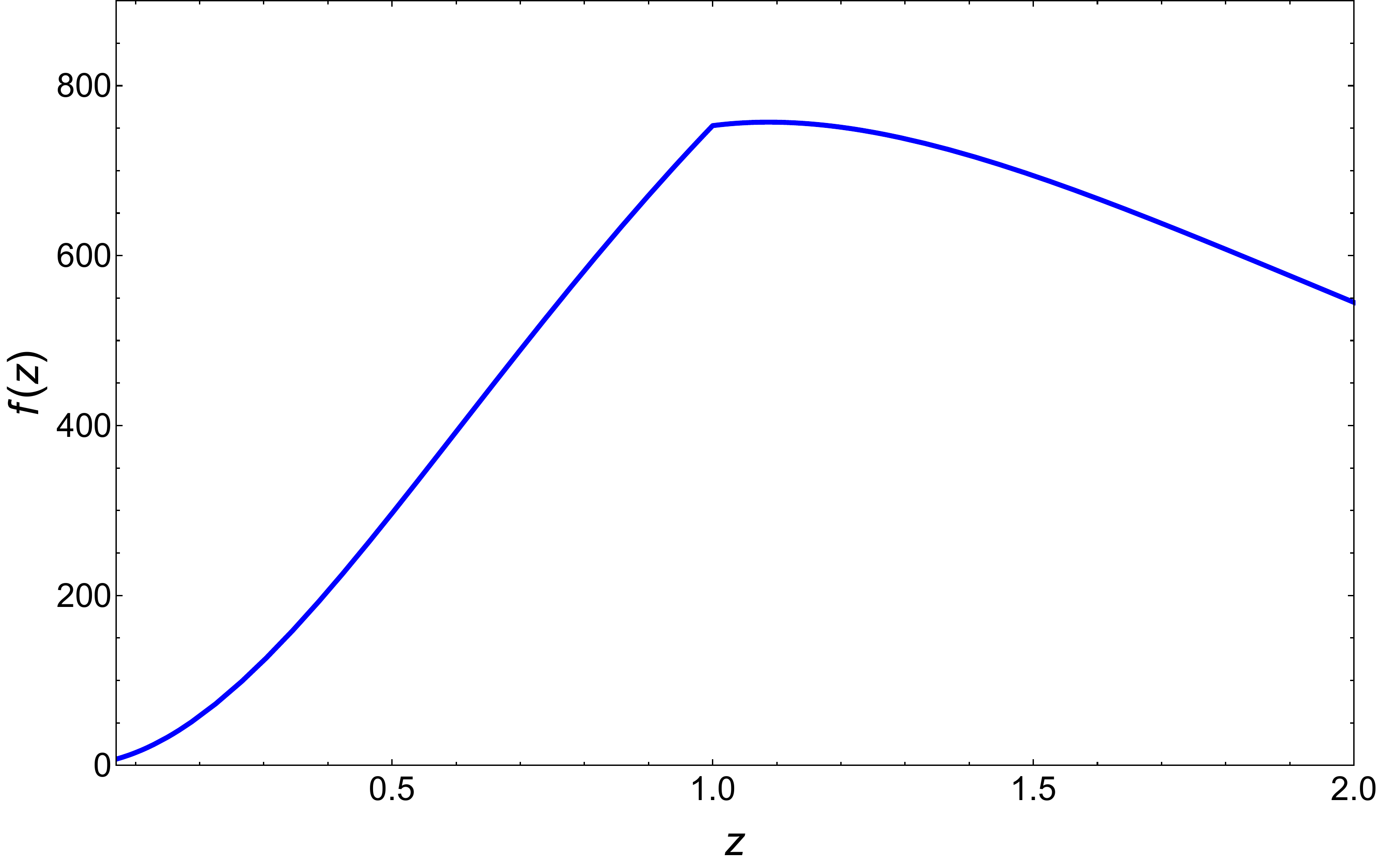}
\caption{The distribution of sources as a function of redshift}
\label{fig:sources_vs_z}
\end{figure}

For $r(z)$ we follow  \cite{Cutler:2009qv,Zhao:2010sz,Cai:2016sby} and we use the form $r(z) = (1 + 2z)$ for $z \leq 1$, $r(z) = (15 - 3z)/4$ for $1 < z < 5$, and $r(z) = 0$ for $z \geq 5$, that is based on a fit  to the observationally determined star formation history  discussed in \cite{Schneider:2000sg}. A sample of the luminosity distance of 1000 sources generated according to these distributions is shown in Fig.~\ref{fig:sources},  while in Fig.~\ref{fig:sources_vs_z} we show their number distribution,  as a function of redshift.

We then run a MCMC, using the  CLASS Boltzmann code \cite{Class}. For the baseline $\Lambda$CDM model we use the standard set of six independent
cosmological parameters:  the Hubble parameter today
$H_0 = 100 h \, \rm{km} \, \rm{s}^{-1} \rm{Mpc}^{-1}$, the physical baryon and cold dark matter density fractions today $\omega_b = \Omega_b h^2$ and $\omega_c = \Omega_c h^2$, respectively, the amplitude  $A_s$ and tilt $n_s$ of the primordial scalar perturbations,   and  the reionization optical depth  $\tau_{\rm re}$. We keep the sum of neutrino masses fixed, at the value $\sum_{\nu}m_{\nu}=0.06$~eV, as in the {\em Planck} baseline analysis~\cite{Planck_2015_CP}.
We then extend  $\Lambda$CDM by adding different combinations of $w_0,w_a$, $\Xi_0$ and $n$, as specified below, assuming for the GW luminosity 
distance the form (\ref{eq:fitz}), and we study to what accuracy we are able to recover the  values
 of our fiducial $\Lambda$CDM model.

As already discussed in \cite{Zhao:2010sz,Cai:2016sby}, because of the degeneracies with $H_0$ and $\oma$, limited information can be obtained on $w_0$ and $w_a$ by using only standard sirens. This also extends to $\Xi_0$. Indeed, using  $10^3$ standard sirens and no other datasets in our MCMC, using $(w_0,\Xi_0)$ as extra parameters with respect to $\Lambda$CDM, fixing for definiteness $n$ to the value $n=5/2$ predicted by the minimal RR model, and computing the corresponding one-dimensional marginalized likelihoods for $w_0$ or for $\Xi_0$, we find  that $w_0$ and $\Xi_0$ can be measured with an error
$\Delta w_0=0.41$ and $\Delta \Xi_0=0.17$, respectively. As expected, this level of accuracy is not very interesting,  particularly for $w_0$, and we need other cosmological datasets to break the degeneracies. In particular we use the same CMB, BAO and SNe datasets that we used in our previous studies of the RR model, namely 
the {\em Planck} temperature and polarization power spectra \cite{Ade:2015rim}, the JLA SNe dataset \cite{Betoule:2014frx} and a set of isotropic and anisotropic BAO data (see Section 3.3.1 of~\cite{Belgacem:2017cqo} for details). We begin by studying how the inclusion of standard sirens improves the estimate of $H_0$ and $\oma$ in $\Lambda$CDM, and we will then
examine extensions of the dark-energy sector parametrized by different combinations of the parameters $w_0,w_a,\Xi_0$ and $n$. 

\subsubsection{$\Lambda$CDM}

\begin{figure}[t]
\includegraphics[width=0.4\textwidth]{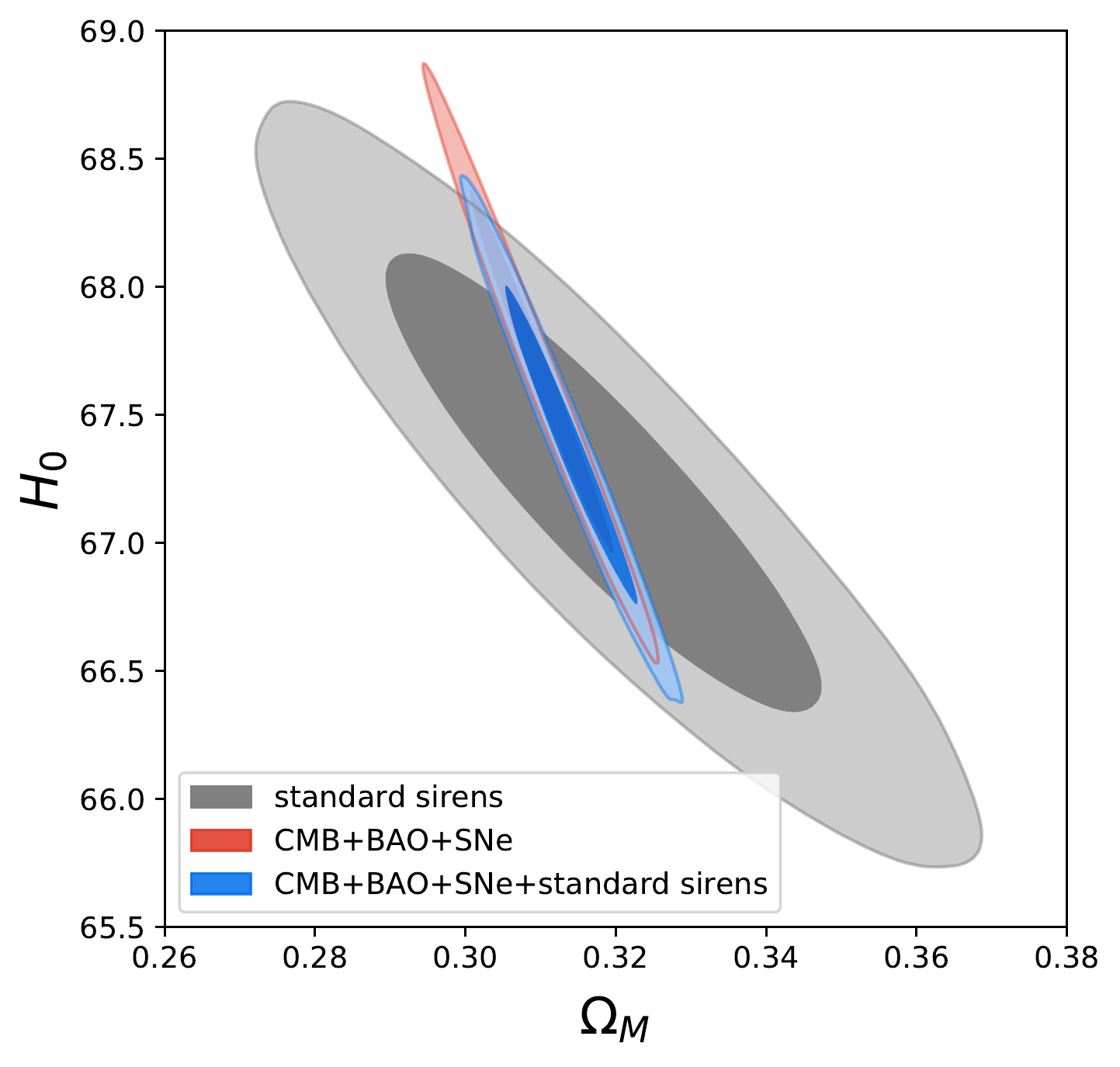}
\caption{The  $1\sigma$ and $2\sigma$
contours  of the two-dimensional likelihood in the $(\oma,H_0)$ plane in $\Lambda$CDM, with the 
contribution from  CMB  + BAO + SNe (red), the contribution from $10^3$ standard sirens at ET (gray) and the overall combined contours (blue).}
\label{fig:H0OmaLCDM}
\end{figure}

We first study the effect of including standard sirens in the parameter estimation of $\Lambda$CDM. 
In this case the luminosity distance  depends only on $H_0$ and $\oma$, so these are the parameters on which the inclusion of standard sirens has the most significant impact. Fig.~\ref{fig:H0OmaLCDM} shows 
the two-dimensional likelihood in the $(\oma,H_0)$ plane in $\Lambda$CDM, comparing the 
contribution from  CMB  + BAO + SNe (red) to the contribution from $10^3$ standard sirens at ET (gray), and the overall combined contours.

From the corresponding one-dimensional likelihoods we find that, using
CMB + BAO + SNe + standard sirens,
 $H_0$ and $\oma$ can be determined to an accuracy
\be\label{DH0suH0totLCDM}
\frac{\Delta H_0}{H_0}=0.6\%\, ,\qquad
\frac{\Delta \oma}{\oma}=1.9\%\, .
\ee
By comparison, using the same CMB + BAO + SNe dataset but without standard sirens, we find 
\be
\frac{\Delta H_0}{H_0}=0.7\%\, ,\qquad
\frac{\Delta \oma}{\oma}=2.1\%\, .
\ee
By itself, the improvement due to $10^3$ standard sirens is therefore relatively small. However, standard sirens have systematic errors completely different from those of electromagnetic probes, and therefore their inclusion significantly increases the confidence in a result.
It is also interesting to observe that, with standard sirens alone, without combining them with other datasets, we already get 
\be\label{DH0suH0sirLCDM}
\frac{\Delta H_0}{H_0}=0.9\%\, ,\qquad
\frac{\Delta \oma}{\oma}=6.5\%\, .
\ee
In particular, the accuracy 
on $H_0$ from standard sirens alone is quite comparable to that from  CMB data+ BAO + SNe but, contrary to SNe data,  it does not make use of any cosmological distance ladder.

\subsubsection{$w$CDM}

\begin{figure}[t]
\includegraphics[width=0.4\textwidth]{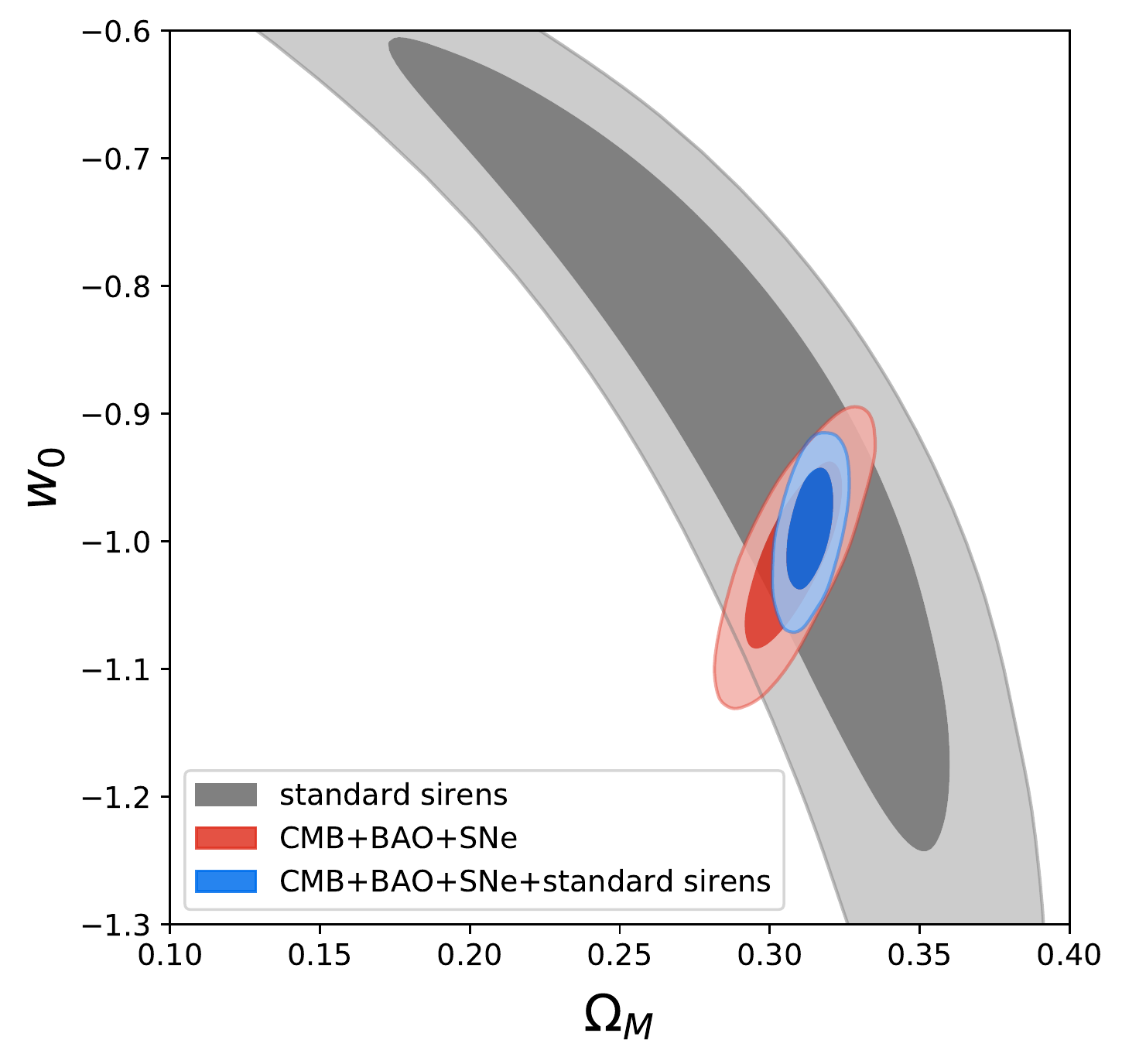}
\caption{The  $1\sigma$ and $2\sigma$
contours  of the two-dimensional likelihood in the $(\oma,w_0)$ plane in $w$CDM, with the 
contribution from  CMB  + BAO + SNe (red), the contribution from $10^3$ standard sirens at ET (gray) and the overall combined contours (blue).}
\label{fig:w0Oma}
\end{figure}

We next consider  the $w$CDM model, where $w_0$ is the only extra parameter. Fig.~\ref{fig:w0Oma} shows the  $1\sigma$ and $2\sigma$
contours  of the two-dimensional likelihood in the $(\oma,w_0)$ plane. We show again the  contours obtained by combining {\em Planck\,} CMB data, BAO and SNe (red), the 
contours from $10^3$ standard sirens at ET (gray), and the total combination 
 {\em Planck\,} CMB data+ BAO + SNe + standard sirens (blue).  From the corresponding one-dimensional marginalized likelihood we get 
\be
\Delta w_0=0.045\, ,
\ee 
with  CMB  + BAO + SNe, and 
\be\label{Deltaw0wCDM}
\Delta w_0=0.031\, ,
\ee 
when adding also $10^3$ standard sirens at ET.

\subsubsection{$(w_0,w_a)$ parametrization}

\begin{figure}[t]
\includegraphics[width=0.4\textwidth]{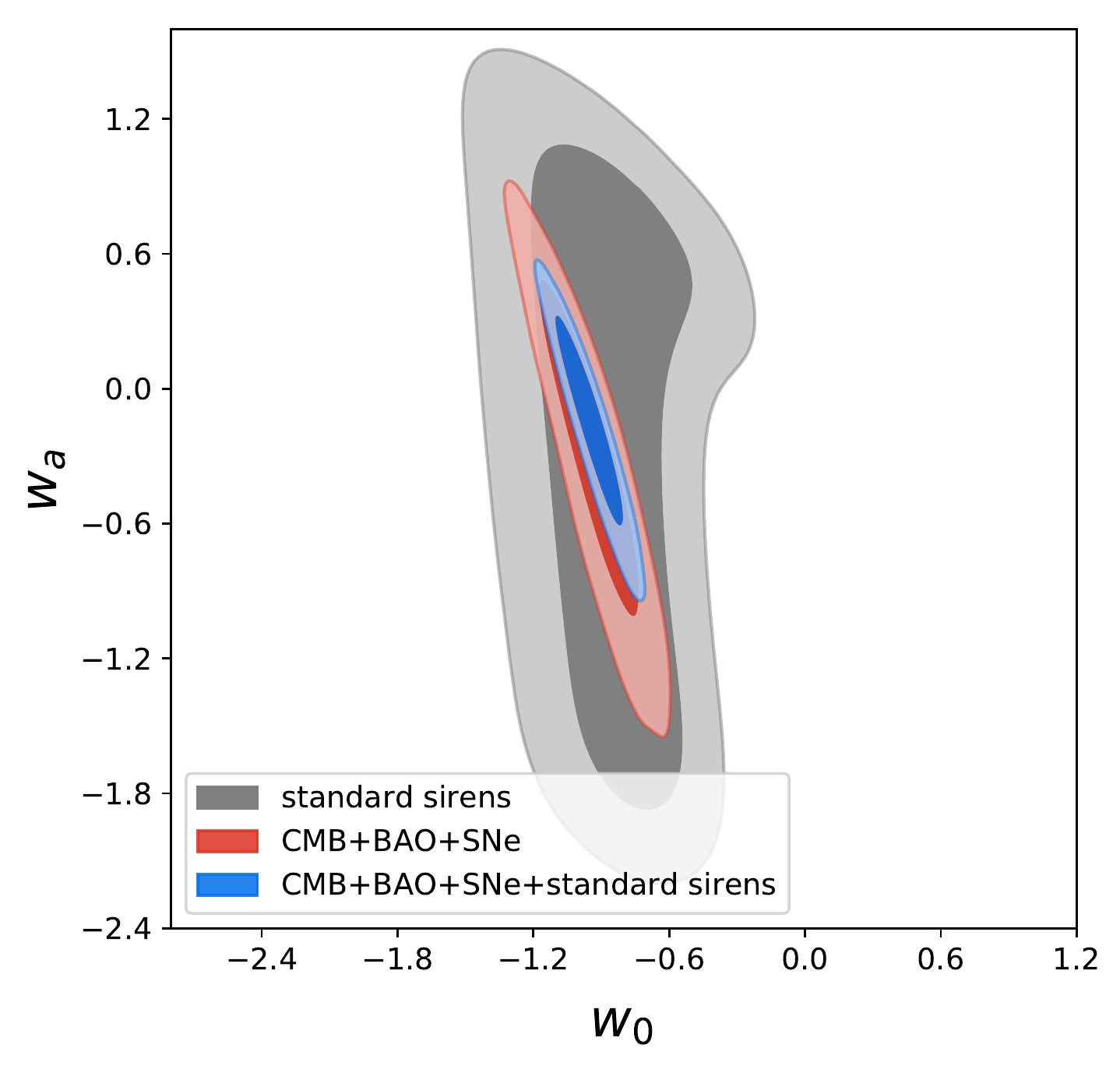}
\caption{The  two-dimensional likelihood in the $(w_0,w_a)$ plane, with the combined  contribution from 
CMB + BAO + SNe (red), the contribution from  $10^3$ standard sirens at ET (gray), and the total combined result (blue).}
\label{fig:w0wa}
\end{figure}

We next consider the $(w_0,w_a)$ parametrization (\ref{w0wa}). Fig.~\ref{fig:w0wa} shows the two-dimensional likelihood in the $(w_0,w_a)$ plane, displaying as before the combined  contribution from 
CMB + BAO + SNe (red), the contribution from  $10^3$ standard sirens at ET (gray), and the total combined result (blue).

From the corresponding one-dimensional likelihoods we find that $w_0$ and $w_a$ can be reconstructed with the accuracy 
\be\label{ourDeltaw0Deltawa}
\Delta w_0=0.140 \, \, ,\qquad \Delta w_a=0.483\, .
\ee
for CMB+BAO+SNe, and
\be\label{ourDeltaw0Deltawa}
\Delta w_0=0.099 \, \, ,\qquad \Delta w_a=0.313\, 
\ee
when adding $10^3$ standard sirens. Following \cite{Huterer:2000mj,Hu:2003pt,Albrecht:2006um,Zhao:2010sz}, 
it is convenient to express the results in terms of the constraint on $w(z)$ at the best pivot redshift $z_p$, defined as the value of redshift for which  $w(z)$ is best constrained. For the $(w_0,w_a)$ parametrization, the pivot scale factor $a_p$ is obtained by minimizing
$\langle (\delta w_0+(1-a)\delta w_a )^2\rangle$, where 
$(\Delta w_0)^2=\langle (\delta w_0)^2 \rangle$ and $(\Delta w_a)^2=\langle (\delta w_a)^2 \rangle$. 
This gives
\be\label{pivota}
1-a_p=-\frac{ \langle \delta w_0\delta w_a \rangle }{(\Delta w_a )^2}\, ,
\ee
and is in the cosmological past if the correlation $ \langle \delta w_0\delta w_a \rangle $ is negative.
One can then show \cite{Albrecht:2006um} that, at the Fisher matrix level,  i.e. assuming that the likelihood is gaussian in all parameters, the error on $w_p\equiv w(z_p)$ is  the same as the error on $w_0$ in the $w$CDM model. 
For the pivot redshift, given by $1+z_p=1/a_p$, \eq{pivota} gives
\be\label{pivotz}
z_p=-\( 1+ \frac{\Delta w_a}{\rho\Delta w_0} \)^{-1}\, ,
\ee
where 
\be
\rho \equiv \frac{ \langle \delta w_0\delta w_a \rangle  }{ \Delta w_0\Delta w_a }
\ee
is the correlation coefficient of $w_0$ and $w_a$. The corresponding error on $w_p$ is then given by
\be\label{Deltawpivot}
\Delta w_p=\Delta w_0\, \sqrt{1-\rho^2}\, .
\ee
Using the values for $\Delta w_0$ and $\Delta w_a$ found in \eq{ourDeltaw0Deltawa}, and the  corresponding value $\rho=-0.909$ from our MCMC, and inserting them into
\eqs{pivotz}{Deltawpivot}, we get
\be
z_p=0.402\, ,\qquad \Delta w_p=0.041 \, \, .
\ee
Observe that this value of $\Delta w_p$ is larger, but consistent, with the value of $\Delta w_0$ given in  \eq{Deltaw0wCDM}.

Our results for the $(w_0,w_a)$ parametrization can be  compared with those of ref.~\cite{Zhao:2010sz}, since we have followed their strategy for generating the catalogue of sources, through \eqs{errorET}{deffz}, and we are using the same number of standard sirens, $N_s=10^3$. With respect to ref.~\cite{Zhao:2010sz} we are performing a full MCMC, rather than a Fisher matrix analysis, and we are using the actual {\em Planck} 2015 likelihoods (rather than the forecasts for {\em Planck} available at the time when  ref.~\cite{Zhao:2010sz} was written), as well as more recent data for SNe and BAO. Combining $10^3$ standard sirens with CMB+BAO+SNe, and using the same distribution of sources,  in ref.~\cite{Zhao:2010sz} was found $\Delta w_0=0.045$, $\Delta w_a=0.174$, and 
$\Delta w_p=0.019$ at a pivot redshift $z_p=0.313$. We notice that our results for  $\Delta w_0$, $\Delta w_a$ and  $\Delta w_p$
are larger by about a factor of two.   This is likely due to the fact that we have used the actual likelihoods of {\em Planck}, rather than the forecast used in  ref.~\cite{Zhao:2010sz}.

\subsubsection{$(\Xi_0, w_0)$ parametrization }

\begin{figure}[t]
\includegraphics[width=0.4\textwidth]{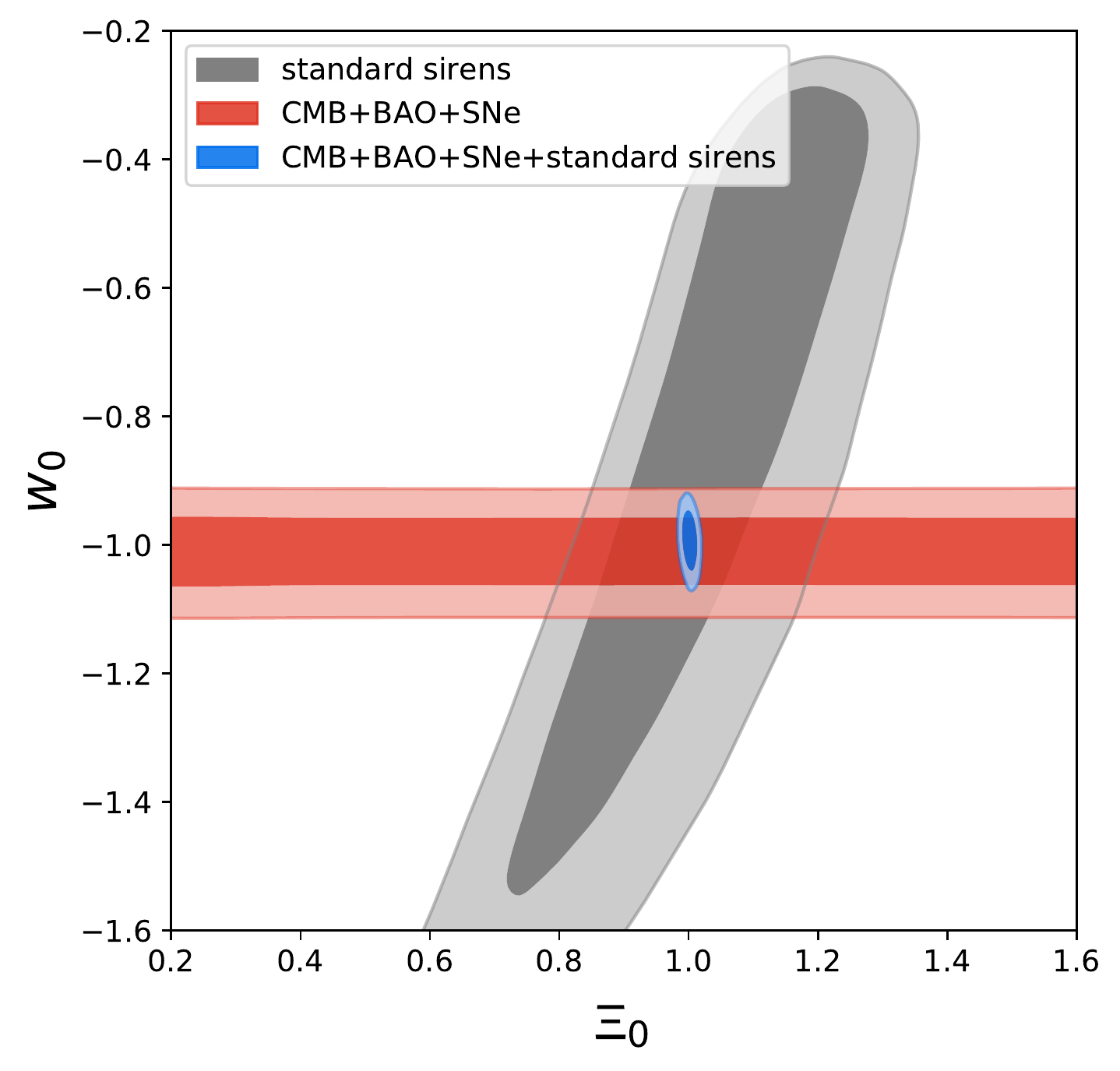}
\caption{The  two-dimensional likelihood in the $(\Xi_0,w_0)$ plane, with the combined  contribution from 
CMB + BAO + SNe (red), the contribution from  $10^3$ standard sirens at ET (gray), and the total combined result (blue).
\label{fig:xi0w0}}
\end{figure}

\begin{figure}[t]
\includegraphics[width=0.4\textwidth]{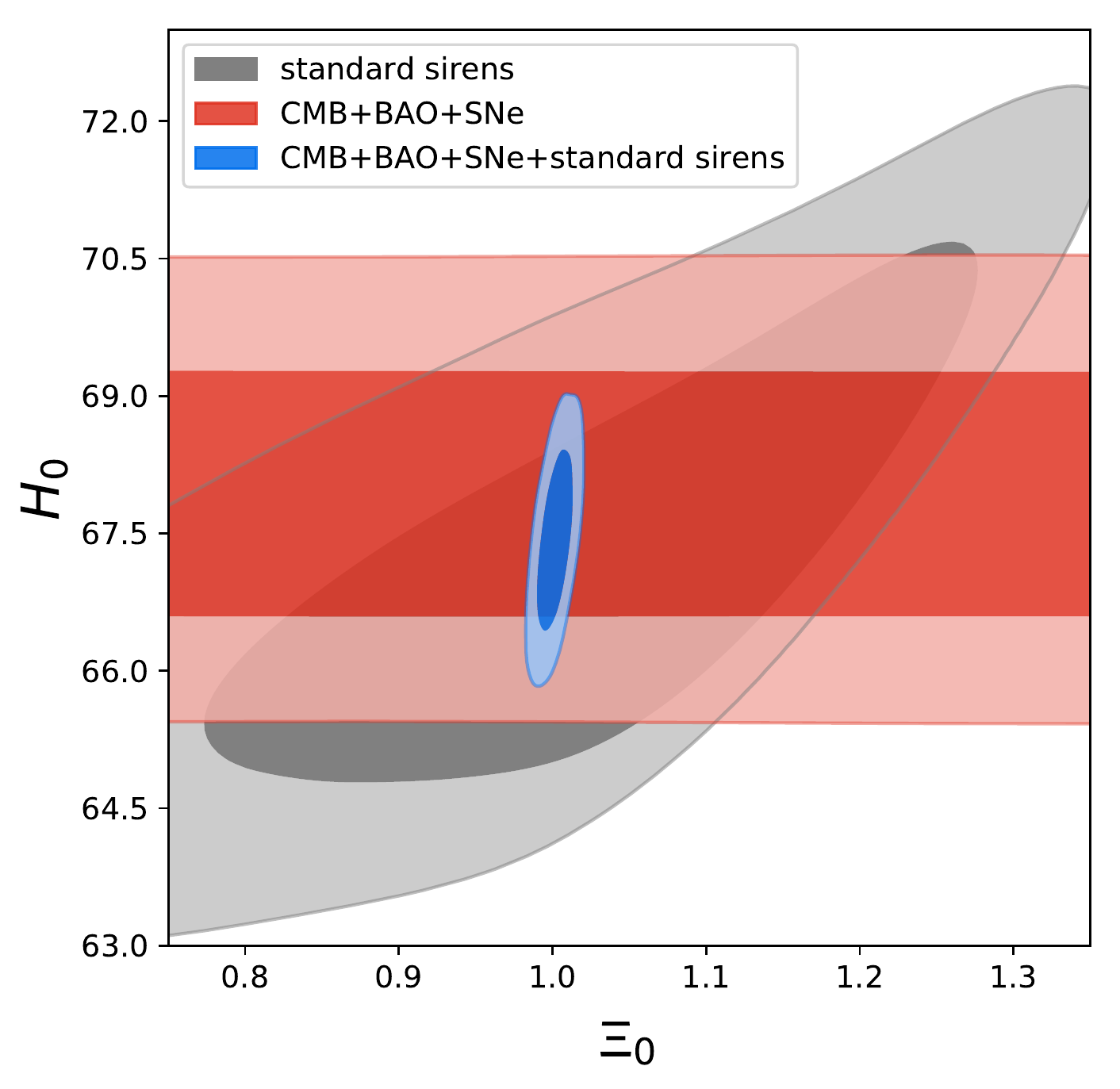}
\caption{The  two-dimensional likelihood in the $(\Xi_0,H_0)$ plane, with the combined  contribution from 
CMB + BAO + SNe (red), the contribution from  $10^3$ standard sirens (gray), and the total combined result (blue). 
\label{fig:xi0H0}}
\end{figure}

\begin{figure}[t]
\includegraphics[width=0.4\textwidth]{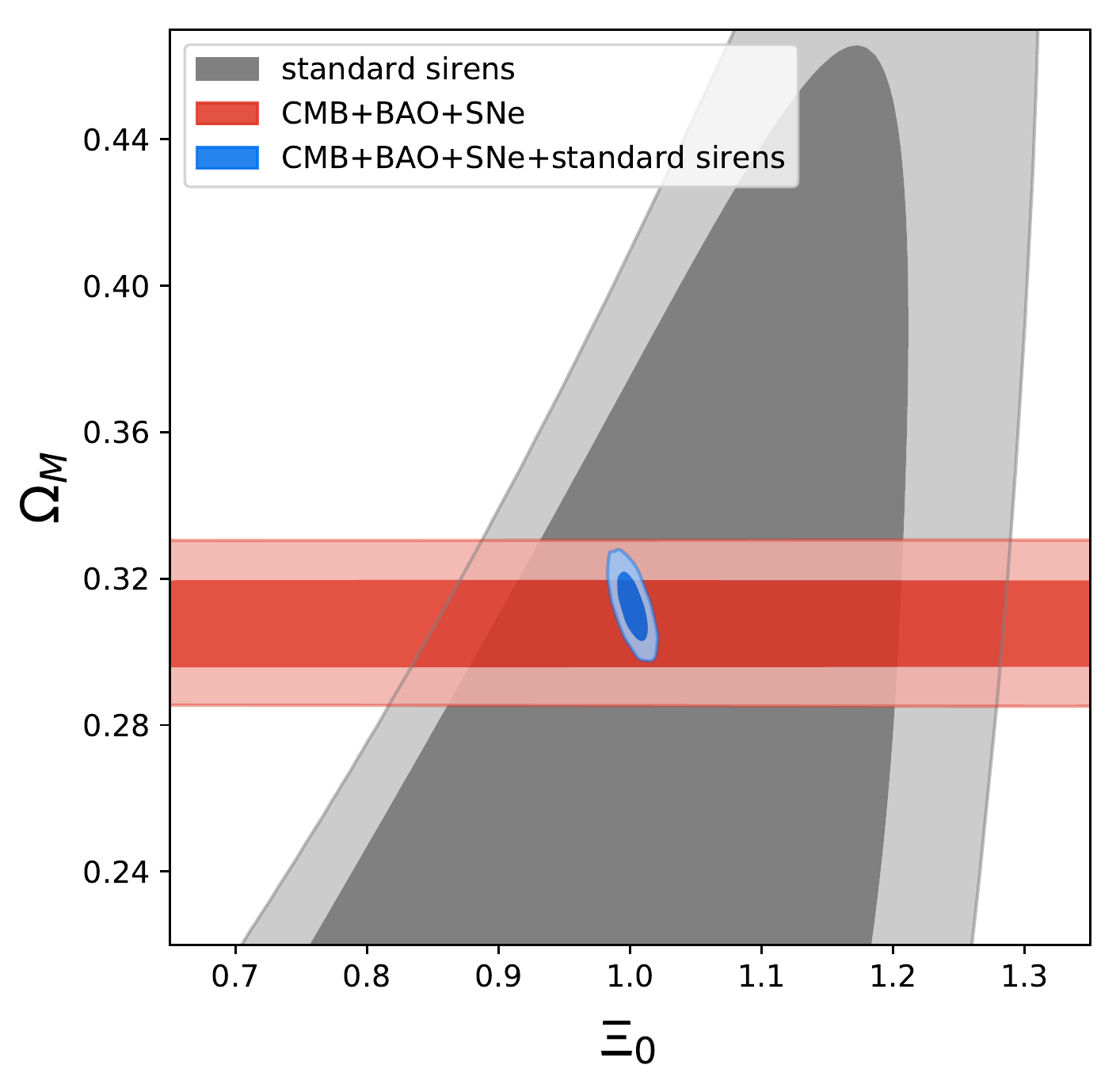}
\caption{The  two-dimensional likelihood in the $(\Xi_0,\oma)$ plane, with the combined  contribution from 
CMB + BAO + SNe (red), the contribution from  $10^3$ standard sirens (gray), and the total combined result (blue). 
\label{fig:xi0Oma}}
\end{figure}

 We finally introduce  the parameter $\Xi_0$ in our extension of $\Lambda$CDM, writing the GW luminosity distance as in \eq{eq:fitz} (with $n=5/2$), and taking $(\Xi_0,w_0)$ as the parameters that describe the DE sector of the theory. Again, we assume $\Lambda$CDM as our fiducial model, so in particular our fiducial values for these parameters are $\Xi_0=1$ and $w_0=-1$.
 
Fig.~\ref{fig:xi0w0} shows the two-dimensional likelihood in $(\Xi_0,w_0)$ plane, 
displaying the  limit from  CMB+BAO+SNe (which is insensitive to modified GW propagation, and hence to $\Xi_0$), the separate contribution from  standard sirens, and the combined limit from  CMB+BAO+SNe+standard sirens.  
From the corresponding one-dimensional likelihoods, we find that the limit from the total combination 
CMB+BAO+SNe+standard sirens is
\be\label{ourDeltaXi0Deltaw0Deltawa}
\Delta\Xi_0=0.008\, ,\quad 
\Delta w_0=0.032\, .
\ee
We see that $\Xi_0$ can be measured to a precision four times better than $w_0$, consistently with the  discussion in Section~\ref{sect:under}. Comparing with \eq{Deltaw0wCDM} we  also observe  that introducing the parameter $\Xi_0$ basically does not degrade the accuracy on $w_0$.

Figs.~\ref{fig:xi0H0} and \ref{fig:xi0Oma} show the analogous two-dimensional likelihoods in the $(\Xi_0,H_0)$ plane and  in the $(\Xi_0,\oma)$ plane, respectively. For the relative error on $H_0$ we find that, using CMB + BAO + SNe, $\Delta H_0/H_0=1.8\%$, while adding also $10^3$ standard sirens at ET this reduces to
\be\label{DH0suH0totXiw0}
\frac{\Delta H_0}{H_0}=1.0\%\, .
\ee
Note that, using only standard sirens, without combining them with other datasets, we get 
\be\label{DH0suH0sirXiw0}
\frac{\Delta H_0}{H_0}=2.8\%\, .
\ee
Once again, this is of course larger than the value obtained combining with CMB + BAO + SNe, but is still an  interesting result, since it is an almost model-independent measure, in which the cosmological model only enters through the possibility of modified GW propagation. Note also that the values in \eqs{DH0suH0totXiw0}{DH0suH0sirXiw0} are larger than the values in \eqs{DH0suH0totLCDM}{DH0suH0sirLCDM}, respectively, because here we have introduced the extra parameters $\Xi_0$ and $w_0$, and the marginalization over these parameters of course increases the error.

These plots also clearly show that standard sirens alone do not give sufficiently strong constraints on $\Xi_0$, because of the degeneracies with $\oma$ and $H_0$, but combining them with CMB + BAO + SNe drastically reduces the degeneracies, and  provides interesting constraints.

\subsubsection{Accuracy on $n$}

We finally study the accuracy that can be obtained on the parameter $n$. Running MCMC chains adding also this parameter to the $(\Xi_0,w_0)$ pairs starts to become computationally  quite expensive.
We will then begin with 
a simple order-of-magnitude estimate. As a first approximation, we can ask what variation $\Delta n$ would induce a change on $d_L^{\,\rm gw}(z)/{d_L^{\,\rm em}(z)}$ in \eq{eq:fitz}, of the same order as a given variation $\Delta \Xi_0$, i.e. we impose 
\bees
&&\hspace*{-10mm}\left| \frac{\pa}{\pa n} \( \Xi_0 +\frac{1-\Xi_0}{(1+z)^n} \)  \right| \Delta n\nn\\
&&=
\left| \frac{\pa}{\pa \Xi_0} \( \Xi_0 +\frac{1-\Xi_0}{(1+z)^n} \)\right| \Delta\Xi_0\, .
\ees
This gives 
\be\label{Deltanzgeneric}
\Delta n =\frac{\Delta \Xi_0}{|1-\Xi_0|}\, \frac{(1+z)^n-1}{\log (1+z)}\, .
\ee
Note that this expression correctly reproduces the fact that, for $\Xi_0\ra 1$, the dependence on $n$ disappears in \eq{eq:fitz}, so $n$ becomes more and more difficult to measure. The above result is valid
for a single source at redshift $z$. 
For a population of sources, we should average the redshift-dependent term in the right-hand side over the redshifts of the sources. As we will see in the next section,  for ET the result is dominated by sources with low redshift, so an expansion of this term to first order in $z$ is a reasonable first approximation. In that case we get
\be\label{Deltan}
\frac{\Delta n}{n}\simeq \frac{\Delta \Xi_0}{|1-\Xi_0|}\, .
\ee
As an example, for the minimal RR model $\Xi_0\simeq 0.970$ and, using  $10^3$ standard sirens combined with CMB+BAO+SNe, we have found  that, when using $(\Xi_0,w_0)$ to parametrize the DE sector,  $\Delta \Xi_0\simeq 0.008$. In this case, 
the estimate (\ref{Deltan}) suggests that we should
be able to measure $\Delta n$ to an accuracy $\Delta n/n\simeq 0.3$. Note however that this is the accuracy that would be obtained by keeping $\Xi_0$ fixed, and using only $n$ as a free parameter. In practice, one will have to vary both parameters simultaneously, and the resulting degeneracies would induce significantly larger errors.

\begin{figure}[t]
\includegraphics[width=0.4\textwidth]{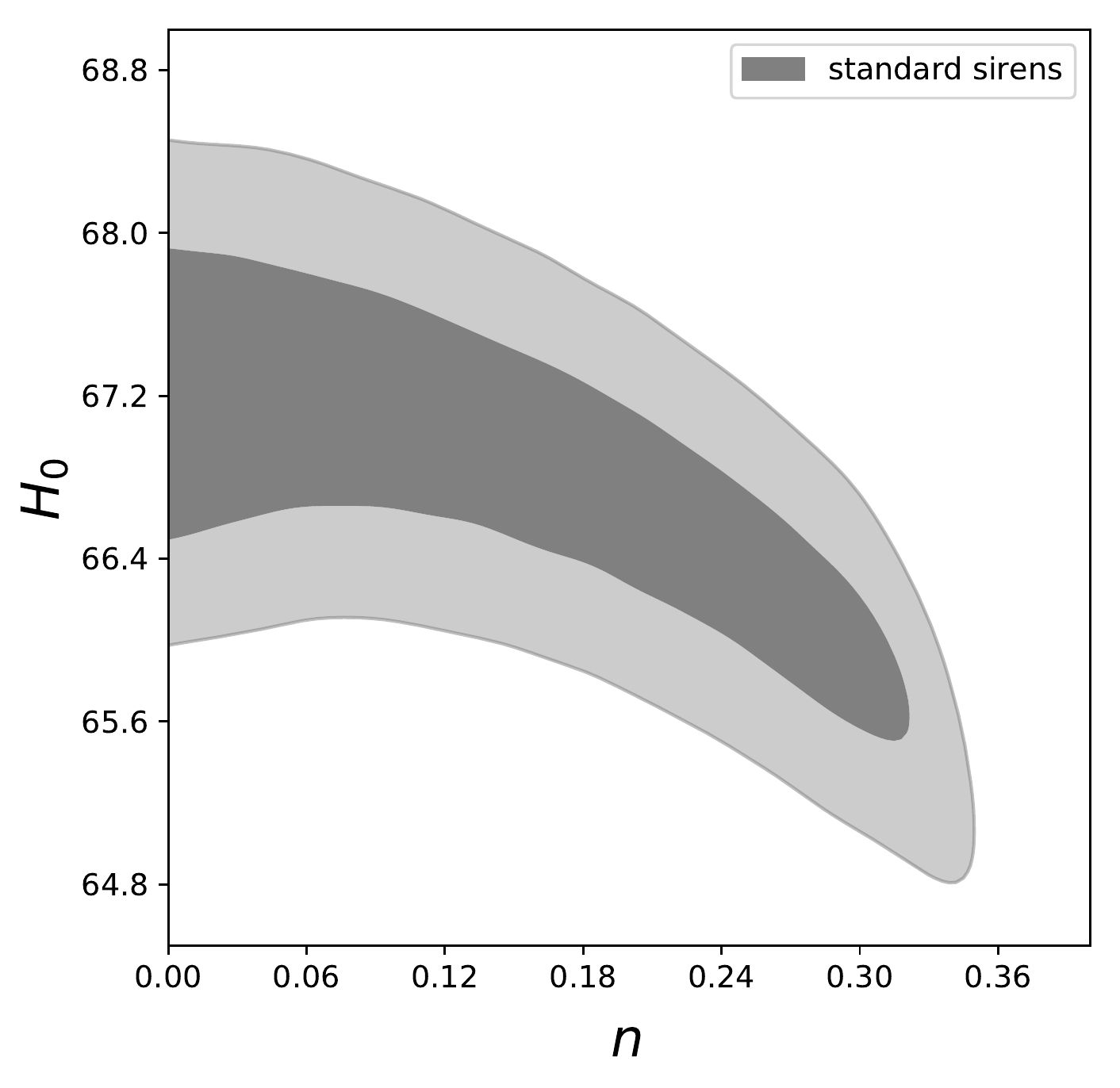}
\caption{The  two-dimensional likelihood in the $(n,H_0)$ plane, from  $10^3$ standard sirens at ET. \label{fig:nH0_xi0}}
\end{figure}

It is interesting to study more in detail the limiting case $\Xi_0\ra 0$, that, as discussed above \eq{paramXi0zero}, corresponds to the case of constant $\delta$. We have then run a further MCMC setting $\Xi_0=0$, $w_0=-1$, $w_a=0$ and using $n$ as the only free parameter in the DE sector, restricted to the range $n\geq 0$. The corresponding two-dimensional likelihood in the $(n,H_0)$ plane is shown in Fig.~\ref{fig:nH0_xi0}, using only standard sirens. 
Given that the mock standard sirens catalog has been generated assuming $\Lambda$CDM, we find that the one-dimensional likelihood for $n$ is peaked very close to $n=0$, and, at the $1\sigma$ level, we get the bound
\be
n\leq 0.221\, .
\ee
Note that, since we have fixed $\Xi_0$, th error on $n$ can no longer be related 
to $\Delta \Xi_0$, as in \eq{Deltanzgeneric}. Rather, the effect of 
the degeneracies with $H_0$ and $\oma$, that when we vary $\Xi_0$ is responsible for $\Delta\Xi_0$, is now responsible for the error on  $n$.

\section{Testing the RR model with ET}\label{sect:RRatET}

In the previous section we considered different subsets of the   $(w_0,w_a,\Xi_0,n)$ parametrization and we  fixed the number of standard sirens to a plausible value, to compute the corresponding accuracy that can be obtained on these parameters. In this section we rather consider a specific modified gravity model, namely the RR nonlocal model, and we  ask what is the minimum number of standard sirens  required to distinguish it from $\Lambda$CDM with ET. We have seen below \eq{eq:fitz} that the model predicts 
$\Xi_0\simeq 0.970$ (when a parameter $u_0=0$, see below), so a measurement at a level $\Delta \Xi_0=0.03$ or better is necessary. From \eq{ourDeltaXi0Deltaw0Deltawa} we already understand that $10^3$ standard sirens should indeed be sufficient. However, having at our disposal a concrete model, with a specific prediction both for $d_L^{\,\rm gw}(z)/d_L^{\,\rm em}(z)$ and for $\wde(z)$, and using the results for the Bayesian parameter  estimation for this model from CMB+BAO+SNe data performed in \cite{Belgacem:2017cqo},
allows us to perform a more detailed study, as a function of the number of sources.

\subsection{The model}

The field-theoretical motivations for the model, the conceptual issues related to the introduction of nonlocal terms, as well as its cosmological predictions, have been discussed in detail in \cite{Maggiore:2016gpx,Belgacem:2017cqo}. Here we simply mention that 
the basic physical idea is that, even if the fundamental action of gravity is local, the corresponding quantum effective action, 
that includes the effect of quantum fluctuations, is  nonlocal. 
These nonlocalities are well understood in the ultraviolet regime, where their computation is by now standard textbook material~\cite{Birrell:1982ix,Mukhanov:2007zz,Shapiro:2008sf}, but are much less understood in the infrared (IR), which is the regime relevant for cosmology.  IR effects in quantum field theory in curved space have been studied particularly in de~Sitter space where strong effects, due in particular to the propagator of the conformal mode \cite{Antoniadis:1986sb}, have been found. It has a priori possible that  quantum fluctuations in the IR generate in the quantum effective action nonlocal terms associated to a dynamically generated  mass scale (see in particular the discussion in Section~2.3 of \cite{Belgacem:2017cqo}), and proportional to inverse powers of the d'Alembertian operator. For instance, in QCD  the strong IR fluctuations generate a term~\cite{Boucaud:2001st,Capri:2005dy,Dudal:2008sp} 
$(m_g^2/2) {\rm Tr}\, \int d^4x\,   F_{\mu\nu} \iBox F^{\mu\nu}$, that corresponds to a dynamical generation of a gluon mass. These considerations have suggested the study of a model whose quantum effective action is
\be\label{RR}
\Gamma_{\rm RR}=\frac{\mplr^2}{2}\int d^{4}x \sqrt{-g}\, 
\[R-\frac{1}{6} m^2R\frac{1}{\Box^2} R\]\, ,
\ee
where $\mplr$ is the reduced Planck mass and  $m$ is a  new mass parameter that replaces the cosmological constant of $\Lambda$CDM. This model was  proposed in \cite{Maggiore:2014sia}, following earlier work in \cite{Maggiore:2013mea}. The nonlocal term in \eq{RR} corresponds to a dynamical mass generation for the conformal mode of the metric~\cite{Maggiore:2015rma,Maggiore:2016fbn}. 
Recently, indications in favor of this term has also  been found from lattice gravity~\cite{Knorr:2018kog}.
The background evolution and cosmological perturbations of the model have been studied in several works~\cite{Maggiore:2014sia,Foffa:2013vma,Barreira:2014kra,Barreira:2015fpa,Kehagias:2014sda,Dirian:2014bma,Dirian:2016puz,Nersisyan:2016hjh,Dirian:2017pwp,Belgacem:2017cqo}, and it has been shown that the model fits the present cosmological data remarkably well, at the same level as $\Lambda$CDM (see \cite{Belgacem:2017cqo} for review).\footnote{A different class of nonlocal models, where the nonlocal terms is not associated to a mass scale, has been discussed in \cite{Wetterich:1997bz,Deser:2007jk,Woodard:2014iga,Nersisyan:2017mgj} and in \cite{Barvinsky:2003kg,Barvinsky:2011hd,Barvinsky:2011rk}.}

A technical point that will be relevant in the following is that the nonlocal model can be formally written in local form by introducing two auxiliary fields $U$ and $S$ 
defined by $U=-\iBox R$ and $S=-\iBox U$, so that eventually one has to deal with a modified Einstein equation, depending on the fields $U$ and $S$, supplemented by the equations $\Box U=-R$ and $\Box S=-U$. At the conceptual level, it is important to understand that these auxiliary fields do not represent genuine new degrees of freedom; rather, their initial conditions are in principle fixed in terms of the initial conditions on the metric (see the discussion in Section~2.4 of \cite{Belgacem:2017cqo}). If one had a first-principle derivation of the nonlocal quantum effective action from the fundamental local theory, one would be able to compute this relation explicitly.\footnote{Indeed, in two-dimensional gravity, the quantum effective action induced by conformal matter fields can be computed exactly, by integrating the conformal anomaly, and gives rise to the well-known Polyakov quantum effective action, proportional to $R\iBox R$. Even in this case one could in principle localize the theory by introducing $U=-\iBox R$. In this case, where everything is well under theoretical control, one can compute explicitly the relation between the initial condition of $U$ and that of the metric, see Section~2.4 of \cite{Belgacem:2017cqo}.} However, in the RR model we do not currently have a derivation of the nonlocal term from the fundamental theory, so in practice we must still face the problem of how to choose the initial conditions for these fields. Luckily it turns out that, out of the four initial conditions on $U,S,\dot{U}$ and $\dot{S}$, at the level of cosmological background evolution three parametrize irrelevant directions in parameter space, i.e. the solution obtained setting these initial conditions to zero is an attractor, and we only remain with the initial conditions $u_0$ on the field $U$, that is set deep in radiation dominance (RD), and that parametrizes a marginal direction in parameter space (i.e. a solution with $u_0\neq 0$ is neither  attracted by the solution with $u_0=0$, nor diverges exponentially from it).  

Thus, $u_0$ is a ``hidden" parameter of the model, that enters through the initial conditions.\footnote{It can also be shown that, at the level of cosmological perturbations, the initial conditions on the perturbations of the auxiliary fields, once taken to be of the order of the initial perturbations of the metric (as expected from the fact that the initial conditions of the auxiliary fields are fixed in terms of those of the metric), give an effect which is numerically irrelevant, and therefore do not introduce further freedom~\cite{Belgacem:2017cqo}.}  We will first study the model defined by $u_0=0$, that  we will call the ``minimal" RR model. However, a large value of $u_0$ at the beginning of RD could be generated by a preceding inflationary phase, which could lead to a value of $u_0$ of order $4\Delta N$, where $\Delta N$ is the number of inflationary e-folds~\cite{Foffa:2013vma,Maggiore:2016gpx}, so for $\Delta N\simeq 60$ we could have $u_0\simeq 240$. It is therefore important to study also the case where $u_0$ is large, say ${\cal O}(250)$.
Unfortunately, the model with large positive values of $u_0$ gets closer and closer to $\Lambda$CDM, and is therefore more difficult to distinguish from it. Still, we will see that even the model with such a large value of $u_0$ could be potentially detectable with a sufficiently large, but not unrealistic, number of standard sirens.\footnote{The RR model with such large values of $u_0$ could also be distinguished from $\Lambda$CDM with future surveys such as Euclid~\cite{Casas:inprep}.} In the following we will first discuss the prospects for 
discriminating the minimal models from $\Lambda$CDM with standard sirens combined with CMB+BAO+SNe, and we will then turn to the case of large $u_0$.

\subsection{Testing the ``minimal" RR model}

We wish to understand what is the minimum number of standard sirens  required to distinguish the RR model  from $\Lambda$CDM. To this purpose, we start by taking $\Lambda$CDM as our fiducial model. 
We have then generated $10^4$  samples each containing   1000 NS-NS binaries,
distributed in redshift according to \eq{deffz} and scattered in $d_L(z)$ according to the estimate (\ref{errortot})--(\ref{errorlensing}) of its  error at ET, with sources  from $z_{\rm min}=0.07$ up to $z_{\rm max}=2$. Given a set of simulated data $d_i\equiv d_L(z_i)$, with  error $\sigma_i\equiv \Delta  d_L(z_i)$, we can form the $\chi^2$ for $\Lambda$CDM 
\be
\chi^2_{\Lambda{\rm CDM}}=\sum_{i=1}^{N_s}
\frac{[d^{\Lambda\rm CDM}_L(z_i;H_0,\oma)-d_i]^2}{\sigma_i^2}\, ,
\ee
where $H_0=67.64$ and $\oma=0.3087$ are the fiducial values  for $\Lambda$CDM, obtained by fitting $\Lambda$CDM to CMB+BAO+SNe data.
Since the data $d_i$ have been extracted from a distribution that assumes that $\Lambda$CDM (with these values of $H_0$ and $\oma$)  is the correct model, by construction for large $N_s$ the reduced chi-square $ \chi^2_{\Lambda{\rm CDM}}/N_s$ will be of order one. Similarly, we can construct the $\chi^2$ for the prediction of the (minimal) RR model
\be\label{chi2RR}
\chi^2_{\rm RR}=\sum_{i=1}^{N_s}
\frac{[d^{\rm gw, RR}_L(z_i;H^{\rm RR}_0,\oma^{\rm RR})-d_i]^2}{\sigma_i^2}\, ,
\ee
where $d^{\rm gw, RR}_L$ is the GW luminosity distance of the RR model and
$H^{\rm RR}_0$ and $\oma^{\rm RR}$ are the best-fit values for the RR model obtained from CMB+BAO+SNe, $\oma=0.2993$ and $H_0=69.44$.\footnote{Of course, more accurately, one should compute a likelihood with the corresponding priors, both for $\Lambda$CDM and for the RR model. However, this would not affect significantly the conclusions.} 
Since the data have been generated according to $\Lambda$CDM, for sufficiently large $N_s$ the difference
\be\label{Dchi2}
\Delta\chi^2=\chi^2_{\rm RR}-\chi^2_{\Lambda{\rm CDM}}
\ee
will become sufficiently large to rule out the RR model. We want to compute the minimum value of $N_s$ for which $\Delta\chi^2$ goes above a threshold for which one can say that 
$\Lambda$CDM fits the data significantly better than RR. For definiteness, we take this threshold value to be  equal to 6. This  choice is motivated by the fact that, when the likelihood is Gaussian and for models with the same number of parameters, a conventional scale, slightly more conservative than the Jeffreys scale, states that $|\Delta \chi^2| \leq  2$ implies  statistical equivalence between the two models, $2\,\lsim\, |\Delta \chi^2|\,\lsim\, 5$ suggests ``weak evidence'' in favor of the model with lower $\chi^2$, and $5\, \lsim \, 
|\Delta \chi^2|\, \lsim 10$ indicates ``moderate to strong evidence'' in favor of the model with lower $\chi^2$~\cite{Trotta:2008qt}.   
We can of course also reverse the process, generating  the data according to the GW luminosity distance of the RR model, and ask what is the minimum value of $N_s$ that is required to rule out $\Lambda$CDM, to the same significance. We have found that the procedure is completely symmetric, within our statistical uncertainty, and for definiteness we show the plots obtained by using $\Lambda$CDM as the fiducial model.

\begin{figure}[t]
\includegraphics[width=0.4\textwidth]{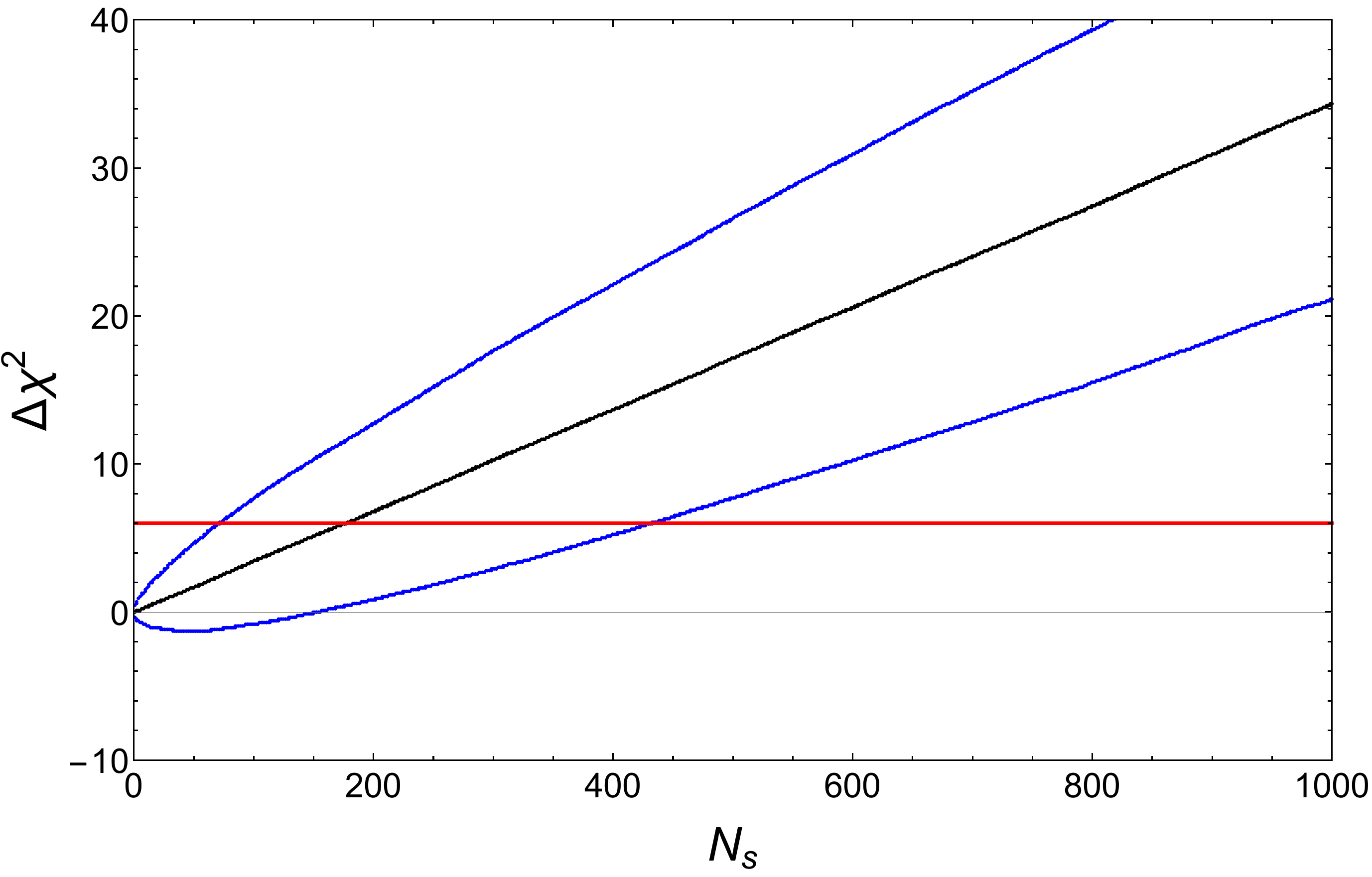}
\caption{Average and standard deviation of $\Delta\chi^2=\chi^2_{\rm RR}-\chi^2_{\Lambda{\rm CDM}}$. The horizontal line corresponds to the threshold value $\Delta\chi^2=6$.}
\label{fig:RRsigma}
\end{figure}

\begin{figure}[t]
\includegraphics[width=0.4\textwidth]{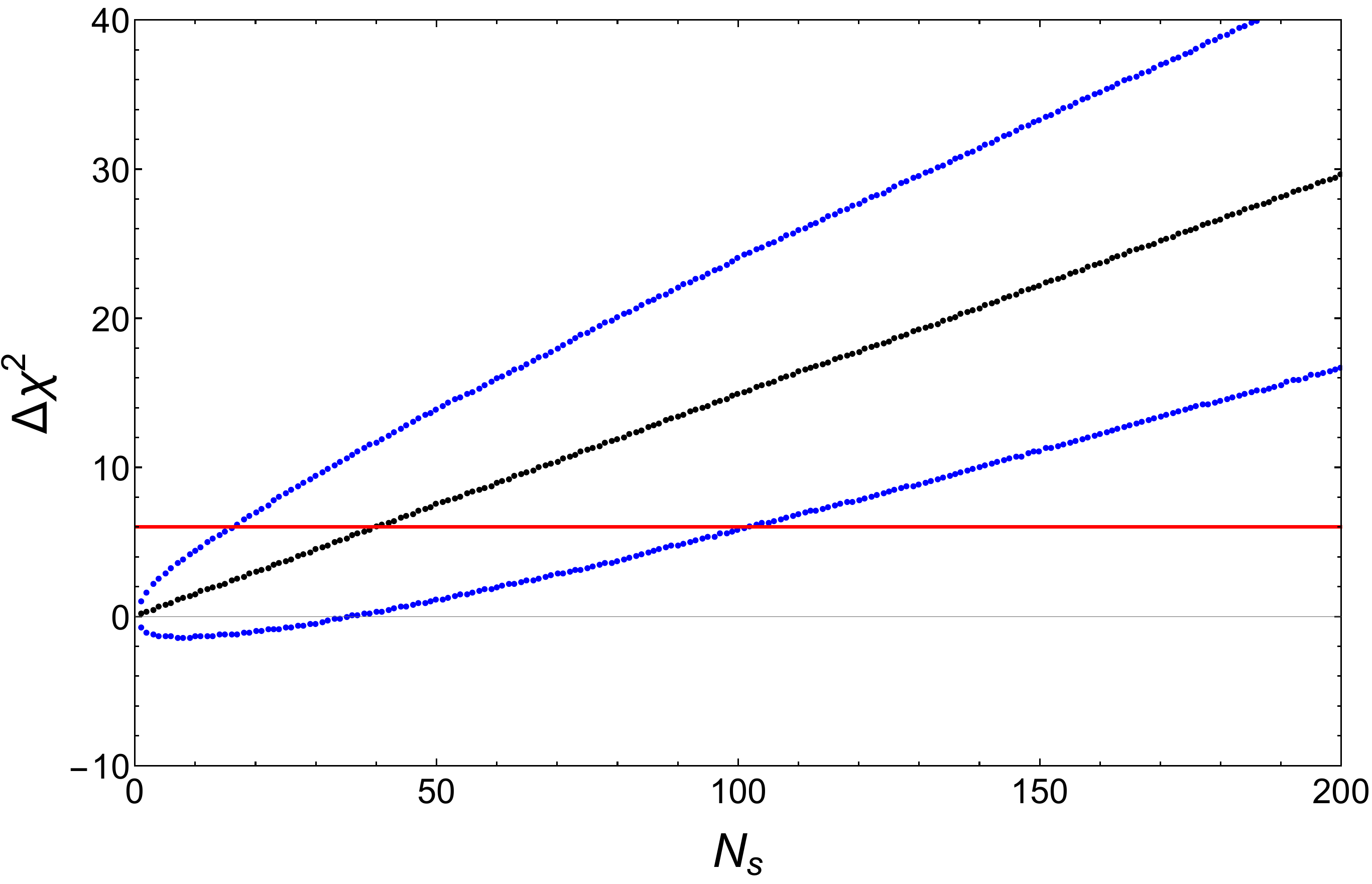}
\caption{As in Fig.~\ref{fig:RRsigma}, restricting to sources with redshift $0.07<z<0.7$.}
\label{fig:RRsigmazlow}
\end{figure}

\begin{figure}[t]
\includegraphics[width=0.4\textwidth]{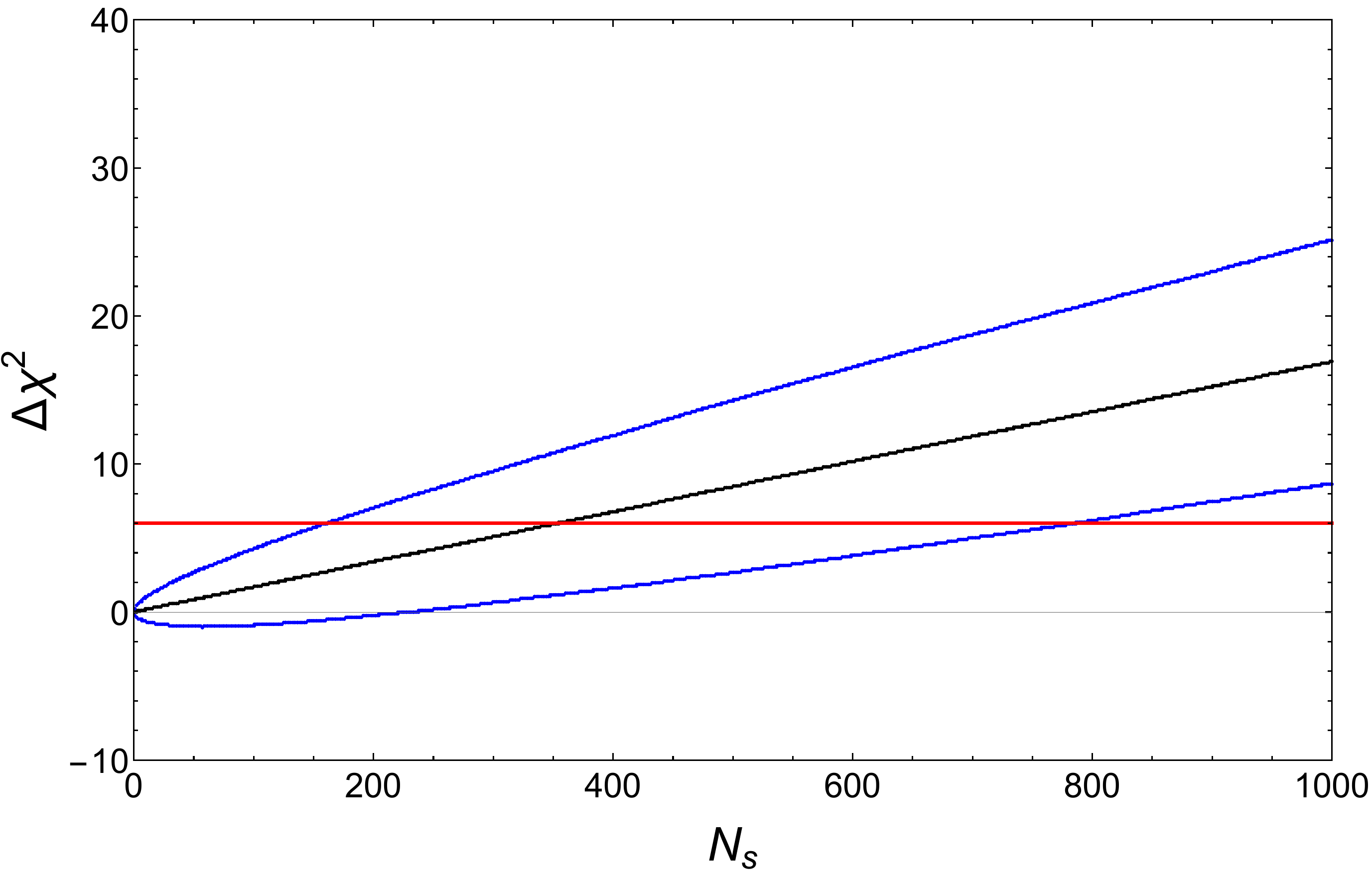}
\caption{As in Fig.~\ref{fig:RRsigma}, restricting to sources with $0.7<z<2$.}
\label{fig:RRsigmazhigh}
\end{figure}

\begin{figure}[t]
\includegraphics[width=0.4\textwidth]{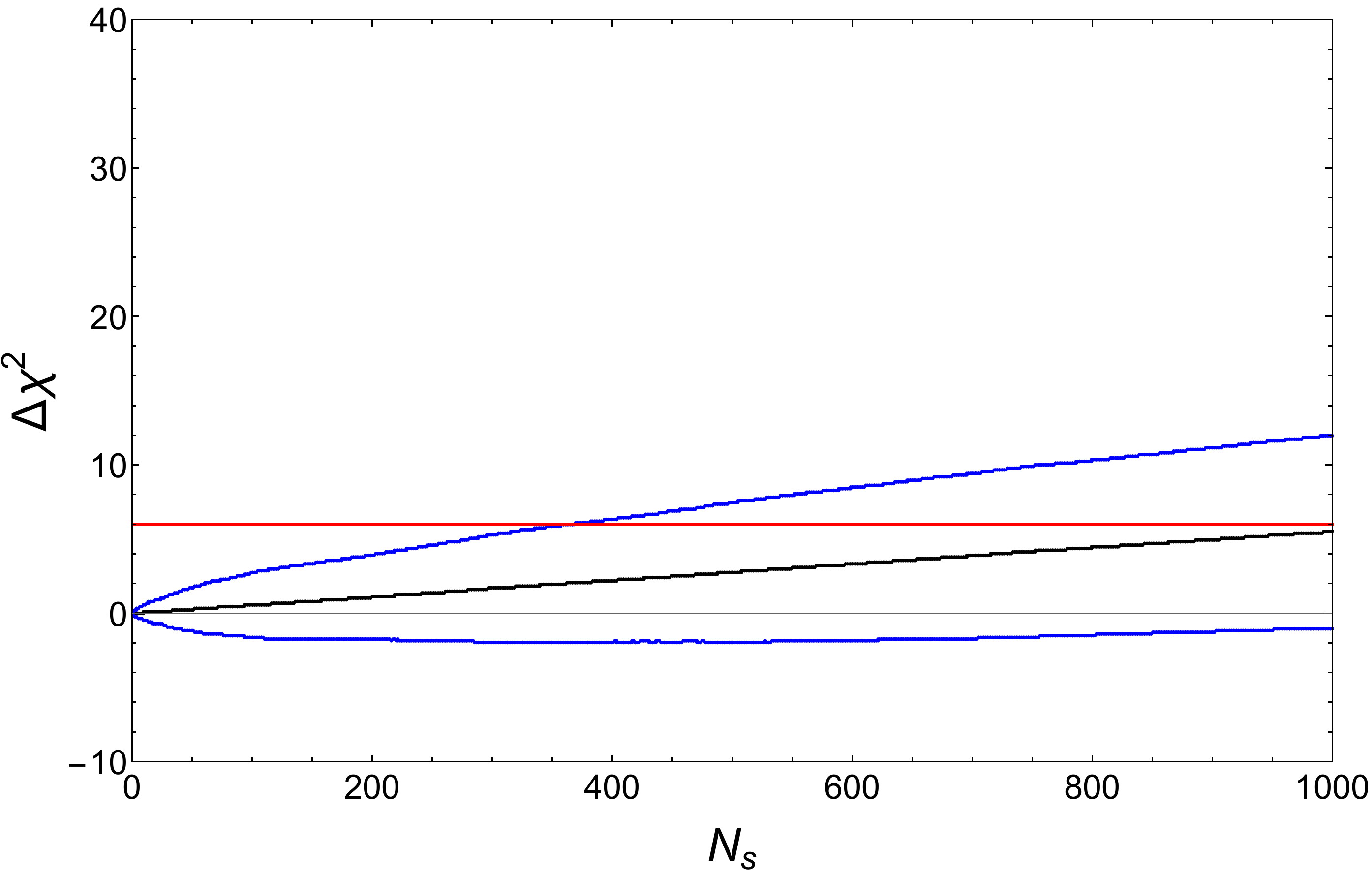}
\caption{As in Fig.~\ref{fig:RRsigma}, using the electromagnetic luminosity distance of the RR model.}
\label{fig:RRem}
\end{figure}

The result is shown in Fig.~\ref{fig:RRsigma}. First of all we observe that  $\Delta\chi^2$ has a significant variability among the $10^4$ realizations of the data  that we have generated. We therefore show, in Fig.~\ref{fig:RRsigma} and in the following figures, the average and the standard deviation of  $\Delta\chi^2$ over these realizations, along with the reference line $\Delta\chi^2=6$. We see that, on average, we need about $200$ standard sirens to tell the two models apart. However, because of the large variability of $\Delta\chi^2$, to exclude that the result is due to statistical fluctuations one would rather need about 400 sources in the pessimistic case. On the other hand, from the variability of $\Delta\chi^2$ over the different realizations, it also follows that in the more optimistic case $O(50)$ standard sirens could already give a highly significant value of $\Delta\chi^2$. 

In order to understand which sources contribute most to this result, we have repeated the analysis limiting ourselves  to sources with redshift $0.07<z<0.7$, and to sources with $0.7<z<2$. The results for 
$\Delta\chi^2$ obtained with sources with  $0.07<z<0.7$ is  shown in Fig.~\ref{fig:RRsigmazlow}, and the result from sources with $0.7<z<2$ is shown in Fig.~\ref{fig:RRsigmazhigh}.
We see that, on average, it is enough to a have about 40 standard sirens at $0.07<z<0.7$, or
about  350 at $0.7<z<2$, to tell the two models apart. Depending on the specific realization, in the most optimistic case it is sufficient to have about 15 standard sirens at $0.07<z<0.7$, or
about  150 at $0.7<z<2$, while in the most pessimistic case we need about 100 standard sirens at $0.07<z<0.7$, or
about  800 at $0.7<z<2$. These results fully confirm the conclusions of the simpler analysis performed in
\cite{Belgacem:2017ihm}. The fact that standard sirens in the `low-$z$' range  $0.07<z<0.7$ gives the dominant contribution can be understood from the fact that, as we see from Fig.~\ref{fig:dLgw_over_dLem},
the ratio $d_L^{\,\rm gw}(z)/d_L^{\,\rm em}(z)$  basically saturates to a constant value beyond 
$z\, \gsim \, 1$ or, equivalently, the function $\delta(z)$ is by now quite close to zero at $z\, \gsim \, 1$, see Fig.~\ref{fig:delta_vs_z}. Thus,  going at larger redshifts the signal from modified GW propagation does not increase further, despite the propagation across longer distances, while the relative error in the luminosity distance 
increases approximately linearly with $z$  (apart from small quadratic and cubic corrections), see \eq{errorET}.

It is  interesting to ask how much this result is affected by the DE EoS of the RR model, and how much by modified GW propagation.  To understand this point, we can artificially switch off the effect of modified GW propagation by using $d^{\rm em, RR}_L$ instead of $d^{\rm gw, RR}_L$
in \eq{chi2RR}. The corresponding result is shown in Fig.~\ref{fig:RRem}. 
We see that now the required number of sources is significantly higher. Indeed,  now
the average $\Delta\chi^2$ goes above the threshold only with about 1000 sources,
while we found in Fig.~\ref{fig:RRsigma} that, including also modified GW propagation,  200 sources are enough. Note also that, without the effect of modified GW propagation,
the lower $1\sigma$ fluctuation does not even have a positive $\Delta\chi^2$  with 1000 sources. This clearly shows the importance of the effect of modified GW propagation.

\subsection{The model for large values of $u_0$}

\begin{figure}[t]
\includegraphics[width=0.4\textwidth]{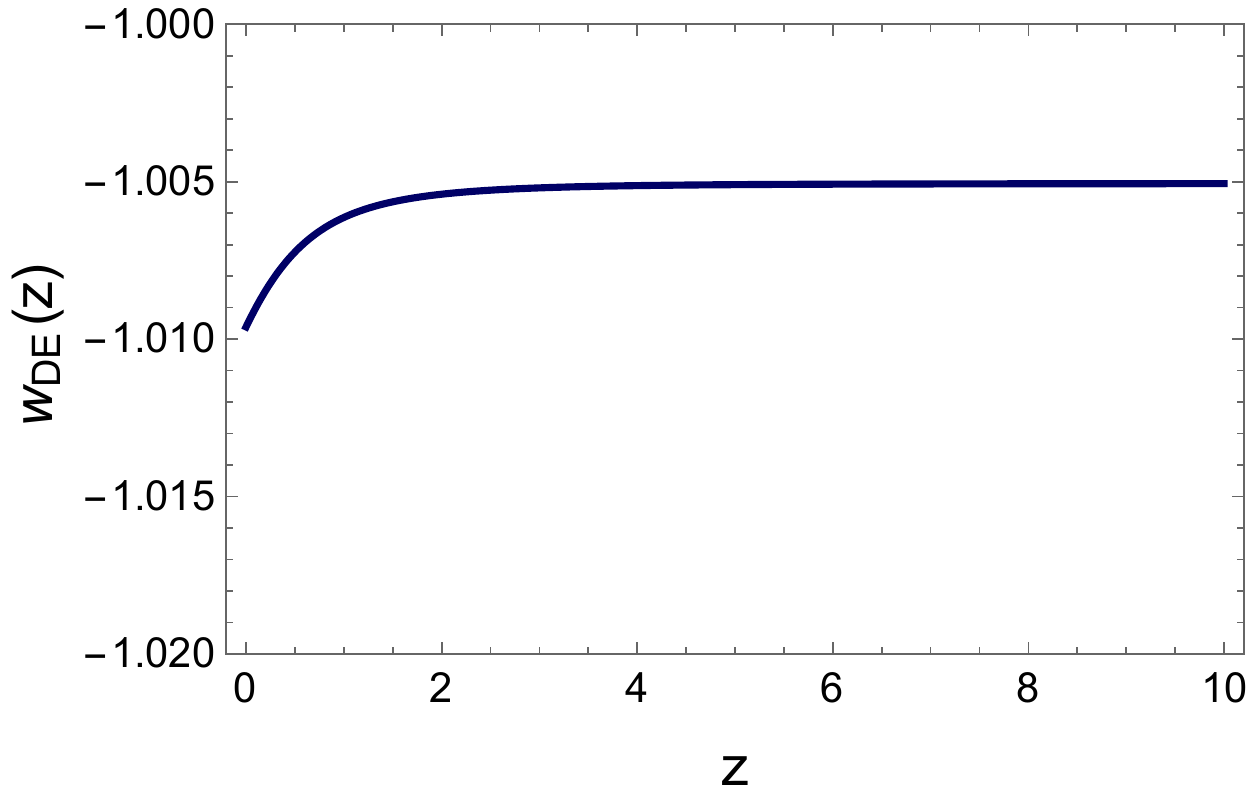}
\caption{The DE EoS in  the RR nonlocal model with $u_0=250$. }
\label{fig:wdeRR250}
\end{figure}

\begin{figure}[t]
\includegraphics[width=0.4\textwidth]{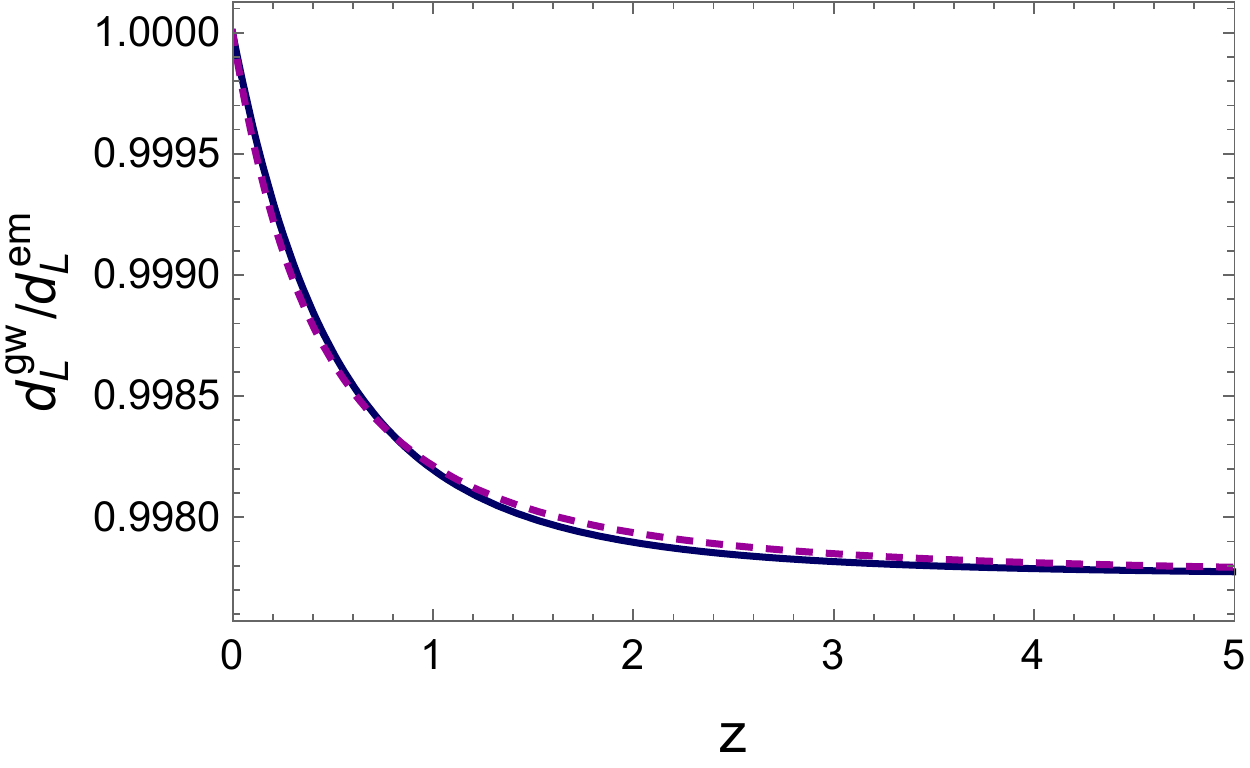}
\caption{The ratio $d_L^{\,\rm gw}(z)/d_L^{\,\rm em}(z)$ in  the RR nonlocal model with $u_0=250$, as a function of redshift (blue solid line) compared to the fitting function (\ref{eq:fitz}) with $n=2.3$ (dashed).}
\label{fig:fit_dLgw-dLem_vs_a_u0250}
\end{figure}

We now repeat the analysis for large values of the parameter $u_0$.  In Fig.~\ref{fig:wdeRR250} we show $\wde(z)$ for $u_0=250$. We see that it is now much closer to the $\Lambda$CDM value $-1$. Similarly,
Fig.~\ref{fig:fit_dLgw-dLem_vs_a_u0250} shows the ratio $d_L^{\,\rm gw}(z)/d_L^{\,\rm em}(z)$, as well as  the fitting function (\ref{eq:fitz}) with $\Xi_0=0.9978$ and 
$n=2.3$. Note that in this case $\delta(0)\simeq 4.3\times 10^{-3}$.

\begin{figure}[t]
\includegraphics[width=0.4\textwidth]{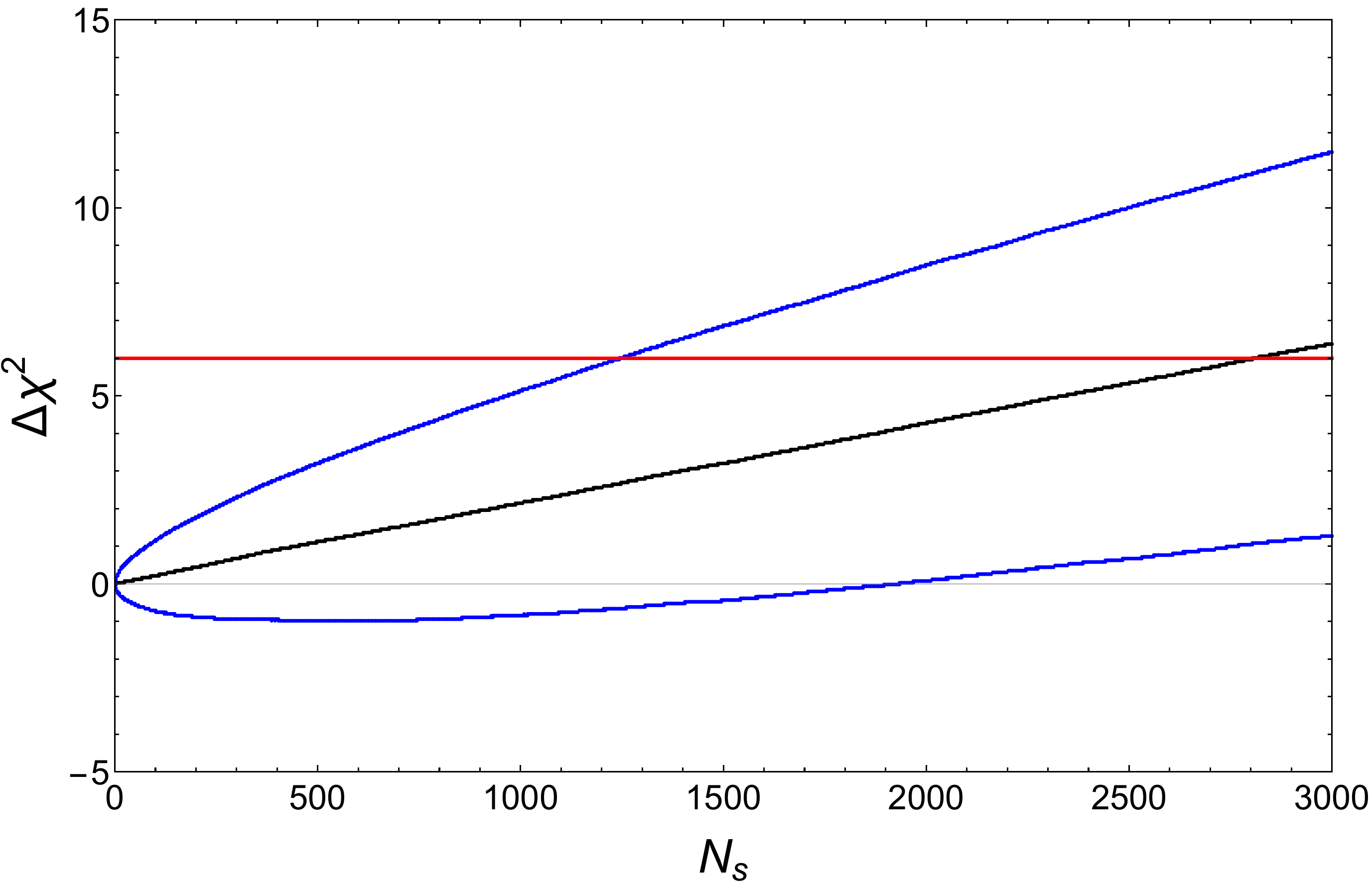}
\caption{As in Fig.~\ref{fig:RRsigma}, for $u_0=250$. }
\label{fig:RRsigma_u0250}
\end{figure}

We can repeat the same analysis as in the case $u_0=0$, and determine, for a given value of $u_0$, the minimum number of standard sirens required to tell the model apart from $\Lambda$CDM, using ET. Fig.~\ref{fig:RRsigma_u0250} shows the result for $u_0=250$. Now, on average, almost 3000 sources are needed (raising to about 6000 in the most pessimistic case). This is a large number of sources, but still within the number of standard sirens with electromagnetic counterpart that could be observed with ET, depending on the precise sensitivity (as well as on the capabilities of  future  $\gamma$-ray networks). Furthermore, in this paper we have limited ourselves to standard sirens with an electromagnetic counterpart, but further
information can be obtained using  statistical methods, even in the absence of counterparts, see footnote~\ref{note:statmeth}.

\section{Primordial GWs and modified transfer function}\label{sect:transfer}

A further consequence of modified GW propagation is that the GW transfer function that connects a primordial GW spectrum to that observed at the present epoch is modified.
Recall that the transfer function is defined by 
\be\label{4defTGW12}
\tilde{h}_A(\eta_0,k)=T_{\rm GW}(k)\tilde{h}_A(\eta_{\rm in},k)\, ,
\ee
where $\eta_0$ is the present value of conformal time and $\eta_{\rm in}$ the initial value at which a primordial GW spectrum is generated. Basically, in GR the transfer function is determined by the fact that, as long as a tensor mode is outside the horizon, it stays constant, while when it enters inside the horizon it scales as $1/a(\eta)$ times oscillating factors (see e.g. Section~19.5 of \cite{Maggiore:2018zz}). Therefore
\be\label{4tildeh2eta0}
\tilde{h}^2(\eta_0,\vk)\simeq\frac{1}{2}\tilde{h}^2(\eta_{\rm in},\vk) \(\frac{a_*(k)}{a_0}\)^2\, ,
\ee
where  $a_*(k)$ is the value of the scale factor when the mode with wavenumber $k$ re-enters the horizon, 
$a_0$ is the present value of the scale factor, and the  $1/2$ comes from the average over the oscillating factors. A more  accurate expression can be obtained following numerically the evolution  across  the  super-horizon and sub-horizon regimes.

If the GW propagation is modified as in \eq{4defhchiproofsRR}, inside the horizon the GW amplitude scales as $1/\tilde{a}$ rather than $1/a$. As a result, the transfer function in modified gravity is related to the GR transfer function by
\bees
T^{\rm mod\,  grav}(k)&=&\frac{\tilde{a}_*(k)}{\tilde{a}_0}\, \frac{a_0}{a_*(k)}\, T^{\rm GR}(k)\nn\\
&=&\frac{\tilde{a}_*(k)}{a_*(k)}T^{\rm GR}(k)
\, ,
\ees
where, as in \eq{dLtilde}
we can set  $\tilde{a}(0)=a(0)=1$ without loss of generality. Similarly to \eq{dLgwdLem}, we can rewrite this as
\be
T^{\rm mod\,  grav}(k)=T^{\rm GR}(k)\exp\left\{\int_0^{z_*(k)} \,\frac{dz'}{1+z'}\,\delta(z')\right\}\, .
\ee
In a model where $\delta (z)$ goes to zero at large redshifts, as the RR model, the integral saturates to its asymptotic value  already at small values of $z$, as in Fig.~\ref{fig:dLgw_over_dLem}, so it is equal to its asymptotic value $1/\Xi_0$, and is independent of $k$, 
\bees
T^{\rm mod\,  grav}(k)&\simeq&T^{\rm GR}(k)\exp\left\{\int_0^{\infty} \,\frac{dz'}{1+z'}\,\delta(z')\right\}
\nn\\
&=&\Xi_0^{-1}\, T^{\rm GR}(k)\, .
\ees
The factor $\Xi_0^{-1}$, that in the RR model is larger than one, therefore enhances the  GW amplitude compared to the GR predictions. 

For computing the evolution of the energy density in GWs we must  also take into account that $G$ is replaced by $G_{\rm eff}(z)$. As discussed in Section~\ref{sect:Geff}, in the RR model, as well as in a broader class of modified gravity theories, once this effect is included  the GW energy density  satisfies  the usual scaling $\rgw\sim 1/a^4$, as in GR. 

Thus, if we consider for instance the  GW stochastic background generated by  inflation in a generic modified gravity theory, there could in principle be two kind of modifications. First, there could be a difference in the production mechanism, because of a possible modified dynamics at the inflationary epoch; and, second, $\rgw$ could evolve differently from the standard $1/a^4$ behavior of GR.
In the RR model, however,  the effect of the nonlocal term at the inflationary scale is utterly negligible~\cite{Maggiore:2016gpx,Cusin:2016mrr}, so there is no modification in the generation of primordial GWs. Furthermore, we have seen  in Section~\ref{sect:Geff} that, thanks to the relation (\ref{dLgwdLemGeff}), the behavior $\rgw(a)\propto 1/a^4$ is preserved. These properties will be shared by  all modified gravity theories where the dynamical dark energy term  responsible for the late-time acceleration is negligible at the inflationary scale, and where furthermore \eq{dLgwdLemGeff} holds. As we have seen, the latter is a general consequence of graviton number conservation. If these conditions are met, the prediction for the present value of the energy density of the  GW stochastic background  generated by inflation will be unaffected.

A  correction due to modified GW propagation however appears  whenever the relevant quantity is the GW amplitude, rather that the GW energy density. In particular,  in the computation of the temperature anisotropies in the direction $\hatn$,   the tensor contribution to ISW effect  is given by
\be\label{isw}
\frac{\d T}{T}(\eta_0,\hatn)=-\frac{1}{2} n^in^j\int_{\eta_{\rm dec}}^{\eta_0}d\eta\, 
\(\frac{\pa\hTTij(\eta, \vx)}{\pa\eta}\)\Big|_{\vx=(\eta_0-\eta)\hatn}\, ,
\ee
where $\hTTij(\eta, \vx)$  is the transverse-traceless metric perturbation at conformal time $\eta$ and position $\vx$,  the prime is the partial derivative with respect to $\eta$, and $\eta_{\rm dec}$ is the conformal time at decoupling. The amplitude $\hTTij(\eta, \vx)$ is computed by evolving the  primordial GW amplitude  up to conformal time $\eta$. If,  in a given modified gravity theory,    the amplitude at the production epoch is unaffected, then its value at conformal time $\eta$ will differ from the value in GR by a factor
\bees
\alpha(\eta)&\equiv&\exp\left\{\int_{z(\eta)}^{z_*(k)} \,\frac{dz'}{1+z'}\,\delta(z')\right\}\nn\\
&&\simeq \exp\left\{\int_{z(\eta)}^{\infty} \,\frac{dz'}{1+z'}\,\delta(z')\right\}\, ,\label{defXidiz}
\ees
so that \eq{isw} can be written as
\bees\label{iswmod}
\frac{\d T}{T}(\eta_0,\hatn)&=&-\frac{1}{2} n^in^j\int_{\eta_{\rm dec}}^{\eta_0}d\eta\, \\
&&\times \(\frac{\pa[\alpha(\eta)h_{ij}^{\rm TT, gr}(\eta, \vx)]}{\pa\eta}\)\Big|_{\vx=(\eta_0-\eta)\hatn}\, ,\nn
\ees
where $h_{ij}^{\rm TT, gr}(\eta, \vx)$ is the value computed in GR, by evolving to conformal time $\eta$ a given primordial perturbation.

\section{Conclusions}\label{sect:concl}

Studies of dark energy and modified gravity have usually focused on the dark energy equation of state $\wde(z)$, that  affects the cosmological evolution at the background level, or  on the modification of  scalar perturbations, that affect for instance   the growth of structure or weak lensing.  The main message of this paper is that there is another potentially very interesting observable, that parametrizes deviations from general relativity at the level of  tensor perturbations, and is given by the ratio
of the GW luminosity distance to the standard electromagnetic luminosity distance, $d_L^{\,\rm gw}(z)/d_L^{\,\rm em}(z)$. This observable is relevant for GW experiments, that can probe the expansion of the Universe through standard sirens.

The fact that GW propagation in modified gravity can be different from GR has already been recognized in a number of papers~\cite{Deffayet:2007kf,Saltas:2014dha,Gleyzes:2014rba,Lombriser:2015sxa,Arai:2017hxj,Nishizawa:2017nef,Belgacem:2017cqo,Belgacem:2017ihm,Amendola:2017ovw,Linder:2018jil,Pardo:2018ipy}. 
In this paper we have performed a detailed quantitative study of this effect. We have seen in
Section~\ref{sect:GW170817} that a first, although not very stringent, limit  on modified GW propagation already comes from the LIGO/Virgo measurement of the  luminosity distance  of the NS binary coalescence GW170817. We have then compared with the capabilities of the proposed Einstein Telescope, both using standard sirens alone and combining  them with other cosmological datasets, such as  CMB, BAO and SNe, in order to break the degeneracies between cosmological parameters. 

We have started our investigation from an explicit modified gravity model, the so-called RR model, recently proposed and developed by our group, and that is motivated by the idea that quantum effects modify the quantum effective action of gravity in the infrared through a specific nonlocal term. This model has been shown to fit remarkably well the current cosmological data, and also provides an example of a concrete, predictive and well-motivated model where GW propagation is modified and the luminosity distance for GWs is different from that of electromagnetic waves (despite the fact that in this model GWs travel at the speed of light, therefore complying with the existing limits).

A significant result, that  we already anticipated in \cite{Belgacem:2017ihm} and we discussed in more detail in Section~\ref{sect:under}, is that in a generic modified gravity model [where the deviation of $d_L^{\,\rm gw}(z)/d_L^{\,\rm em}(z)$ from 1 is of the same order as the deviation of $\wde(z)$ from $-1$], the effect of 
$d_L^{\,\rm gw}(z)/d_L^{\,\rm em}(z)$ on standard sirens dominates on the effect of $\wde(z)$. This makes the detection of signs of the DE sector at GW interferometers easier than previously thought. We have confirmed this result  through explicit computations using MCMC, see in particular Fig.~\ref{fig:xi0w0}.

The explicit result of the RR model also suggested us a more general parametrization of  modified GW propagation, in terms of  the pair of parameters $(\Xi_0,n)$ given in \eq{eq:fitz}. This parametrization fits extremely well the explicit form of $d_L^{\,\rm gw}(z)/d_L^{\,\rm em}(z)$ in the RR model, while still being very simple, and is fixed naturally by the condition that $d_L^{\,\rm gw}(z)/d_L^{\,\rm em}(z)\ra 1$ as $z\ra 0$, and by the assumption that $d_L^{\,\rm gw}(z)/d_L^{\,\rm em}(z)$ goes to a constant at large redshift. The latter condition is expected to hold in a large class of models, where the function $\delta(z)$ defined in \eq{prophmodgrav} goes to zero at large redshift (as expected in models where DE starts to affect the dynamics only in the recent cosmological epoch). For these reasons, as well as for its simplicity, it can be considered as a natural  analogous of the $(w_0,w_a)$ parametrization for the DE EoS. Our analysis indicates that, as far as standard sirens are concerned, $\Xi_0$ and $w_0$ are the two most important parameters (with $\Xi_0$ actually being the single most important one), so, for standard sirens, a simple but still meaningful truncation of the parameter space of the DE sector can be to the pair $(\Xi_0,w_0)$.

Using as an example the estimated sensitivity of the Einstein Telescope, we have studied the accuracy that can be obtained on these parameters by combining standard sirens at ET with CMB, BAO and SNe data. The  corresponding two-dimensional likelihoods for 
$(\Xi_0,w_0)$, $(\Xi_0,H_0)$ and $(\Xi_0,\oma)$ are shown in Figs.~\ref{fig:xi0w0}--\ref{fig:xi0Oma}.
These results are quite encouraging. In particular, as shown in Section~\ref{sect:RRatET},
the predictions of the  ``minimal" RR model could be verified, or falsified, with just a few hundreds standard sirens with electromagnetic counterpart, a figure that is well within the potential sensitivity of the Einstein Telescope. Actually, even the version of the model where a ``hidden" parameter $u_0$ has a value ${\cal O}(250)$, which is much closer to $\Lambda$CDM, is testable with a number of standard sirens of order $10^3-10^4$, which is still consistent with expectations for ET.

The overall conclusion is that, as GW detectors will begin to detect standard sirens at higher and higher redshifts, as already possible with LIGO/Virgo at design sensitivity, and later with third generation ground-based detectors such as the Einstein telescope and with the space interferometer LISA,
modified GW propagation, as encoded e.g. in the parameter $\Xi_0$, will become a key observable.

\vspace{15mm}\noindent

{\bf Acknowledgments.} We thank  Nicola Tamanini for useful comments and Pierre Fleury and Lucas Lombriser for a useful discussion. We are particularly grateful to the referee for very useful comments.
The work  of the authors is supported by the Fonds National Suisse and  by the SwissMap National Center for Competence in Research.

\bibliography{myrefs_massive}

\end{document}